	\documentclass[twocolumn, amssymb,nobibnotes,aps,prb,nopacs]{revtex4}

	\usepackage{amsmath}
	\usepackage{amssymb}
	\usepackage{graphicx}
	\usepackage{subfig}
	\newcommand \dd[1]  { \,\textrm d{#1} }   
	\usepackage[justification=RaggedRight,font=footnotesize]{caption}
	\graphicspath{ {figs/} }
	\usepackage{multirow}
	\usepackage[usenames, dvipsnames]{color}
	\usepackage{textcomp}

\begin{document}

	\title{Head-on collision of nonlinear solitary solutions to Vlasov-Poisson equations}
	\author{S. M. Hosseini Jenab\footnote{Email: Mehdi.Jenab@umu.se}}
	\author{G. Brodin  \footnote{Email: Gert.Brodin@umu.se}}
	\affiliation{Department of Physics, Faculty of Science and Technology, Ume\aa \ University, 90187, Ume\aa , Sweden}
	\date{\today}

\begin{abstract}
	Nonlinear solitary solutions to the Vlasov-Poisson set of equations are
	studied in order to investigate their stability by employing a fully-kinetic
	simulation approach. 
	The study is carried out in the ion-acoustic regime for a collisionless, electrostatic 
	and Maxwellian electron-ion plasma.
	The trapped population of electrons is modeled based on 
	well-known Schamel distribution function.
	Head-on mutual collisions of nonlinear solutions are performed 
	in order to examine their collisional stability.
	The findings include three major aspects: 
	(I) These nonlinear solutions are found to be divided into three categories based on 
	their Mach numbers, i.e. stable, semi-stable and unstable.
	Semi-stable solutions indicates a smooth transition from stable to unstable solutions
	for increasing Mach number.
	(II) The stability of solutions is traced back to a condition imposed on
	averaged velocities, i.e. net neutrality.
	It is shown that 
	a bipolar structure is produced in the 
	flux 
	of electrons,
	early in the temporal evolution.
	This bipolar structure acts as the seed of the net-neutrality instability,
	which tips off the energy balance of nonlinear solution during collisions.
	As the Mach number increases, the amplitude of bipolar structure grows and 
	results in a stronger instability.
	(III) It is established that during mutual collisions,
	a merging process of electron holes can happen to a variety of degrees, 
	based on their velocity characteristics. 
	Specifically, the number of rotations of electron holes around each other
	(in the merging phase) varies.
	Furthermore, it is observed that in case of a non-integer number of rotations, two electron holes exchange 
	their phase space cores. 
\end{abstract}
\maketitle

\section{Introduction} \label{Sec_Introduction}
		Electron holes have been proven to be ubiquitous 
		in the Earth's magnetosphere via multiple satellite 
		observations~\cite{bale1998bipolar,franz1998polar,matsumoto1994electrostatic,ergun1998fast,deng2006observations,kojima1997geotail}.
		These space-borne  experiments display the existence of 
		electrostatic solitary waves (ESWs) which 
		implies holes in phase space.
		In 1956, Bernstein, Greene, and Kruskal\cite{bernstein1957exact} derived the solution of 
		the time-independent Vlasov-Poisson equations
		and proved the existence of general solutions for nonlinear
		plasma  waves  in  the  electrostatic  regime, which is known as BGK modes.
		BGK modes include localized (solitary), shock and double layers solutions.
		Their study established that \emph{trapped population} diversifies 
		nonlinear solutions, way beyond the limits of fluid theory predictions. 
		A broad class of nonlinear solutions, BGK modes, 
		can be constructed based on the BGK approach.
		This method, basically, divides the distribution function into two 
		parts, namely free and trapped populations. 
		The existence of trapped population makes the solution fundamentally nonlinear, 
		with no linear counterparts. 
		Hence, even in small-amplitude limit BGK solutions
		are not reduced to linear solutions~\cite{bohm1949theory, schamel2000hole,schamel2012cnoidal}. 
		In the years after BGK's seminal work, it was suggested that two more conditions
		had to be added to the Vlasov-Poisson equations in order to have 
		physically meaningful solutions, i.e. net-neutrality condition\cite{smith1970steady}
		and positiveness of distribution function\cite{schamel1971stationary}.
		These two conditions provide critical limitations 
		which make it easier to pinpoint the physically possible solutions, among the infinite number of BGK modes.
		Hence, the set of equations to be solved simultaneously can be listed as:
		\begin{align}
			v \frac{\partial f_s(x,v)}{\partial x} 
			+  \frac{q_s}{m_s} E(x,t) & \frac{\partial f_s(x,v)}{\partial v} 
			= 0, \ \ \  s = i,e \\
			\frac{\partial^2 \phi(x)}{\partial x^2} & = \sum_s -\frac{1}{\epsilon_0} q_s n_s(x). 
		\end{align}

		In general (when the fully time-dependent system is considered), the Vlasov
		equation for each species imply charge conservation, which for our case of
		one-dimensional spatial dependence reads $\partial \rho _{c}/\partial t+$ $%
		\partial j/\partial x=0$, where $\rho _{c}=\sum q_{s}\int f_{s}dv$ and $%
		j=\sum q_{s}  \overline{n_s v_s} $ which basically 
		a sum over flux of particles ($\overline{n_s v_s} = \int vf_{s}dv$).
		For time-independent solutions this implies $j=%
		\mathrm{const}$. However, considering cases where the constant current
		deviates largely from zero is unphysical, as this would imply a substantial
		magnetic field generation, invalidating an analysis based purely on the
		Vlasov-Poisson system. The most strict choice is to limit ourselves to $j=0$%
		, which for a hydrogen plasma ($q_{e}=-q_{e}$) reduces to 
		\begin{equation}
			\int vf_{e}dv=\int vf_{i}dv \label{eq_net_neutrality}.
		\end{equation}
		We do not need net neutrality to hold exactly, as a small deviation only
		implies a weak magnetic field that can be neglected, but Eq.~(\ref{eq_net_neutrality}) must apply
		at last approximately.
		
		Finally, we cannot allow for the distribution
		function to be negative , i.e. we must have 
		\begin{equation}
			f_{s}>0\ \ \  s = i,e.
		\end{equation} 
		  
		Thus, the full set includes the time-independent Vlasov equations  for each species (1),
		Poisson equation (2),
		the net-neutrality condition (3)
		and the positiveness of distribution function (4).
		Since we are focusing on the solitary (localized) solutions here,
		it is assumed that the electric potential approaches zero 
		far enough from the peak of the solution, $(\phi \longrightarrow 0$ when $x \longrightarrow \infty)$.

		Solving the above equations can be translated into finding two  functions, according to:
		\begin{enumerate}
			\item Distribution functions of each species $(f_i, f_e)$:
			Each should be a function of the total energy (kinetic and potential energy) 
			and has to be always positive.
			This applies to both free and trapped populations.
			It also should be continuous in the whole phase space domain. 			
			\item Sagdeev pseudo-potential function $(S)$: It originates from the Poisson's equation.
		\end{enumerate}

		The Sagdeev pseudo-potential approach\cite{davis1958structure,sagdeev1966cooperative} 
		(also known as the phase plane approach\cite{infeld2000nonlinear}) 
		transforms the Poisson's equation into an equation of the same form as for a particle moving in a potential. 

		Additionally, the flux of each species $(\overline{n_i v_i}, \overline{n_e v_e})$ 
		should almost cancel out, in accordance 
		with the net-neutrality condition (3).

		The BGK method approaches the problem by guessing a 
		solution for $\phi(x)$ and then finding the desired trapped distribution function.
		The BGK approach has been used to find shock solutions~\cite{montgomery1969shock, smith1970exact} 
		as well as solitary waves~\cite{turikov1984electron,muschietti1999analysis}.
		The BGK approach predicts the width and potential range that supports the nonlinear solitary solution.

		In contrast, Schamel~\cite{schamel_3} suggested a known trapped distribution function as the starting point,
		which results in the electric potential profile ($\phi(x)$). 
		Specifically, Schamel\cite{schamel1971stationary,schamel_2}
		proposed a trapped distribution function
		which is positive everywhere, and
		continuous in both velocity and spatial directions:
		\begingroup\makeatletter\def\f@size{8.3}\check@mathfonts
		\def\maketag@@@#1{\hbox{\m@th\large\normalfont#1}}%
		\begin{equation*}
		f_{s}(v) =  
		  \left\{\begin{array}{lr}
		    A \ exp \Big[- \big(\sqrt{\frac{\xi_s}{2}} v_0 + \sqrt{\varepsilon(v)} \big)^2 \Big]   &\textrm{if}
		      \left\{\begin{array}{lr}
		      v<v_0 - \sqrt{\frac{2\varepsilon_{\phi}}{m_s}}\\
		      v>v_0+\sqrt{\frac{2\varepsilon_{\phi}}{m_s}} 
		      \end{array}\right. \\
		    A \ exp \Big[- \big(\frac{\xi_s}{2} v_0^2 + \beta_s \varepsilon(v) \big) \Big] &\textrm{if}  
		    \left\{\begin{array}{lr}
		      v>v_0-\sqrt{\frac{2\varepsilon_{\phi}}{m_s}} \\
		      v<v_0 + \sqrt{\frac{2\varepsilon_{\phi}}{m_s}} 
		      \end{array}\right.
		\end{array}\right.
		\label{Schamel_Dif}
		\end{equation*}\endgroup
		Here $A = \sqrt{ \frac{\xi_s}{2 \pi}} n_{0s}$
		and $\xi_s = \frac{m_s}{T_s}$ are the amplitude and normalization factor respectively.
		Moreover, $\varepsilon(v) = \frac{\xi_s}{2}(v-v_0)^2 + \phi\frac{q_s}{T_s}$ 
		represents the (normalized) energy of particles. 
		The shape of the trapped population is determined by the \emph{trapping parameter ($\beta$)}.
		It can take three forms, e.g. hollow ($\beta<0$), flat ($\beta=0$) or hump ($\beta>0$). 
		Other types of distribution function has been proposed to model trapped population.
		For example, Smith\cite{smith1970steady} suggested monotonic transitions and distribution functions.
		This selection leads to monotonic particle densities.
		More recent studies for space plasmas have treated
		the trapped distribution function by considering a kappa distribution
		for the ambient plasma.\cite{aravindakshan2018effects,aravindakshan2018bernstein}.

		Kinetic simulation studies of BGK modes (which may be generated
		for example by the Schamel approach or the BGK approach),
		can be divided into two main genres, i.e. production studies and propagation studies.  
		There is a large body of numerical studies about different methods of exciting 
		solitary BGK modes (referred to here as production studies).
		Interested readers might follow this topic in review papers such as 
		Ref.[8]\nocite{eliasson2006formation} and [16]\nocite{hutchinson2017electron} and the references within.
		
		Both for production studies and stability studies, ions are typically assumed to  be immobile (with infinite mass). 
		Immobile ions result in merging (coalescence) 
		behavior of electron holes~\cite{rasmussen1982effects,fijalkow2003phase,ghizzo1988stability,lynov1980interaction,shoucri2017formation}.
		When the ion dynamics is included in the simulation box, 
		a splitting process of the merged hole is reported,
		which suggests that the BGK modes can be considered as solitons of the Vlasov-Poisson set of 
		equations~\cite{jenab2017fully,zhou2018dynamics,kakad2017formation}.

		When it comes to the propagation studies of BGK modes, two questions 
		is in focus, the nonlinear dispersion relation (NDR) and the stability of the solutions. 
		The NDR refers to a relationship between the key aspects of the BGK modes, i.e. width, amplitude and velocity. 
		Turikov~\cite{turikov1984electron} used the BGK approach to produce electron holes (nonlinear solitary waves)
		in the simulation box, 
		by assuming two types of bell shaped functions for the electrical potential.
		They found that the width of the solitary wave increases with
		increasing amplitude, unlike the KdV soliton.
		Concerning the stability of solitary waves, 
		they reported 
		that for Mach numbers bigger than two $(M>2)$ the
		electron holes decay during several plasma periods, and this
		time decreases with increasing Mach number.
		Other simulation studies were mostly concerned with NDR rather than stability~\cite{muschietti1999analysis,lynov1985phase}.

		In this study, we adopt the Schamel approach in order to produce BGK modes. 
		Hence, the shape of the electric potential is determined self-consistently 
		through the Sagdeev pseudo-potential equation. 
		Amplitude, width depends on the trapping parameter and velocity of nonlinear solution.
		In our numerical approach both electrons and ions are described by the  
		Vlasov equation. 

		The main point of inquiry for this study can be formulated as follows:\\
		\emph{Are nonlinear solutions (produced by the Schamel approach) stable against collisions?}\\
		We answer this question by temporal simulations of nonlinear solutions 
		with different velocities.
		In other words, this study explores the effect of velocity (as a major factor in the NDR) 
		on the collisional stability of nonlinear solutions.

		Our simulation results indicate that the nonlinear solution (provided by the Schamel distribution function)
		can be divided into three categories, e.g. stable, semi-stable  and unstable solutions. 
		Stable solutions are clustered slightly above the ion-thermal velocity (Mach number$>1$).
		As the velocity (Mach number) increases, the solutions transform from stable to unstable in a 
		smooth transition. 
		Hence, there exist solutions which are semi-stable.
		 Some fluid simulations have suggested the instability of Sagdeev solutions. 
		 However our study stands as a the first attempt to explore this in the fully-kinetic regime\cite{lotekar2017generation}.

		Our study reveals that net-neutrality violation is a cause of the instability of the 
		nonlinear solutions in our study.
		In all the cases of simulations, a bipolar structure in flux of electrons is observed to form 
		early in the simulation, due to an initial violation of equation (3). 
		A larger deviation from (3) results in a stronger bipolar structure and a more rapid instability. 
		The study of nonlinear solutions to the Vlasov-Poisson system
		shows that during collisions, the electron holes merge and exchange trapped populations. 
		However, the merged holes break up due to the presence of the ion dynamic. 
		Our study shows that the details of the merging and breaking up process varies for different
		electron hole set-ups. 
		We focused on the number of rotations of the core of the electron holes during collisions,
		and the exchange of the trapped population, in order to parameterize different behaviors.
		It was observed that the number of rotations grows with the relative velocity of the electron holes.
		In the case when each of the electron hole core makes half turns during a collision 
		(e.g. if the number of turns are $2.5$ or $3.5$), the electron holes   
		exchange their cores with each other.



\section{Basic Equations and the Numerical Schemes} \label{Sec_Model}
		The normalization of the equations and variables are carried out 
		with respect to the ion parameters. 
		The spatial direction (length) and time are normalized by 
		the ion Debye length ($\lambda_{Di} = \sqrt{ \frac{\varepsilon_0 K_B T_i}{n_{i0} e^2}   }$) and
		the ion plasma frequency ($\omega_{pi}  = {\big(\frac{n_{i0} e^2}{m_i \varepsilon_0}\big)^{\frac{1}{2}} }$),
		respectively. 
		The velocity, energy, and potential are 
		rescaled by the ion thermal velocity ($v_{th_i} = \sqrt{\frac{K_B T_i}{m_i}}$),
		ion thermal energy ($K_B T_i$)
		and ($\frac{K_B T_i}{e}$ ),
		respectively.
		The mass and charge of the species are normalized by the ion mass ($m_i$) and the elementary charge ($e$).

	\subsection{Initialize the nonlinear solutions} \label{SubSec_Initial_condition}
		As mentioned in the introduction, we are following the approach proposed by 
		Schamel~\cite{schamel_3}.
		Consequently, we start with choosing the distribution function of the trapped population, 
		i.e. setting the value of the trapping parameter ($\beta$).
		Note that the Schamel distribution function guarantees positiveness and continuity of the distribution function,
		unconditionally for any value of the trapping parameter ($\beta$) in the whole phase space.

		In the next step,
		one can determine the dependence of 
		the distribution function on the electric potential ($f(\phi)$), 
		as implied by the Schamel distribution function. 
		Next, the number density dependency on the electric potential $n(\phi)$ can be determined
		by integration over the velocity:
		\begin{equation}
		    n_s(\phi) = n_{0s} \int f_s(\phi) \dd v. \label{Eq_density_calculation}
		\end{equation}
		Next, the Sagdeev pseudo-Potential function $S(\phi)$ should be established 
		by integration over $\phi$: 
		\begin{equation}
		    S(\phi) = \int_0^{\phi} \sum_s n_s(\phi) \dd \phi - 1 + \textrm{const} \quad s=i,e \label{Eq_Sagdeev_potentail}
		\end{equation}
		Finally, the potential profile in the spatial direction $\phi(x)$ can constructed 
		from the Sagdeev pseudo-potential $S(\phi)$ by integrating:
		\begin{equation}
		    \frac{1}{2} \Big(\frac{d \phi}{d x}\Big)^2 + S(\phi) = 0.
		    \label{Eq_Sagdeev}
		\end{equation}
		The constructed electric potential and the Schamel distribution function associated with it,
		is used then to initialize the Vlasov-Poisson simulations, i.e. following the temporal
		evolution of the distribution functions.
		The details of this method can be found in Ref.[15].
 
	\subsection{The fully kinetic Vlasov-Poisson method} \label{SubSec_Equations_Vlasov}
		In order to perform fully kinetic simulations, 
		both species are studied numerically using the Vlasov equations:
		\begin{multline}
			\frac{\partial f_s(x,v,t)}{\partial t} 
			+ v \frac{\partial f_s(x,v,t)}{\partial x} 
			\\ +  \frac{q_s}{m_s} E(x,t) \frac{\partial f_s(x,v,t)}{\partial v} 
			= 0, \ \ \  s = i,e.
			\label{Eq_Vlasov}
		\end{multline}
		The Poisson's equation provides the force (electric) field:
		\begin{equation}
			\frac{\partial^2 \phi(x,t)}{\partial x^2}  = \sum q_s n_s(x,t) , \ \ \  s = i,e. 
			\label{Eq_Poisson}
		\end{equation}
		where $s = i,e$ represents the corresponding species.
		The Vlasov and Poisson equations 
		are coupled by density integrations 
		for each species to form a closed set of equations:
		\begin{equation*}
		  n_s(x,t) = n_{0s} N_s(x,t), 
		\end{equation*}
		\begin{equation}
		N_s(x,t) = \int f_s(x,v,t) dv.
		\label{Eq_number_density}  
		\end{equation}
		In which $N_s$ stands for the number density.
		$n_{0s}(=N_{0s}q_s)$ is the (normalized) unperturbed value of the charge density.
		The background charge density is taken to be zero, i.e.
		\begin{equation*}
		\sum_s n_{0s} = \sum_s N_{0s} q_s = 0.
		\end{equation*}

		Here, the Vlasov-Hybrid Simulation (VHS) method
		is adopted, which is closely related to the Particle-In-Cell (PIC)
		simulation method. 
		In VHS, the distribution function in phase space of each species 
		is sampled by enough phase points, 
		i.e. all the details of the trapped population are 
		covered~\cite{nunn1993novel,kazeminezhad2003vlasov,jenab2011preventing,kilian2018afterlive}.
		At each time step, the arrangement of phase points in phase space
		produces the associated distribution function for that time step.
		By knowing the distribution function for each step, 
		all the useful quantities at the fluid level, e.g. the momentum density
		of the distribution function, can be easily deduced. 
		Quantities such as kinetic energy density and entropy density are used for 
		monitoring the conserved quantities, i.e. the total energy and entropy. 
		Finally, the number density (and thereby charge density) is used to calculate the electric field 
		through Poisson's equation, which
		in turn is fed back to the Vlasov equation to push the phase space
		for the next time step.
		Note that the initial value of the distribution function associated with each of
		the phase points stays intact
		during the simulation.
		This property of the simulation method guarantees the positiveness of the distribution function
		under any circumstances during the temporal evolution~\cite{kazeminezhad2003vlasov}.


	\subsection{Variables and parameters} \label{SubSec_Equations_variables}
		A number of parameters remain fixed through all
		of our simulations. Specifically this include: 
		the mass ratio $\frac{m_i}{m_e} = 1836$,
		$\Delta x = 1.0$ (the grid size on the spatial direction)
		and $\Delta v_x = 0.0025$ ($\Delta v_x = 0.0025/v_{th_{e}}$) 
		which is the grid size in the velocity direction for ions (electrons).
		Periodic boundary conditions are adopted in the spatial direction. 
		Furthermore,
		the input parameters for each set of the simulations consist of two
		variables,  the trapping parameter ($\beta$) and  the velocity of the nonlinear solution ($v_p$).
		Note that by our choice of normalization,
		the ion-acoustic velocity, the electron plasma frequency
		and the electron thermal velocity are 
		$v_C = \sqrt{1 + \frac{T_e}{T_i}}$,
		$\omega_{pe} = \sqrt{\frac{m_i}{m_e}}$
		and $v_{th_{e}} = \sqrt{\frac{m_i}{m_e}\frac{T_e}{T_i}}$;
		respectively.
		In what follows, we will express the velocity of solitary waves 
		using the \emph{Mach number}: $M = \frac{v_{p}}{v_C}$,
		in order to follow the notation of many theoretical papers.

		The length of the simulation box is chosen as $L=1024$.
		The number of grid points in the simulation grid is 
		$(S_x, S_v) = (1024,4800)$ for each species. 
		Initially there are 16 phase points per cell in phase space, which are chosen randomly~\cite{jenab2011preventing}. 
		Hence, the total number of phase points for each species 
		is $\Sigma=78\times10^6$.
		The velocity cut-off is (-6$V_{ths}$ ; 6$V_{ths}$): 
		for electrons ($-6V_{th_e},  6V_{th_e}$) with $V_{th_e}=\sqrt{T_e/M_e}$ and ions ($-6, 6$).

		The Vlasov equation, as a collisionless Boltzmann equation,
		should conserve entropy and other forms of Casimir invariants \cite{elskens2014vlasov} as well as the 
		total energy.
		These conservation quantities are used as a yardstick for the simulation results. 
		All simulation results reported here have a deviation strictly less than one percent in total energy and entropy.


\section{Results and Discussion} \label{Sec_Results}

	\subsection{Three categories of the solutions:}
		In order to study the stability of the 
		nonlinear solutions, we have made a simulation for each combination of
		$\beta$ (trapping parameter) and $M$ (Mach number).
		In each of these simulations, two identical nonlinear solutions 
		(with opposite velocity, i.e. $\pm M$)
		are positioned on each side of the simulation box in the spatial direction.
		The temporal evolution of these simulations result in 
		symmetric mutual head-on collisions. 
		Since periodic boundary conditions are adopted, 
		collisions happen repeatedly as long as the simulation is executed. 
		\begin{figure*}
			\subfloat[$M = 1.1,\beta = 0$]{\includegraphics[width=0.49\textwidth]{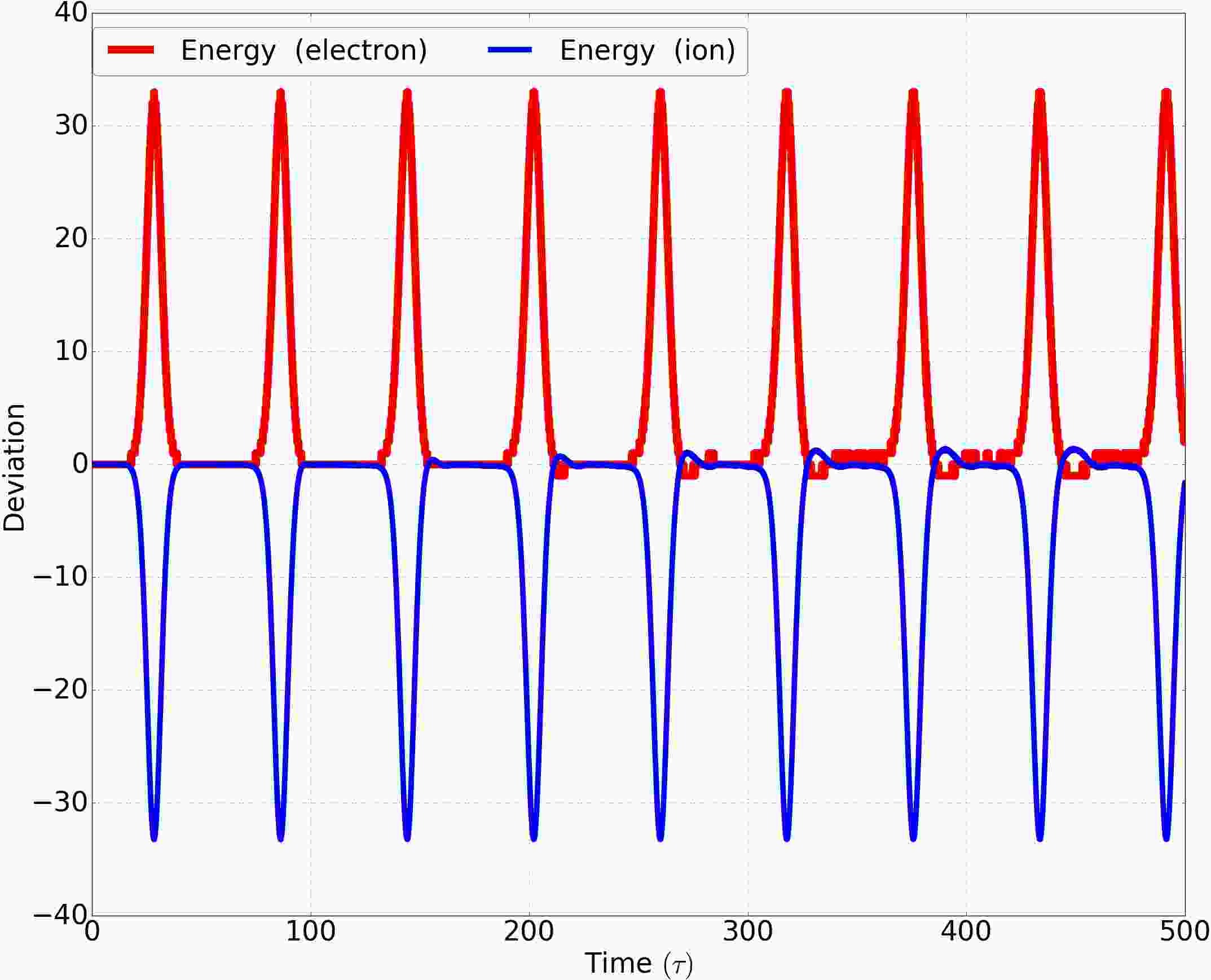}} \hspace{0.1cm}
			\subfloat[$M = 1.25,\beta = -1.25$]{\includegraphics[width=0.49\textwidth]{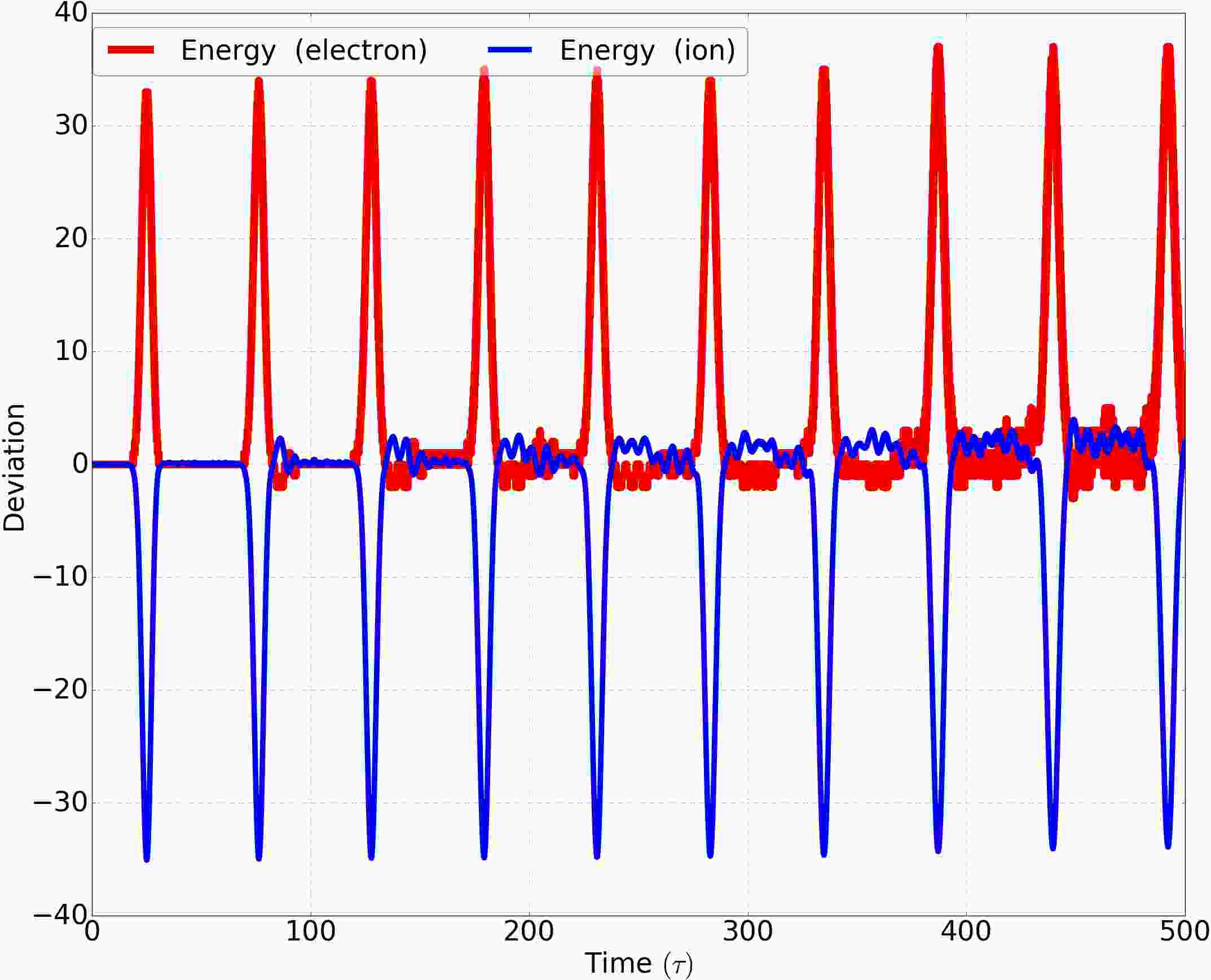}}\\
			\subfloat[$M = 1.5,\beta = -2.5$]{\includegraphics[width=0.49\textwidth]{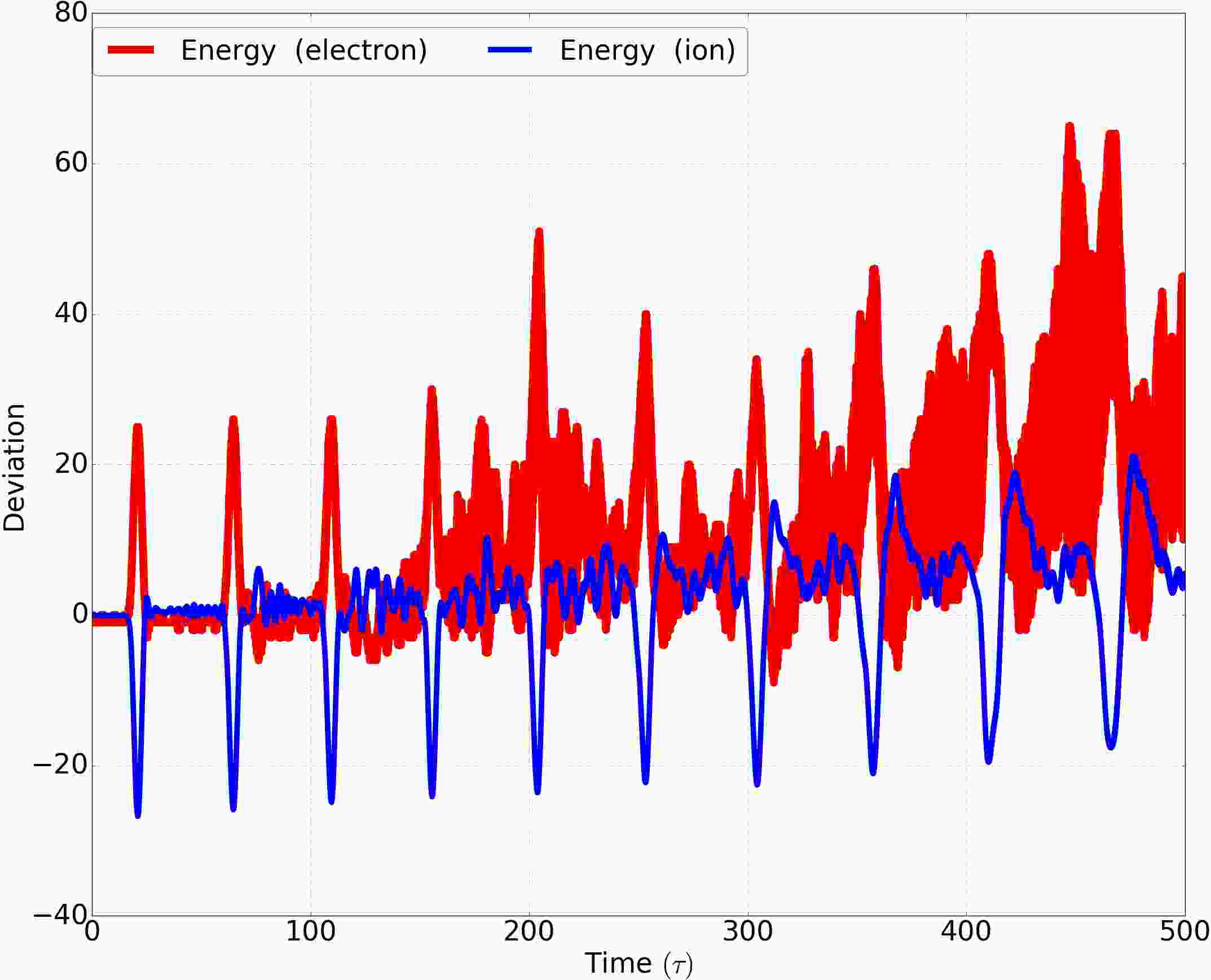}}  \hspace{0.1cm}
			\subfloat[$M = 1.75,\beta = -2.75$]{\includegraphics[width=0.49\textwidth]{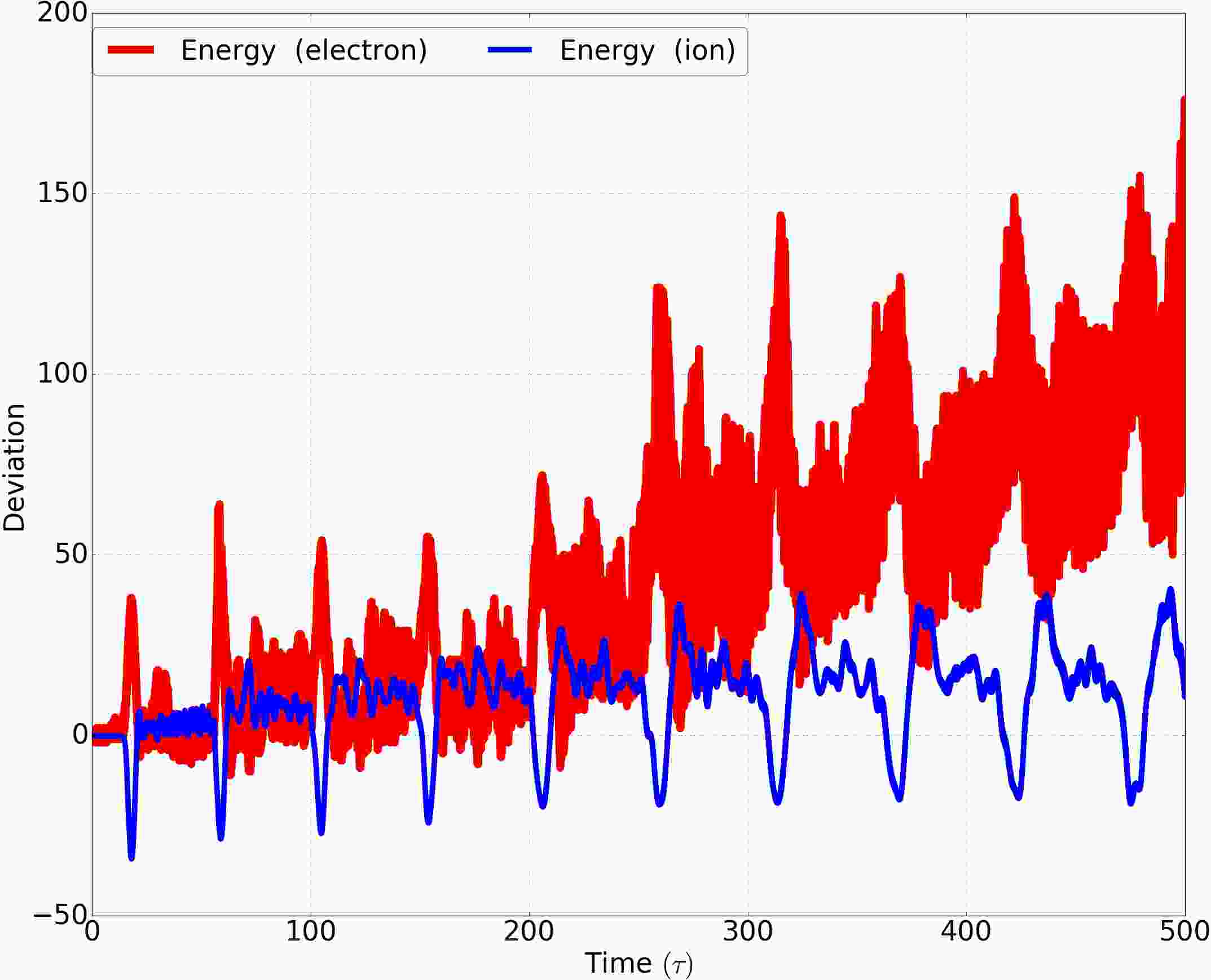}}	
			\caption{The temporal evolution of kinetic energy of electrons and ions during multiple head-on collisions are 
			presented for four different cases. $M$ stands for Mach number and $\beta$ represents the value of the trapping parameter.
			For a stable solution, the energy exchange between electrons and ions 
			is symmetric and both species can recover their pre-collision level of kinetic energy, i.e (a) and (b).
			Semi-stable solutions can survive a few collisions before destruction such as the case of (c).
			Unstable solutions can not survive even one collision (d).
			Note that the deviation in total energy of the simulation stands strictly below $0.1\%$ for the all the above simulations.}
			\label{Fig_collision_Mach_effect}
		\end{figure*}
		Fig.~\ref{Fig_collision_Mach_effect} presents the results regarding 
		the electron-ion energy exchange for $(M,\beta)=$ $(1.1,0)$, $(1.25,-1.25)$ $(1.5,-2.5)$ and  $(1.75,-2.75)$.
		The ion-acoustic speed for these simulations is $v_C = 8$.
		The exchange of kinetic energy between electron and ions provides a 
		reasonable estimator for the stability of the solutions. 
		For stable solutions, the exchange of energy is perfectly symmetrical in time.
		Electrons gain kinetic energy from ions during the first half of the collision,
		and opposite flow of energy happens in the second half. 
		After the collision, if the solutions are stable, both kinetic energies 
		return to the same level as before the collision.

		In case of $(M,\beta) =$ $(1.5,-2.5)$, the deviation from the initial value starts to diverge after each collision.
		This accumulates, and after a few number of collisions the solution can not recover its 
		original level of energy balance even approximately, and the initial structure is lost. 
		Such solutions are considered as semi-stable solutions. 
		As for stable solutions, they can recover their original balance of energy 
		after numerous collisions, for example the case of $(M,\beta) =$ $(1.1,0)$ and $(1.25,-1.25)$.
		The case of $(M,\beta) =$ $(2, -3)$ represents an unstable solution. 
		After just one collision, the energy balance is destroyed and two electron holes
		merge with each other and they lose their initial characteristics.

		Therefore, we have distinguished three categories of solutions which can be 
		considered as a continuum, starting from stable solutions up to completely unstable solutions.
		On one side of this continuum, collisionally stable solutions are situated slightly above the ion-acoustic speed and survive multiple 
		mutual head-on collisions.
		As Mach number increases, solutions start to get more unstable, i.e.
		they can survive a smaller number of collisions before destruction. 
		On the other side of the continuum, for large Mach numbers, solutions 
		are so unstable that they do not survive even one collision.

		The conservation of energy and entropy of all these simulations were closely monitored and the deviation from exact conservation 
		stood below $0.1 \%$.
		The deviation of a variable ($X$) is defined here as the deviation from its initial value:
		$\Delta X= X(t)-X(0)$.
		In case of unstable nonlinear solutions, each collision creates some ion-acoustic and Langmuir noise in the simulation box. 
		The Langmuir and the ion-acoustic modes are identified by using the Fourier transform 
		and by checking their respective dispersion relation.\cite{jenab2016IASWs}.
		When high frequency noise are produced, with a wavelength smaller than the grid spacing in the spatial direction, the 
		small scale structures can no longer be tracked in the simulation and this causes a small deviation in the energy conservation. 
		Electron (due to their light weight) reaches these level faster than the ions, and hence, one can observe an artificial
		increase in their energy level for unstable solutions. 
		The more unstable the solution, the more prominent this effect appears in the results. 
		However, the kinetic energy of electrons is in order of $10^5$ and even in the most unstable results reported here
		the deviation stands around $10^2$ which is around $0.1\%$.

		\begin{figure*}
			\subfloat{\includegraphics[width=0.45\textwidth]{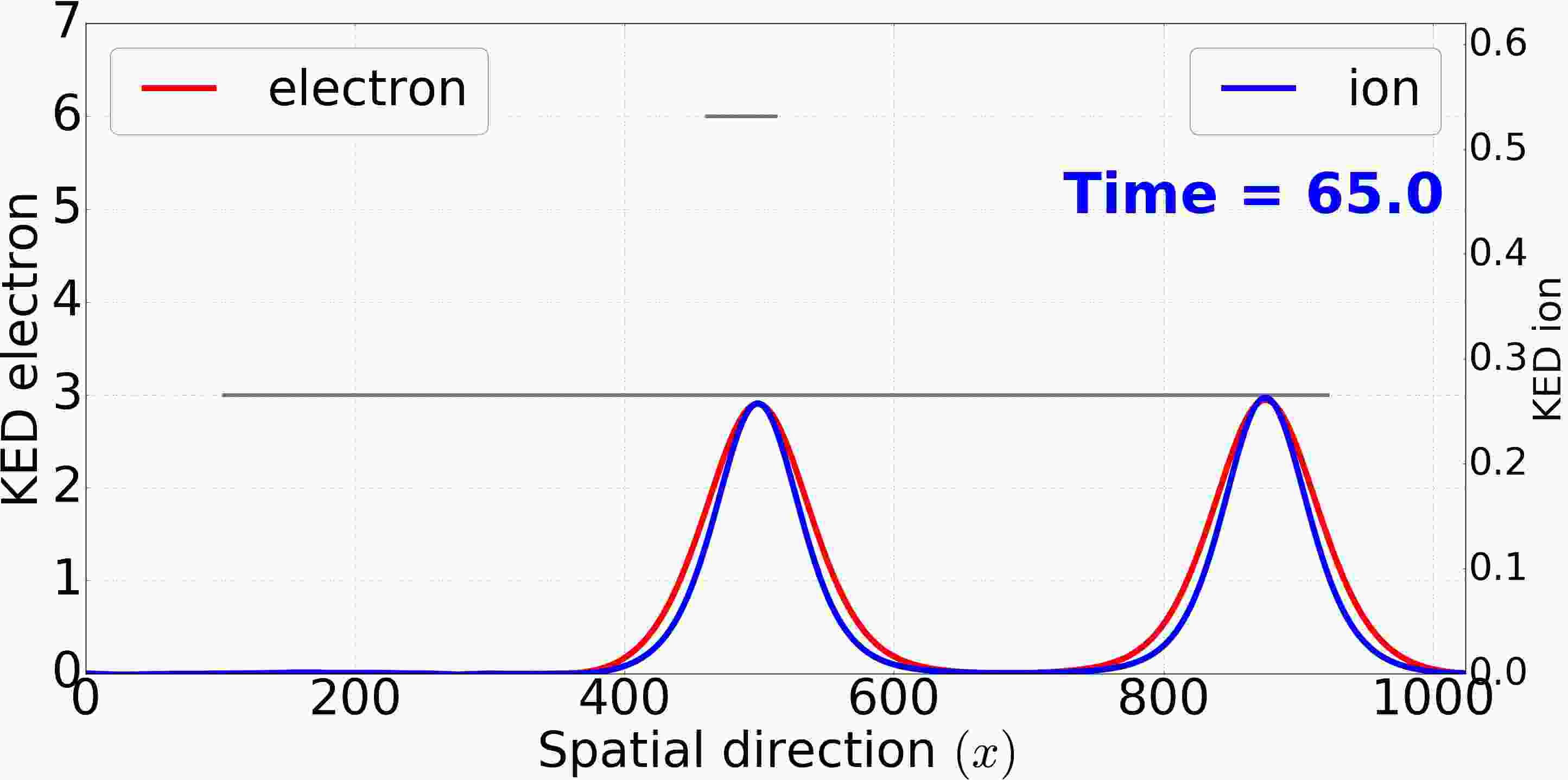}}\hspace{0.1cm}
			\subfloat{\includegraphics[width=0.45\textwidth]{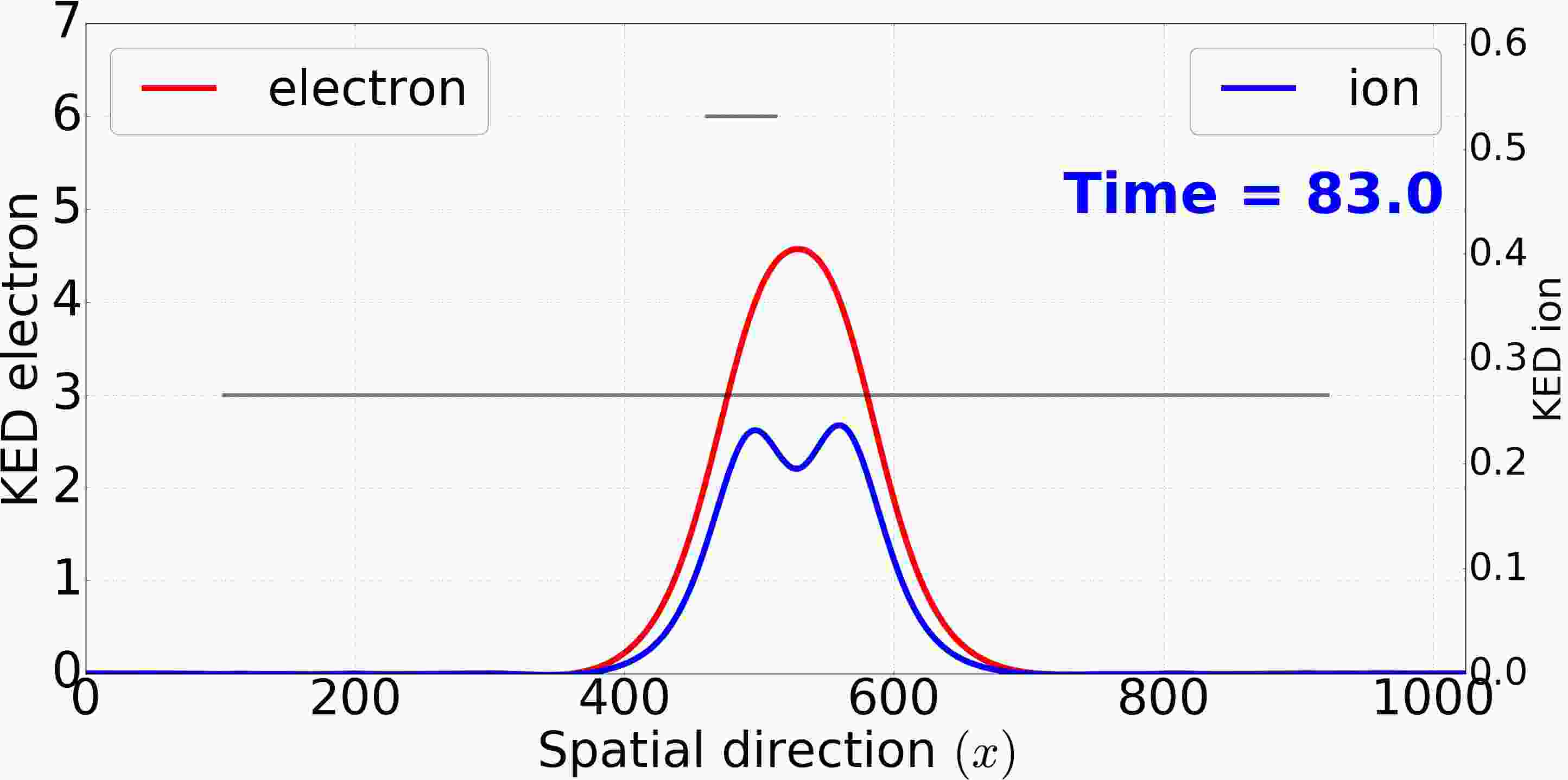}} \\
			\subfloat{\includegraphics[width=0.45\textwidth]{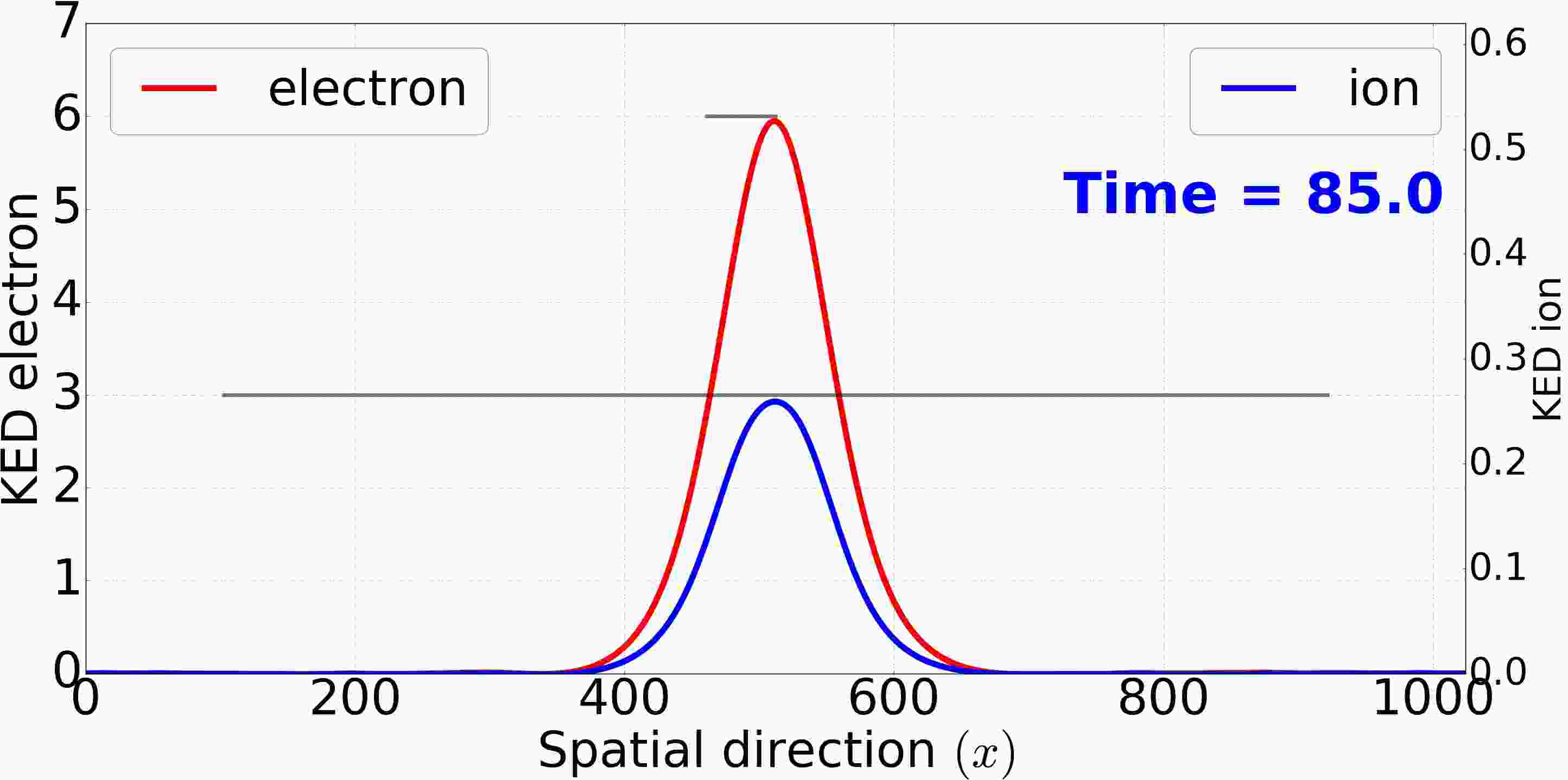}} \hspace{0.1cm}
			\subfloat{\includegraphics[width=0.45\textwidth]{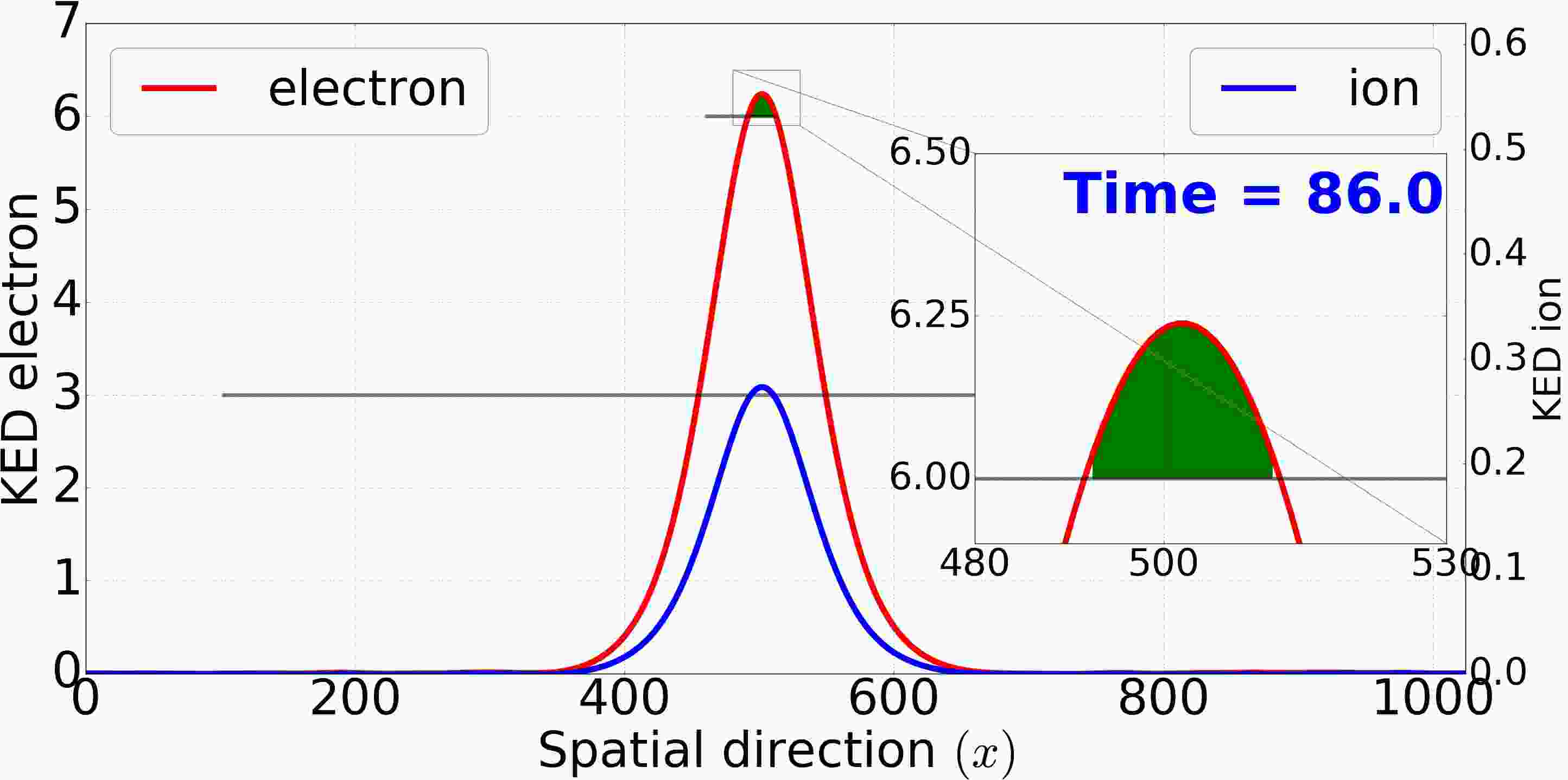}} \\
 			\subfloat{\includegraphics[width=0.45\textwidth]{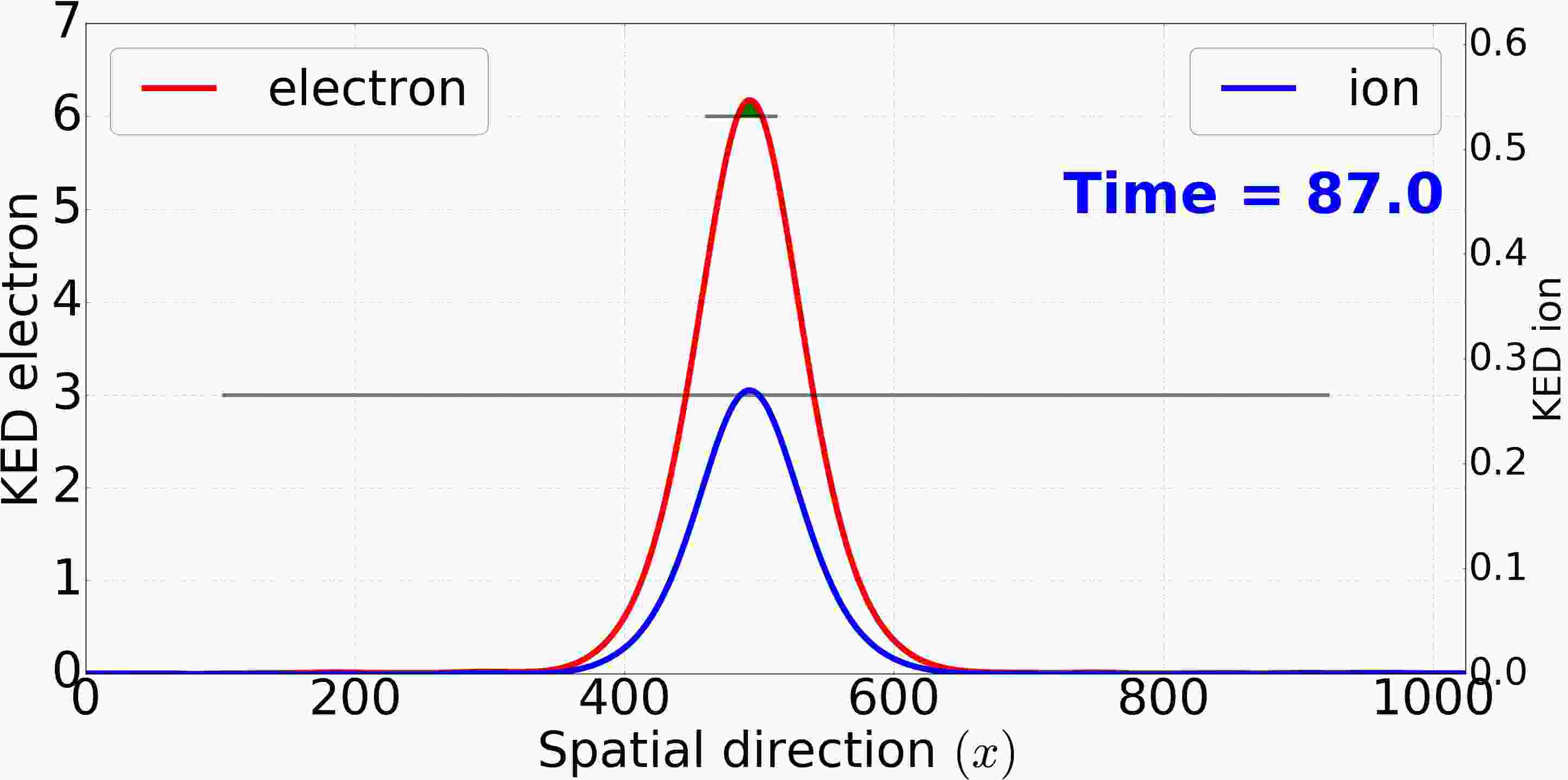}}\hspace{0.1cm}
 			\subfloat{\includegraphics[width=0.45\textwidth]{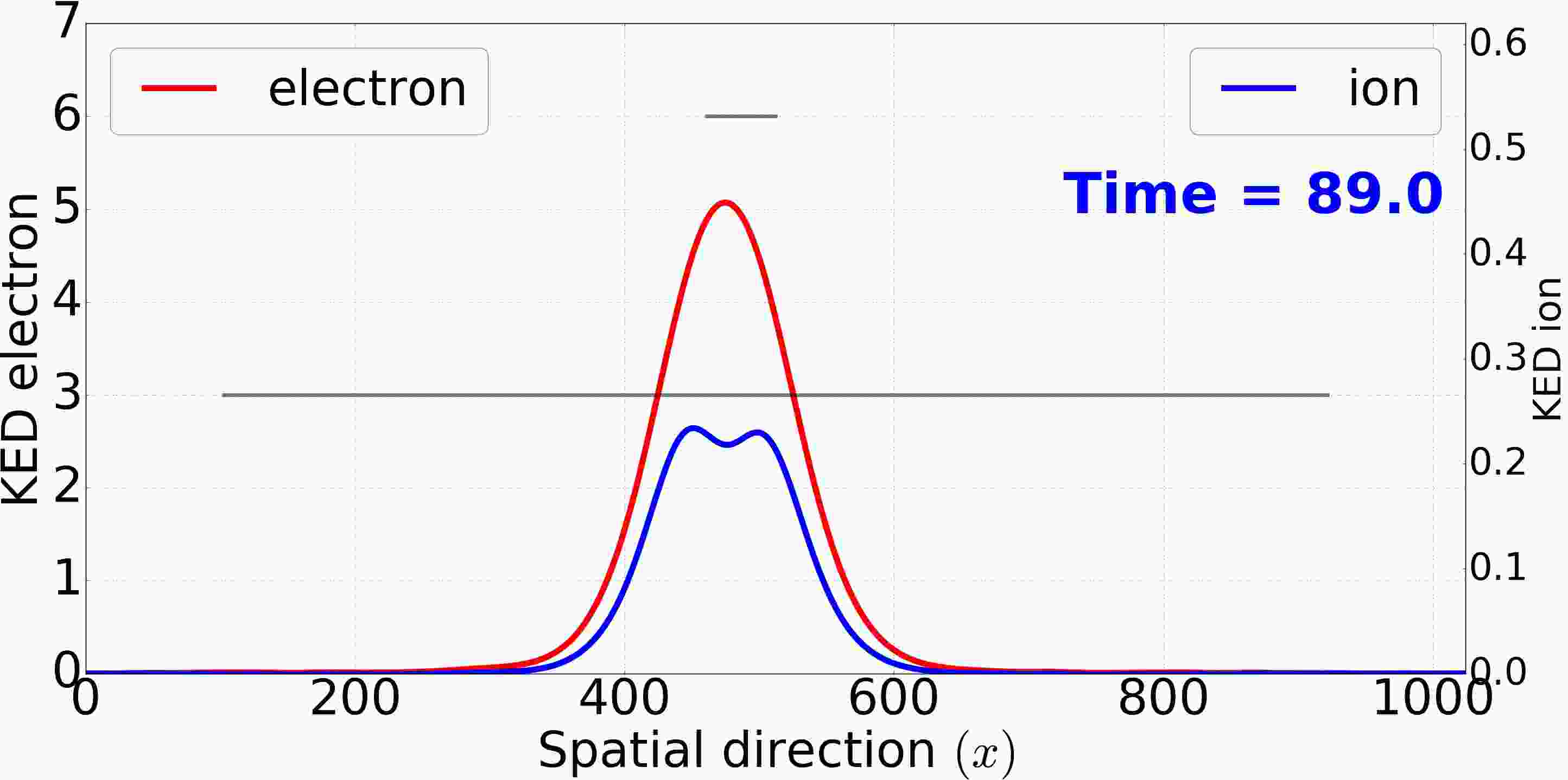}}\\
			\subfloat{\includegraphics[width=0.45\textwidth]{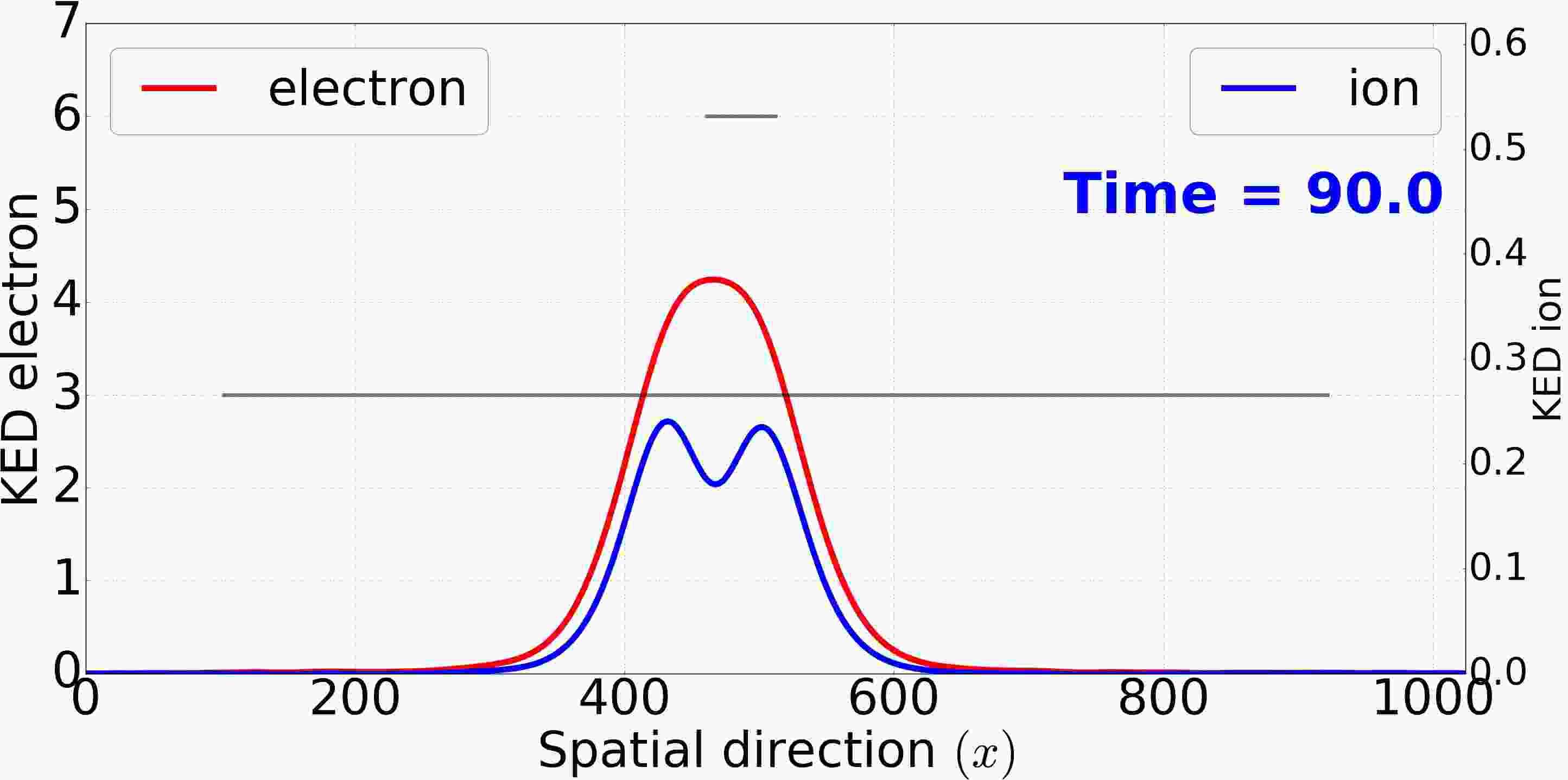}}\hspace{0.1cm}
			\subfloat{\includegraphics[width=0.45\textwidth]{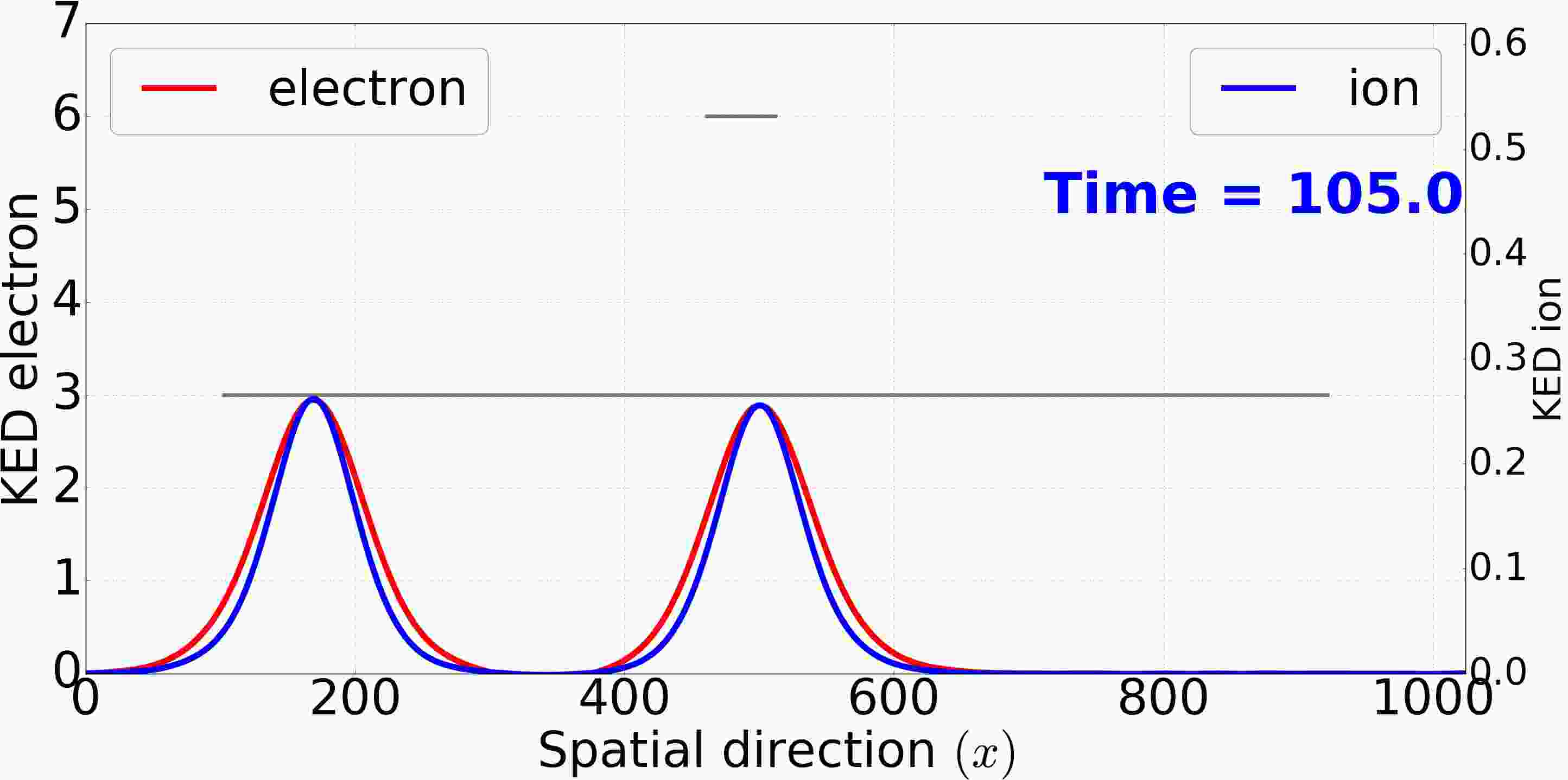}}			
			\caption{The temporal evolution of the electron and ion kinetic energy density (KED) is shown for a few time steps including:
			before, during and after the second collision
			in case of $M=\pm1.1, \beta=0$. The ion KED does not obey the superposition principle.
			The electron KED, on the other hands, follow a simple linear superposition. 
			In addition to the superposition of electron energies from the two holes, a small energy exchange of electrons and ions 
			can also be observed. 
			During $85<\tau<89$ the total electron energy density goes above $6$ (each of the holes has maximum of $3$) 
			which is due to the exchange of 
			energy between electron and ions which is also observed in the Fig.~\ref{Fig_collision_Mach_effect}.}
			
			\label{Fig_temporal_Ene_kin}
		\end{figure*}
		Fig.\ref{Fig_temporal_Ene_kin} explores the temporal evolution of the 
		kinetic energy density (KED) of ions and electrons during head-on mutual collision.
		The figure shows results for the second collision between two nonlinear solutions with $M=$\textpm$1.1$.
		As the two nonlinear solutions collide, the electron KED of the two solutions add up by following the superposition principle. 
		However, during the collision, the ion KED stays the same as before or after the collision.
		This signals the duality that exists in the nature of these nonlinear solutions. 
		They can be considered as ``Coupled electron Holes with ion-acoustic Solitons (CHS)''\cite{saeki1998electron,zhou2018dynamics}.
		This theoretical framework is established when considering the electron- and ion-dynamic separately. 
		On one hand, in the ion-acoustic regime and in the absence of ions' motions,
		electron holes are considered as vortexes in phase space with a nonlinear structure prone to merging.  
		When they merge, the electron KED adds up with little or no signs of a nonlinear interaction.
		On the other hands, considering ions without independent electron-dynamics (electrons act as Boltzmannian fluid)
		one can produce solitons, described by the KdV equation (for a weakly nonlinear amplitude).
		During the collision of two solitons, due to the nonlinearity, the superposition principle does not apply, 
		and hence, it is not a surprise that the ion KED of two solitons do not add up. 
		
		Detailed analysis, however, revealed that during a short time in the middle of the merging stage and before the 
		start of splitting, the energy of the electron holes goes a bit above the sum of the kinetic energy of the two electron holes. 
		This is what can be observed in the Fig.~\ref{Fig_collision_Mach_effect} as an symmetric exchange of energy between electrons and ions.
		As the two ion-acoustic solitons collide, the extra energy of ions is converted to electrons. 
		

	\subsection{net-neutrality instability:}
		Inspired by the symmetric energy exchange between electron and ions during collisions (in case of stable solutions),
		we have studied the balance of energy between electrons and ions, and the relation to the stability properties. 
		For stable solutions, the potential energy (directly related to the electric potential profile) should stay
		stable.
		The changes in the potential energy is directly related to the 
		flux of species:
		\begin{equation}
		\frac{\dd \epsilon_{\phi}}{\dd t}  =  E \sum_s q_s \int v f_s \dd v, \ \ \ s=i,e
		\end{equation}
		In order to have a stable solution, there should be 
		no flow of potential energy ($\frac{\dd \epsilon_{\phi}}{\dd t} = 0$). 
		This condition (in its strongest form) results in
		\begin{equation}
		(\overline{n_e v_e} =) \int v f_e \dd v = \int v f_i \dd v (= \overline{n_i v_i})
		\end{equation}
		which is called ``net neutrality'' condition in some of the theoretical studies 
		of Vlasov-Poisson equation~\cite{schamel_3, smith1970steady,montgomery1969shock}.
		However, we found numerically that if one wants to adhere to the strong form of this condition,
		the Sagdeev pseudo-potential will not converge. 
		For the case of $\overline{n_e v_e} = \overline{n_i v_i}$, $S(\phi)$ appears as a straight line with a small slope. 
		Hence, for the Sagdeev pseudo-potential to converge, a small deviation in the net neutrality should exist.
		The amplitude of this deviation depends on the details of the shape of the electric field and the potential. 
		However, the strongest form of deviation from the net-neutrality condition happens when the electron average velocity
		has a bipolar structure, with opposite signs on the frontal side and the backside of the pulse.

\begin{figure*}
			\subfloat[$M = 1.1,\beta = 0$]{
			\includegraphics[width=0.3\textwidth]{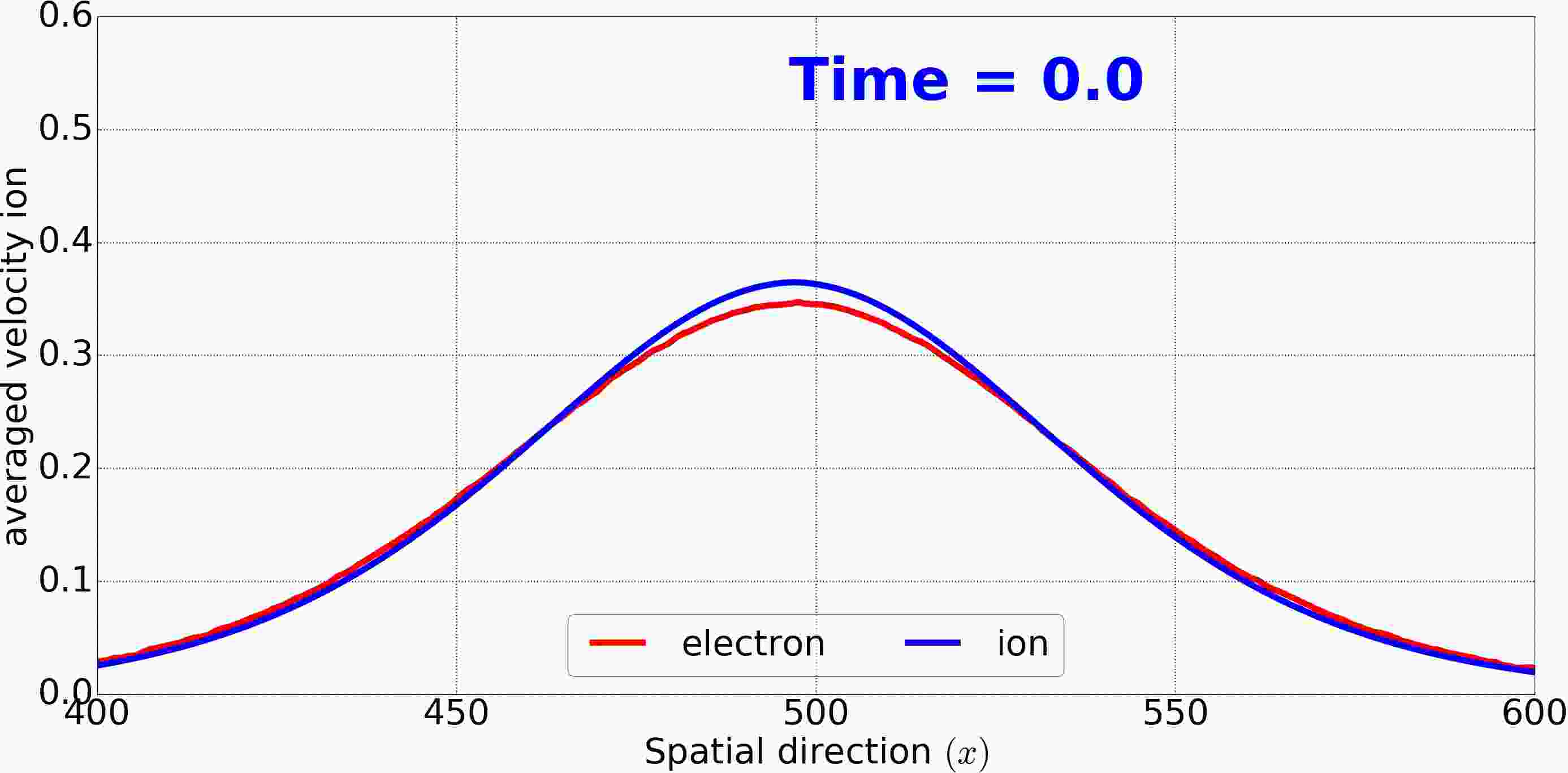} \hspace{0.1cm}
			\includegraphics[width=0.3\textwidth]{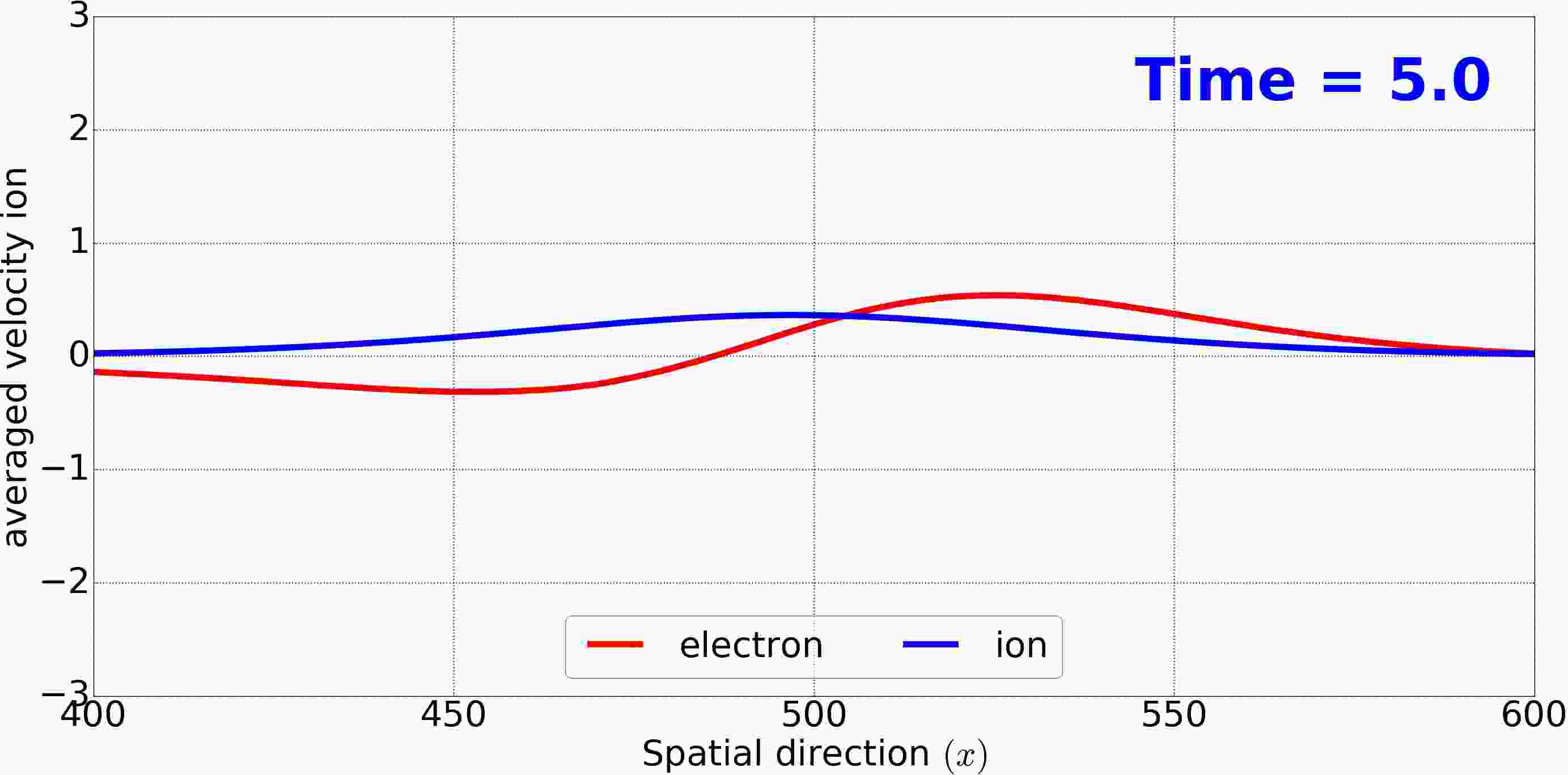} \hspace{0.1cm}
			\includegraphics[width=0.3\textwidth]{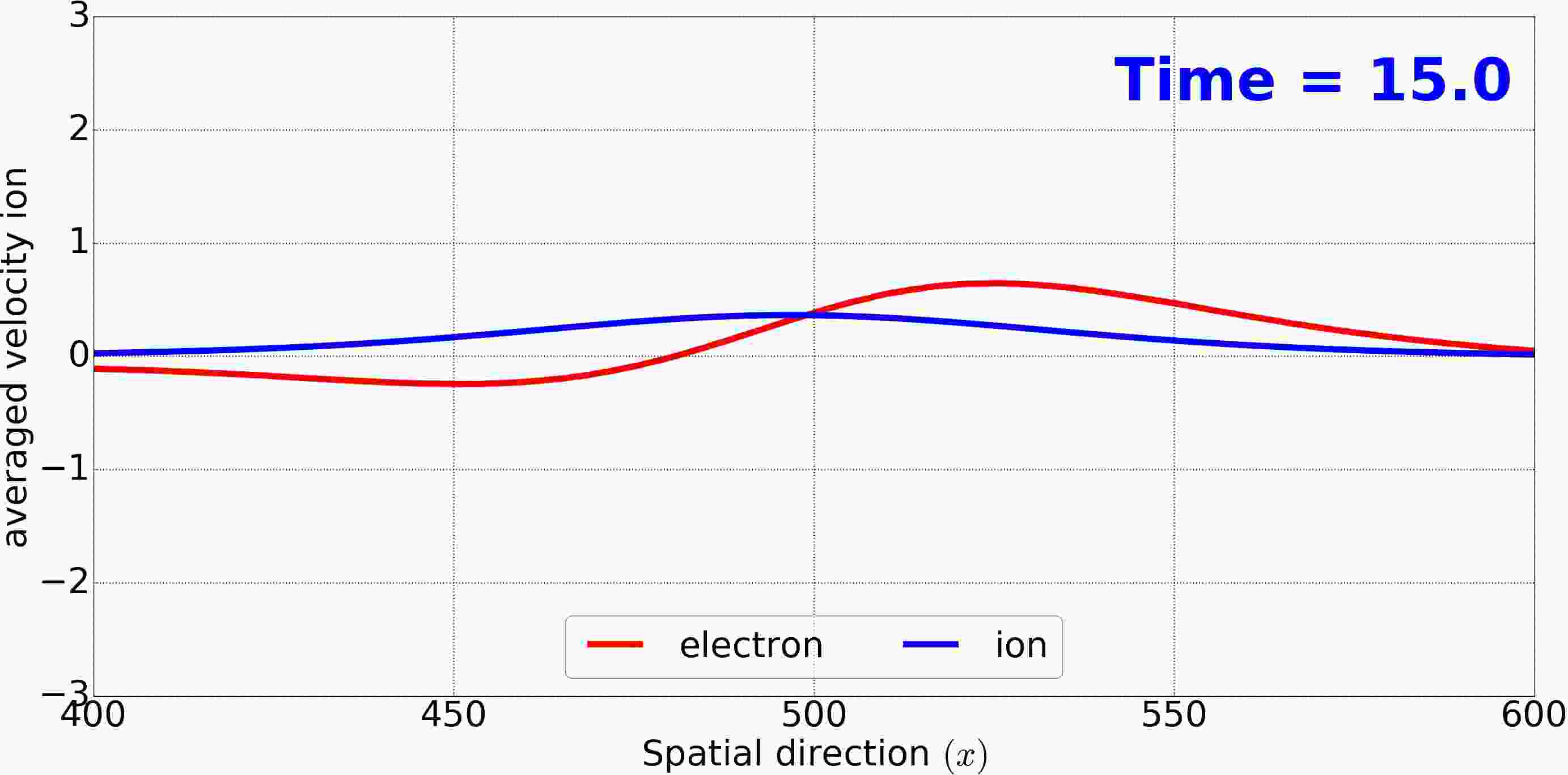}}\\
			\subfloat[$M = 1.25,\beta = -1.25$]{
			\includegraphics[width=0.3\textwidth]{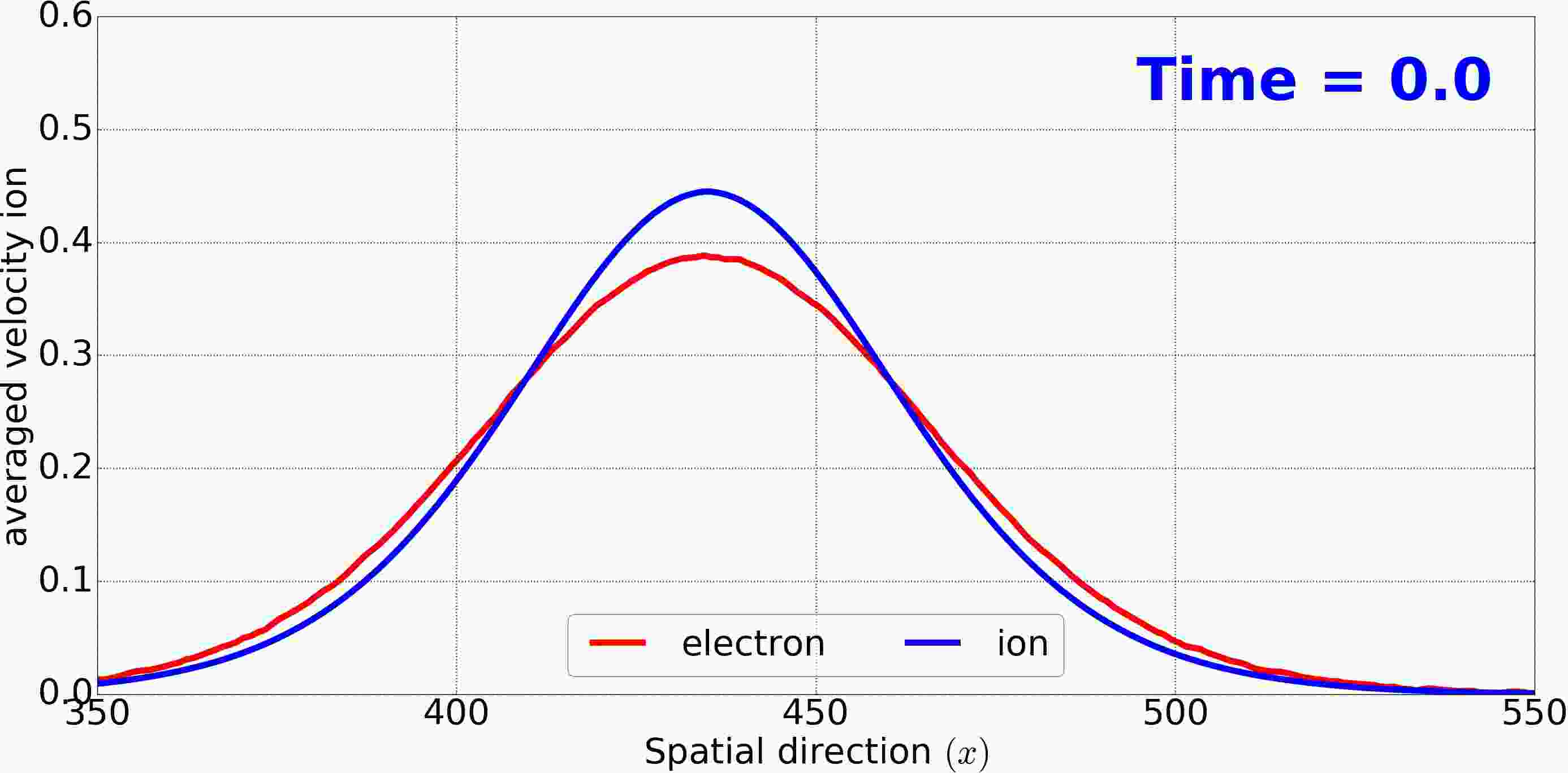} \hspace{0.1cm}
			\includegraphics[width=0.3\textwidth]{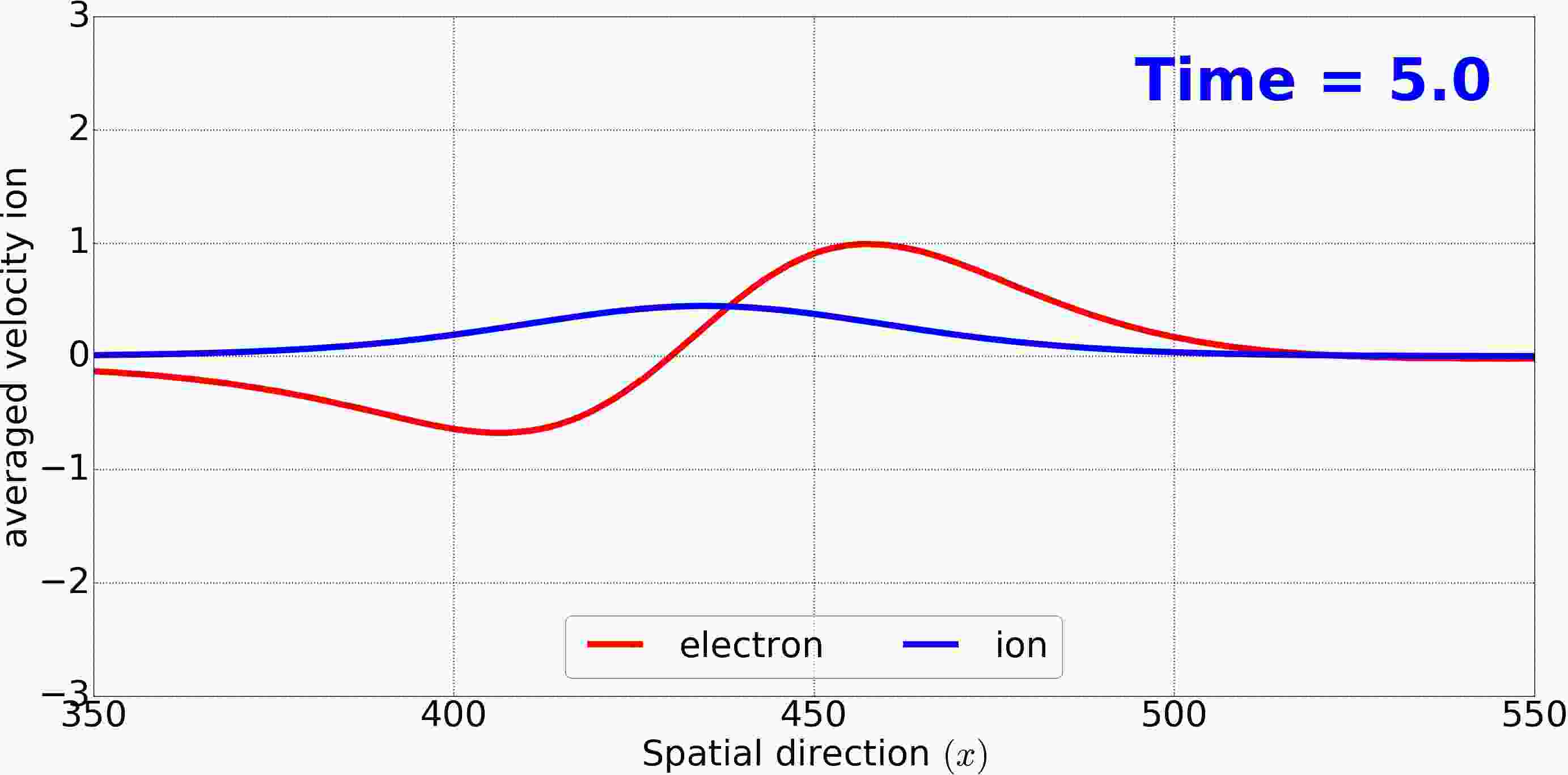} \hspace{0.1cm}
			\includegraphics[width=0.3\textwidth]{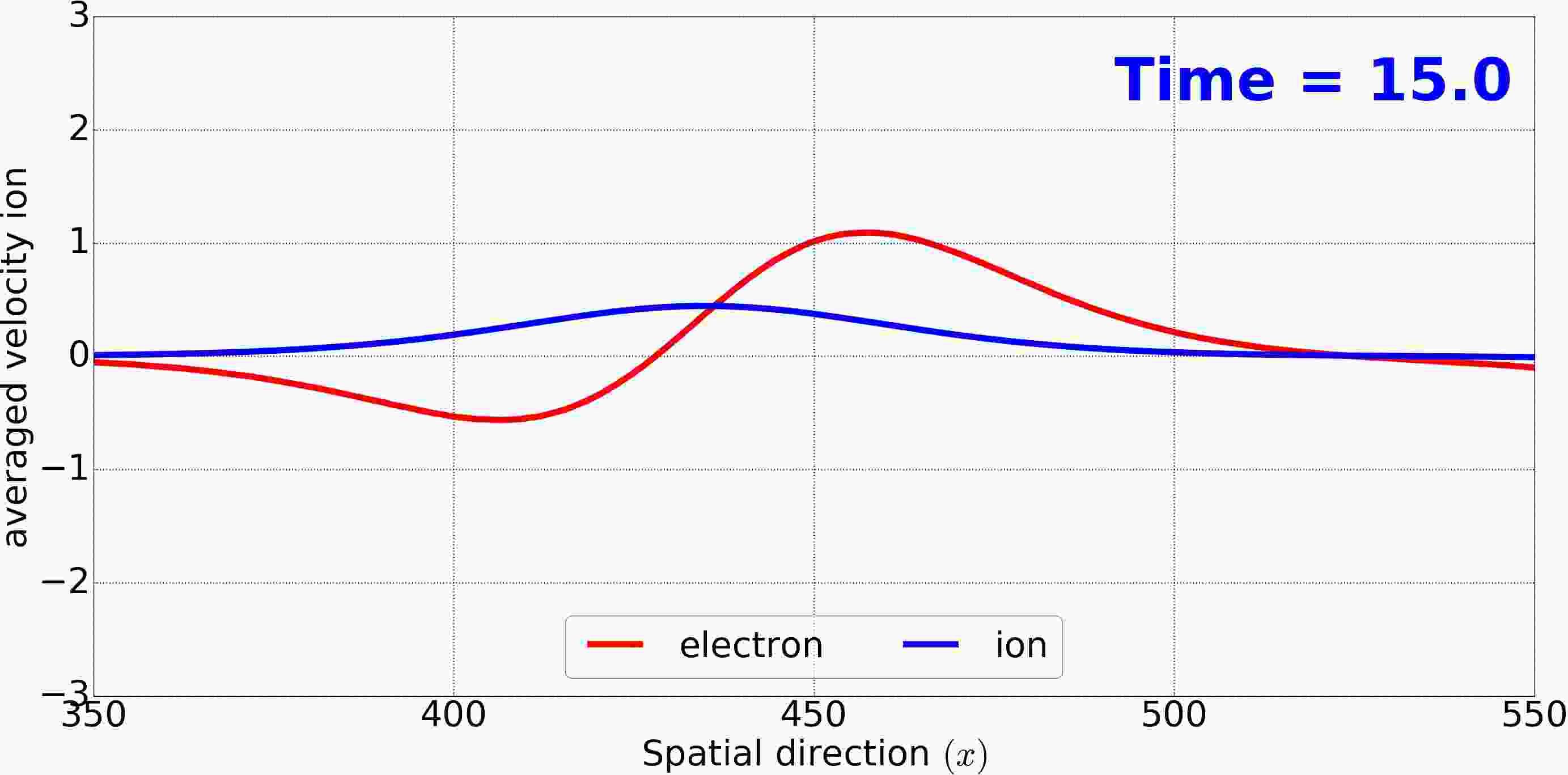} }\\
			\subfloat[$M = 1.5,\beta = -2.5$]{
			\includegraphics[width=0.3\textwidth]{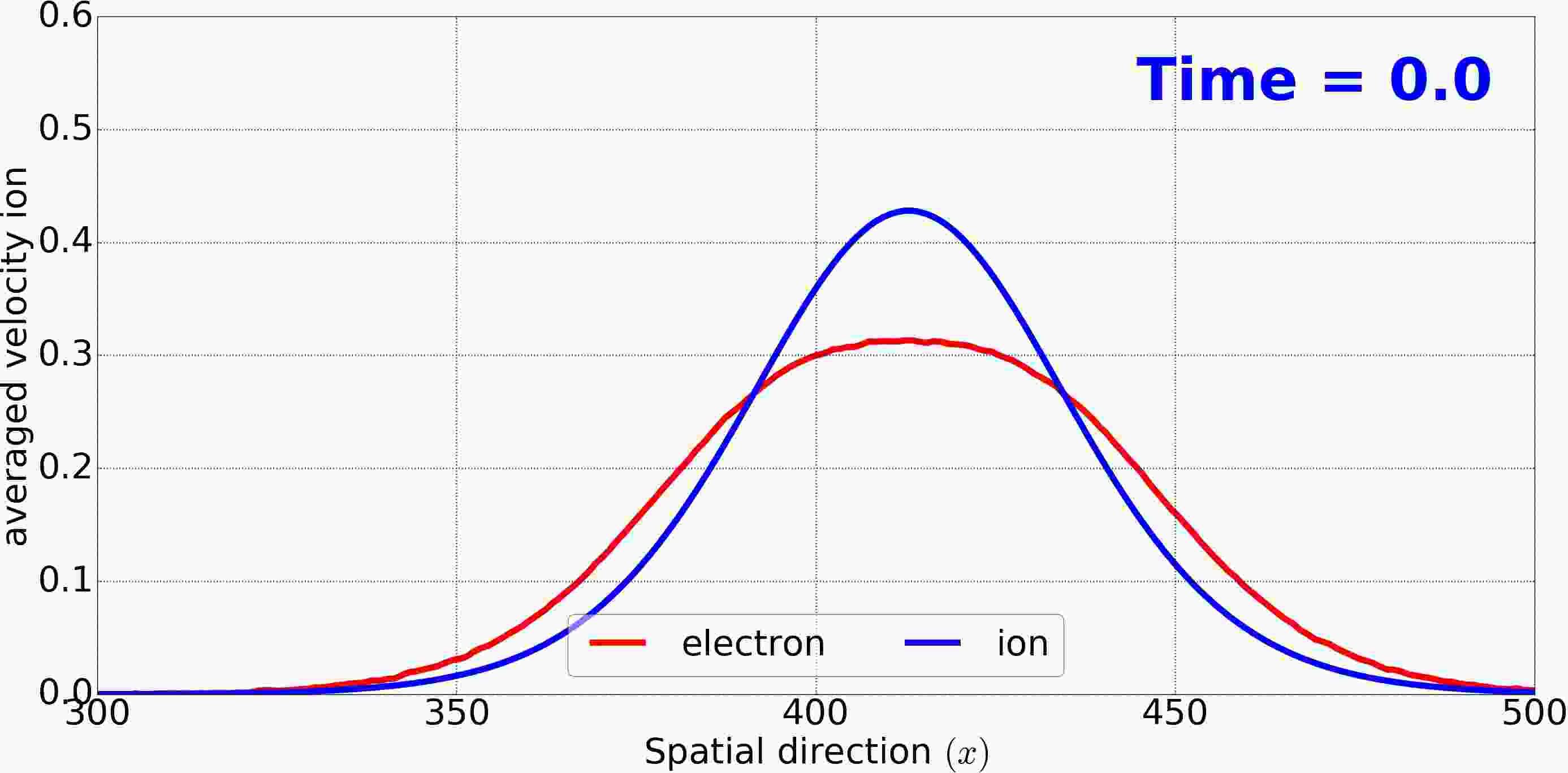} \hspace{0.1cm}
			\includegraphics[width=0.3\textwidth]{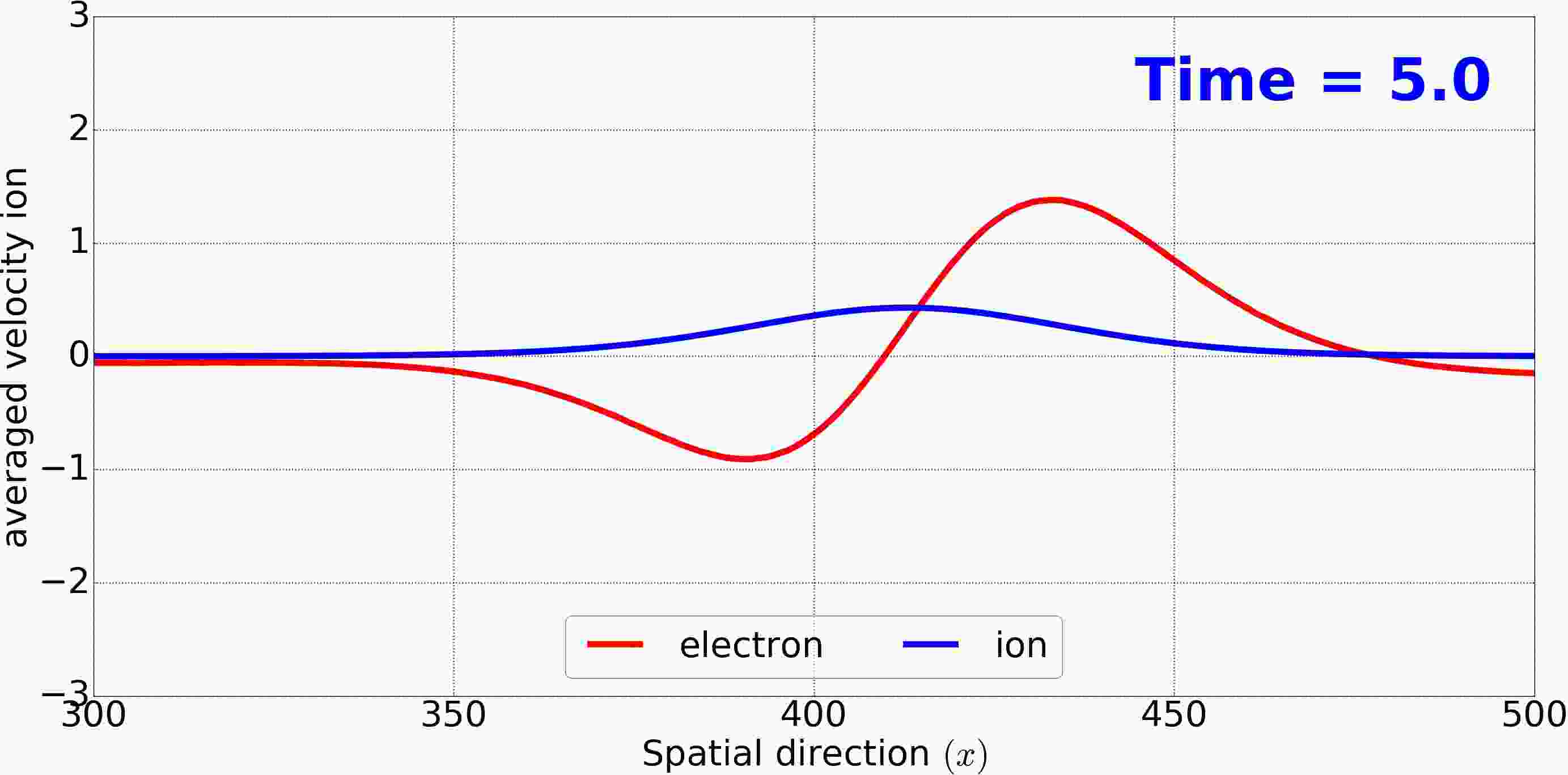} \hspace{0.1cm}
			\includegraphics[width=0.3\textwidth]{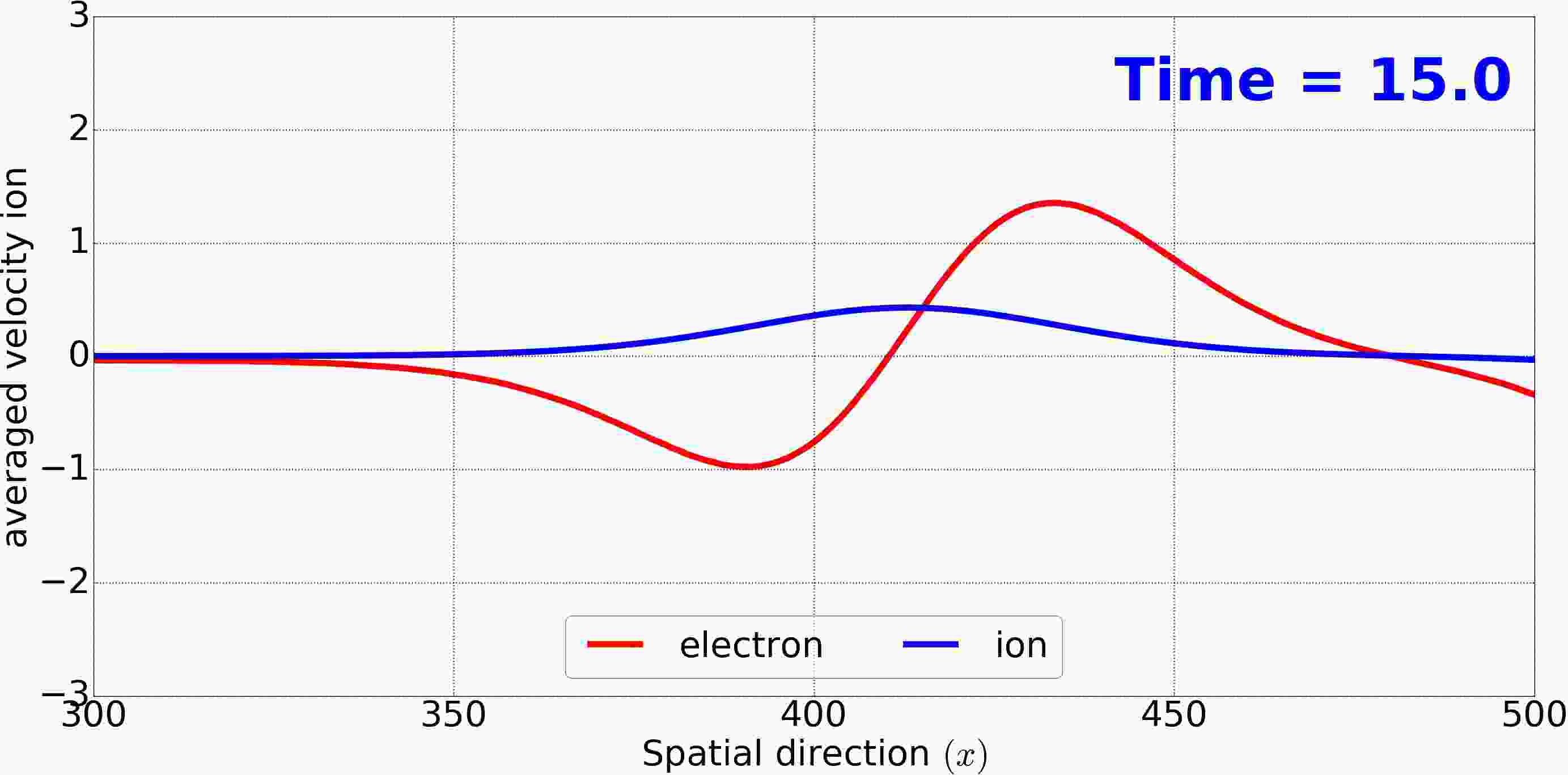}}\\
			\subfloat[$M = 1.75,\beta = -2.75$]{
 			\includegraphics[width=0.3\textwidth]{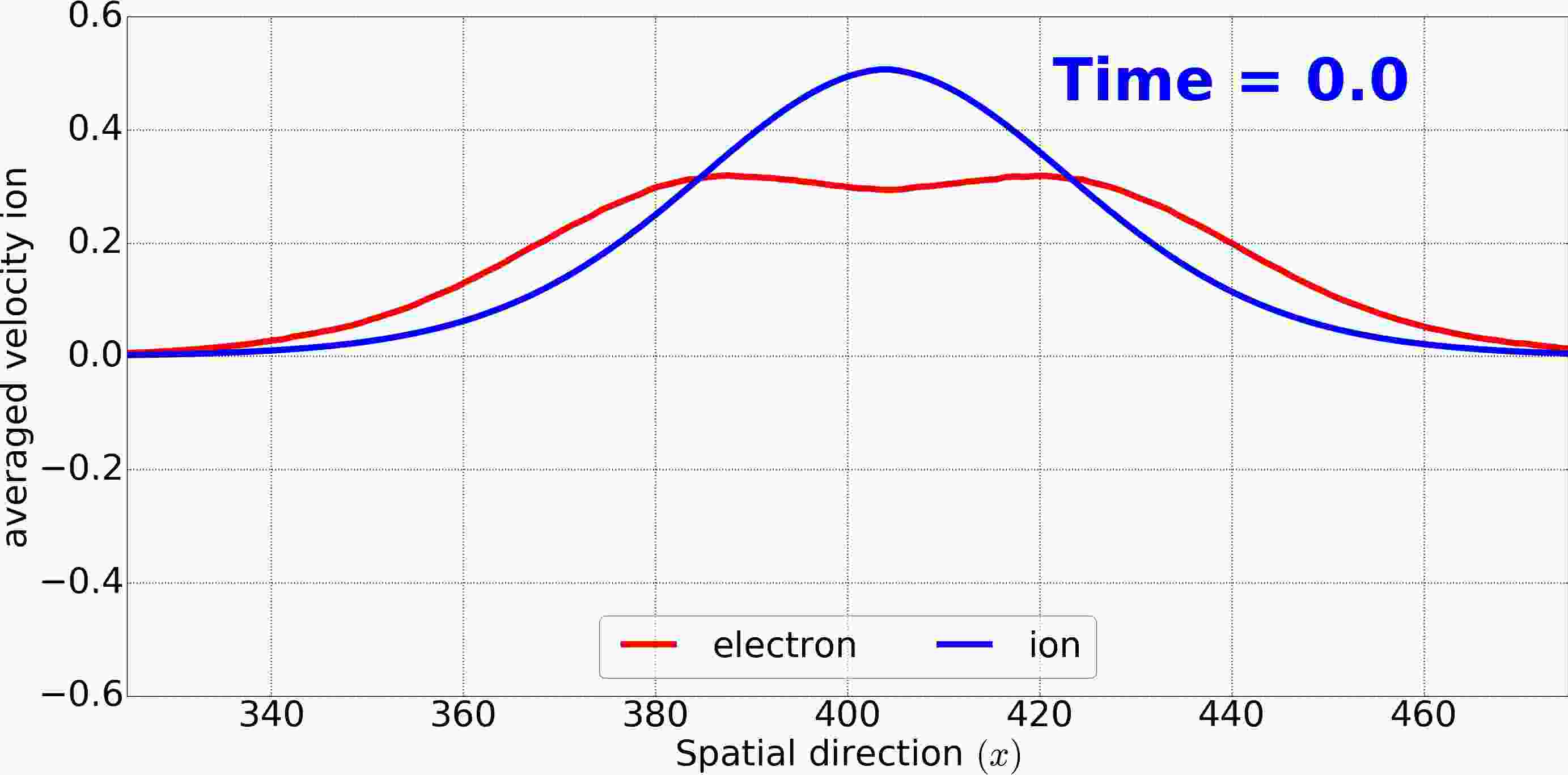} \hspace{0.1cm}
 			\includegraphics[width=0.3\textwidth]{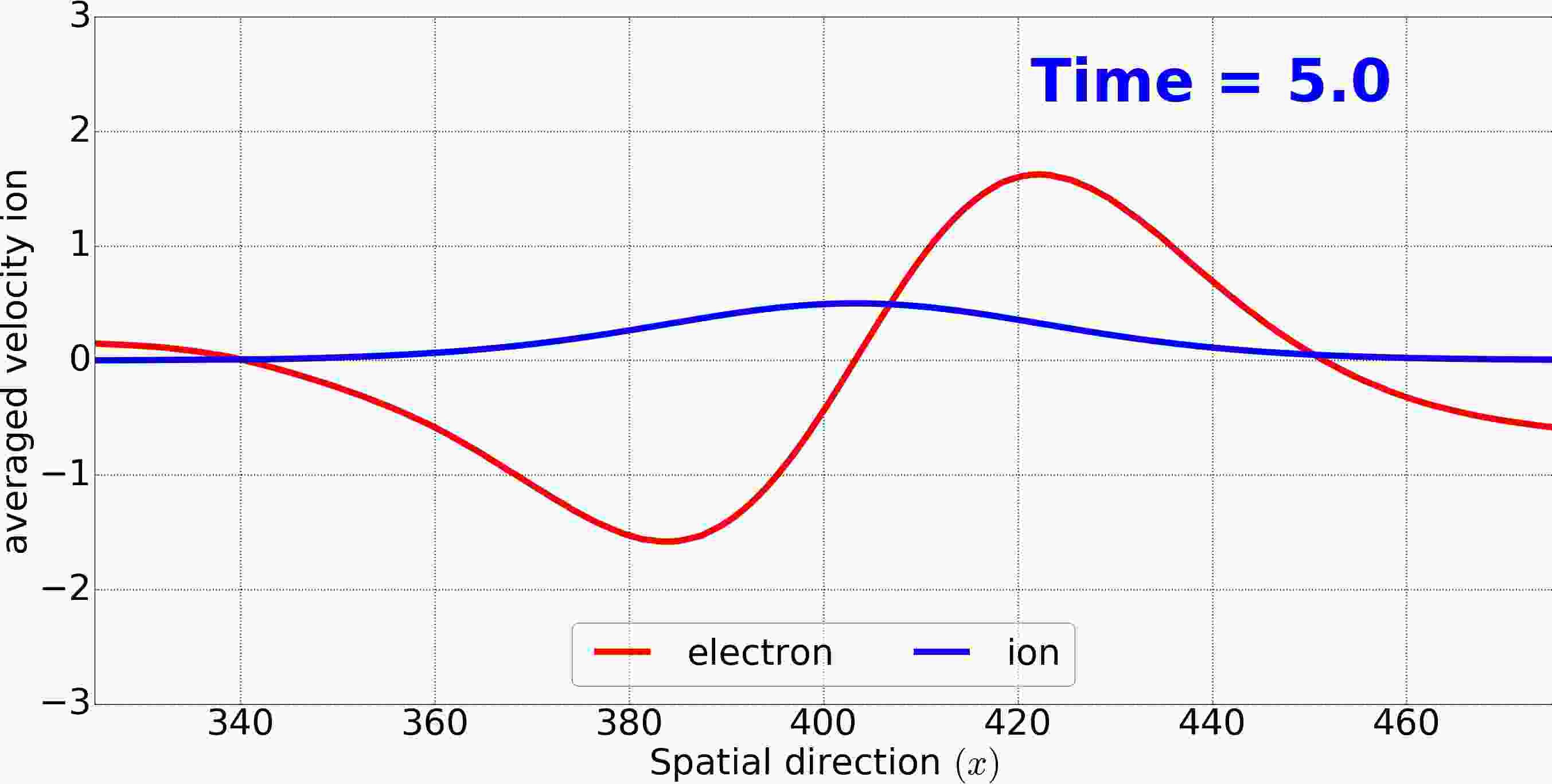} \hspace{0.1cm}
 			\includegraphics[width=0.3\textwidth]{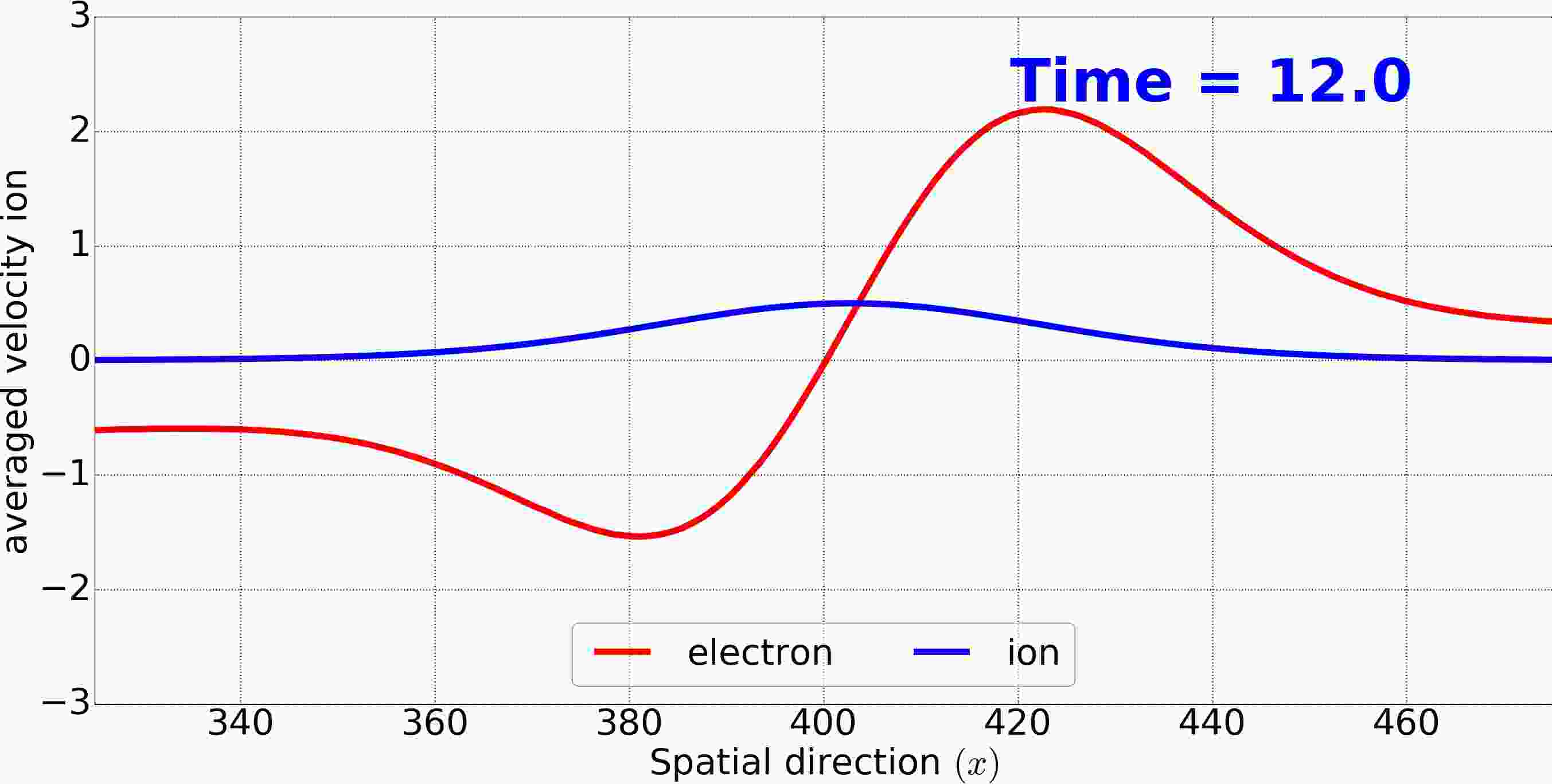}}
			\caption{The flux of electrons and ions along the x-direction is shown at three time steps, i.e. for initial condition ($\tau=0$),
			early in the temporal evolution ($\tau=5$) and before the first collision ($\tau\approx 15$). Four different cases are shown, 
			i.e. $(M,\beta)=$ a $(1.1,0)$, b $(1.25,-1.25)$, c $(1.5,-2.5)$ and d $(1.75,-2.75)$ 
			starting from the top. The bipolar structure develops fast in the early steps of the simulation. The amplitude of the bipolar structure 
			has a direct effect on the stability of the nonlinear solutions, the stronger the bipolar structure, the more unstable the solution becomes.
			Note that the ions flux have almost the same amplitude for all the cases.  }
			\label{Fig_DF_average_v_Mach_effect}
		\end{figure*}

		Fig.~\ref{Fig_DF_average_v_Mach_effect} represents the profile of the flux of the electrons and ions
		early in the simulations, presented in Fig.~\ref{Fig_collision_Mach_effect}. 
		Here the nonlinear solution has a positive velocity and hence moves to the right.
		One can observe that the flux of electrons forms a bipolar structure early in the simulation ($\tau \simeq 5$). 
		As the Mach number increases from $M=1.1$ to $M=2$, so is the amplitude of the
		bipolar structure and the instability of the solutions (compare this with Fig.~\ref{Fig_collision_Mach_effect}). 
		Note that the trapping parameter is selected in order to have the same profile for the flux of ions.

	\subsection{Internal dynamics of electron holes during collisions:}
	
		\begin{figure*}
			\subfloat[$M=1.1, \beta = 0$]{
			\includegraphics[width=0.25\textwidth]{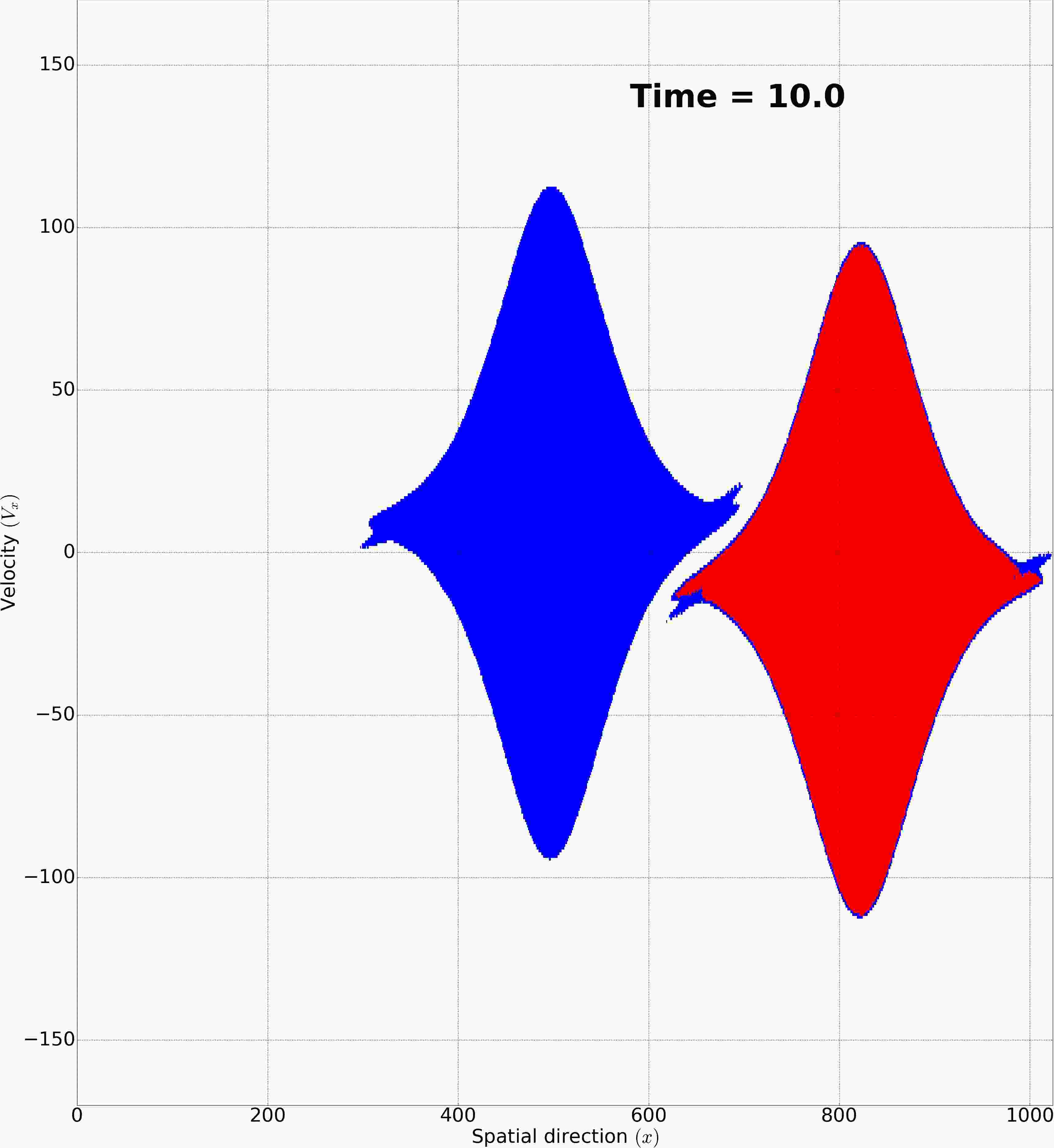}
			\includegraphics[width=0.25\textwidth]{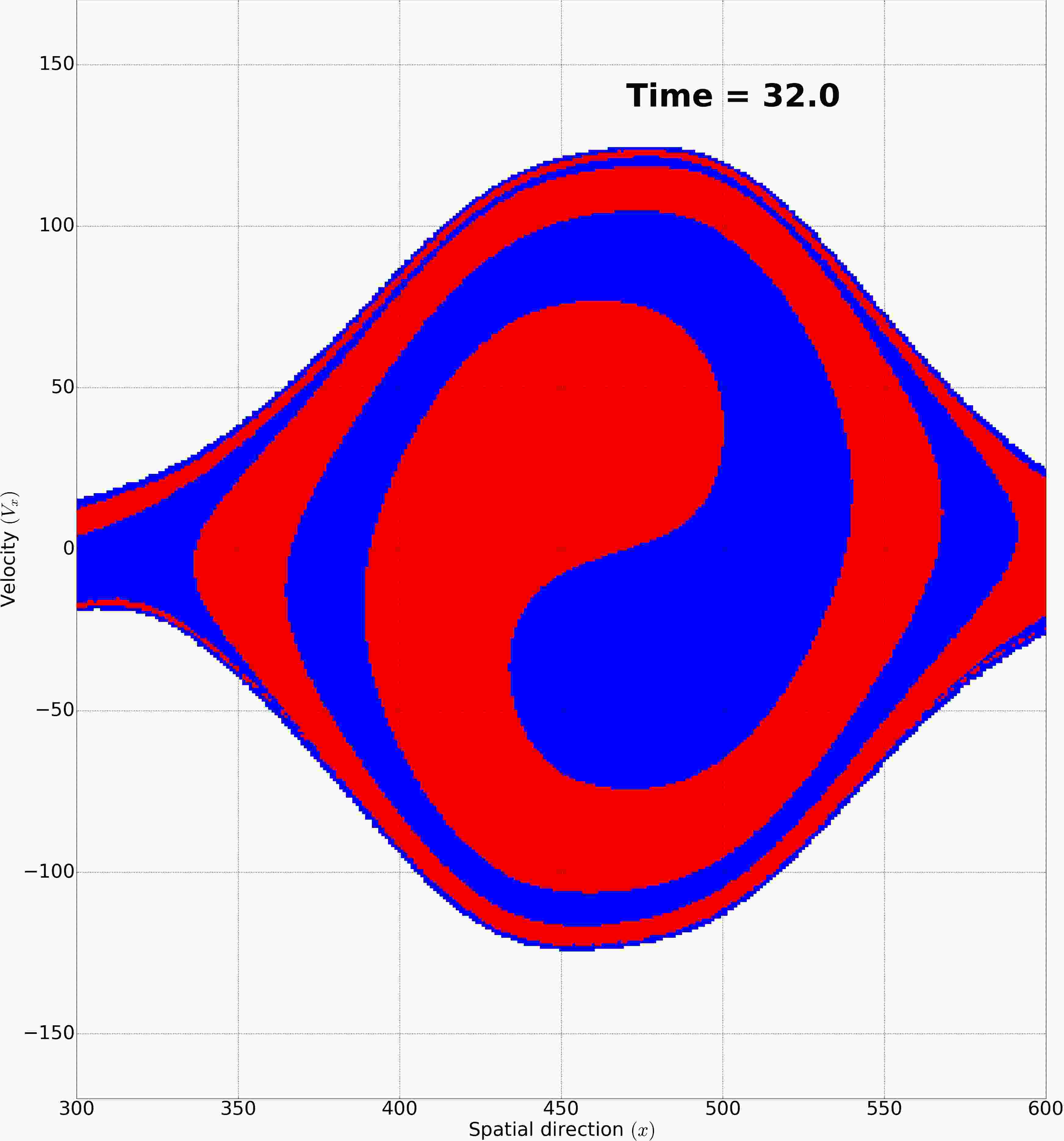}
			\includegraphics[width=0.25\textwidth]{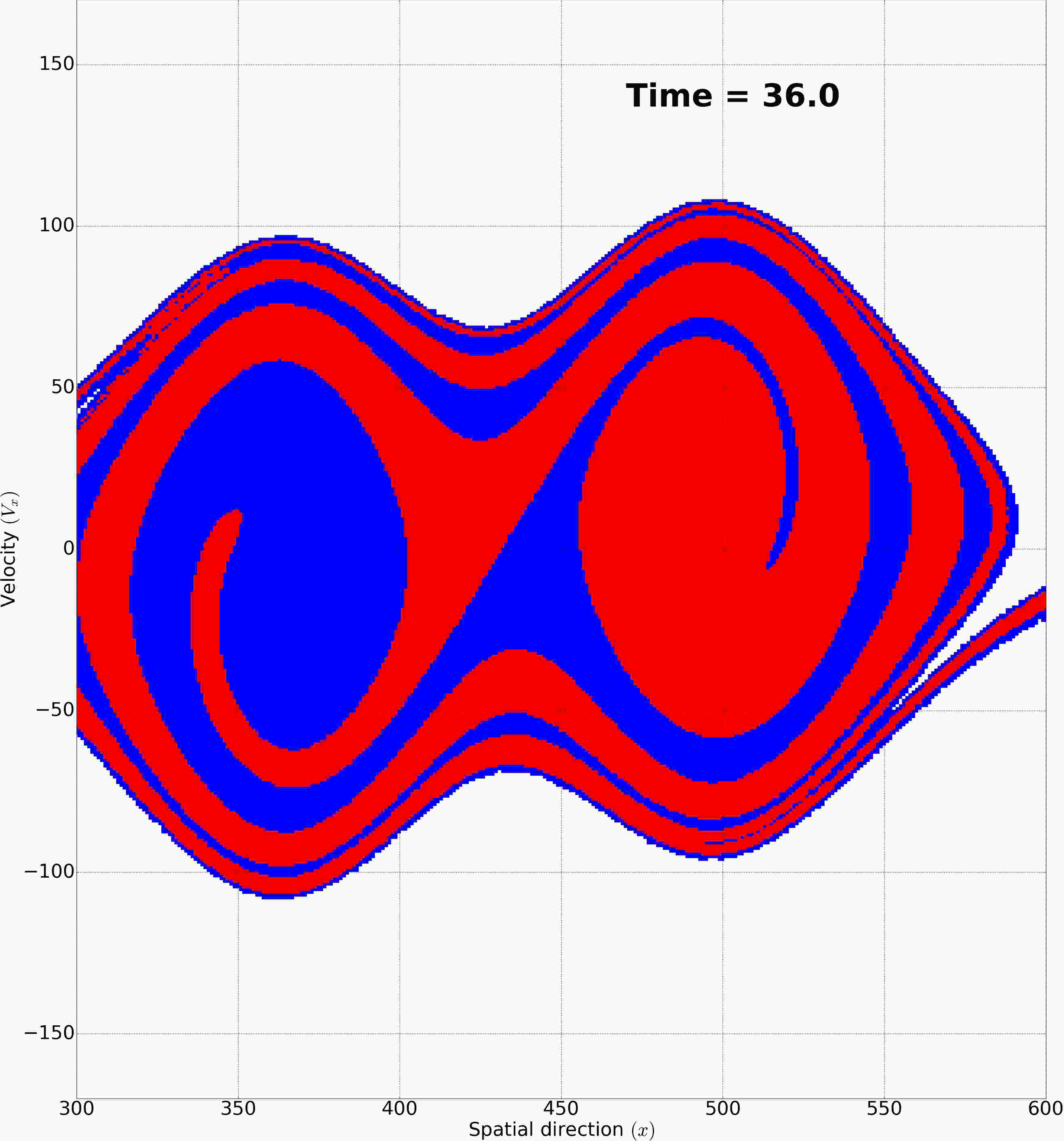}
			\includegraphics[width=0.25\textwidth]{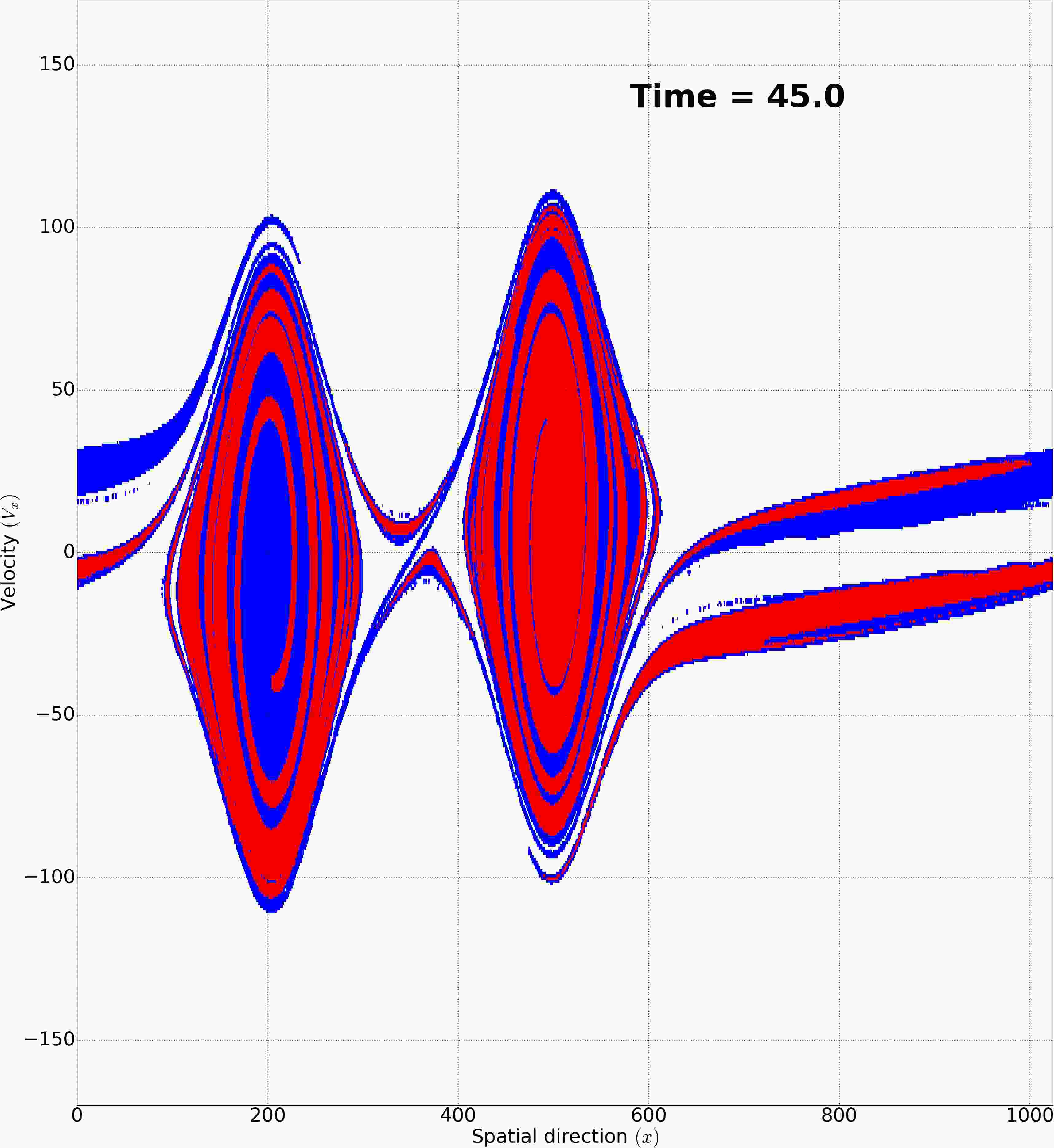}			
			} \\
			\subfloat[$M=1.25, \beta = -1.25$]{
			\includegraphics[width=0.25\textwidth]{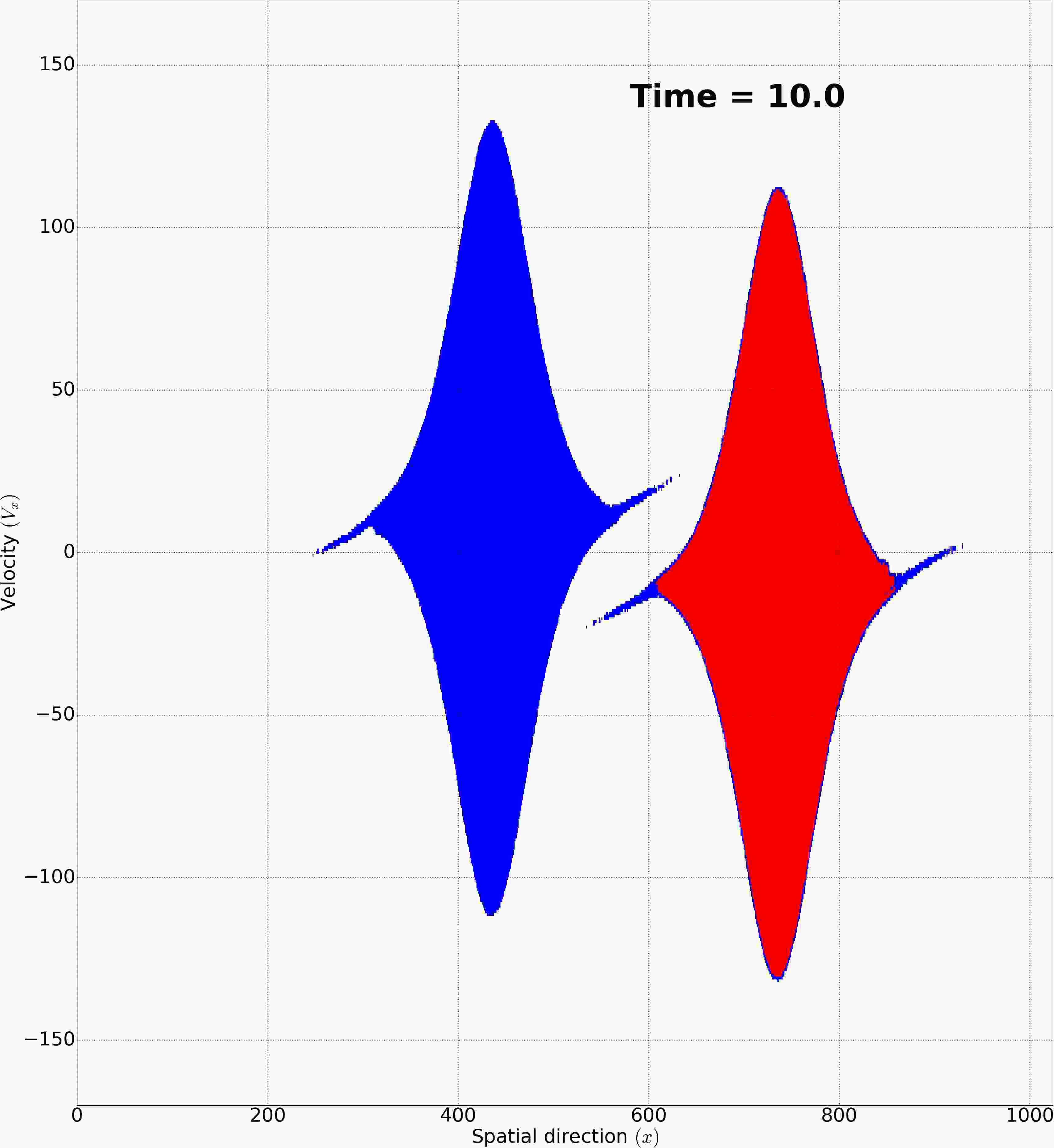}
			\includegraphics[width=0.25\textwidth]{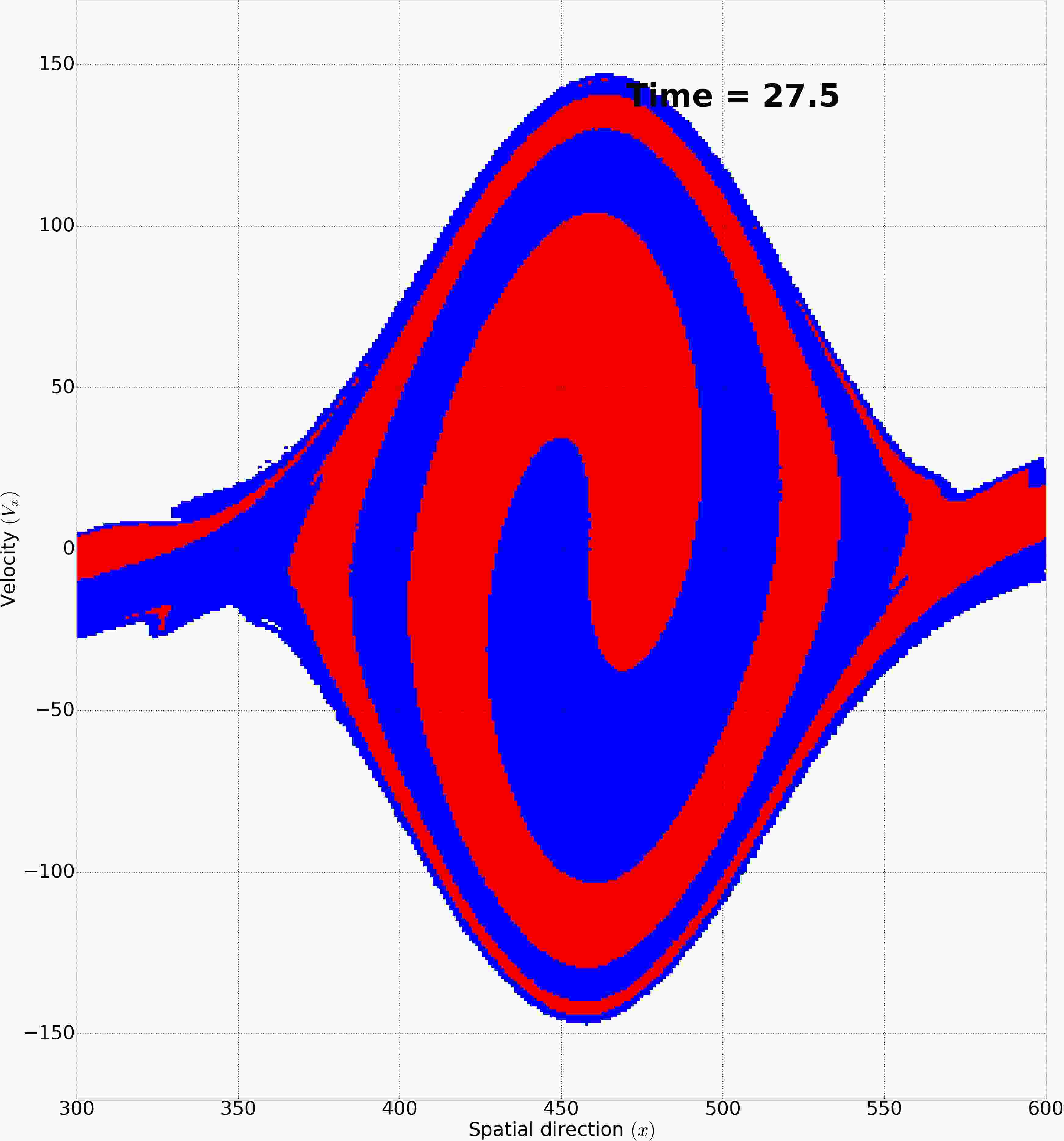}
			\includegraphics[width=0.25\textwidth]{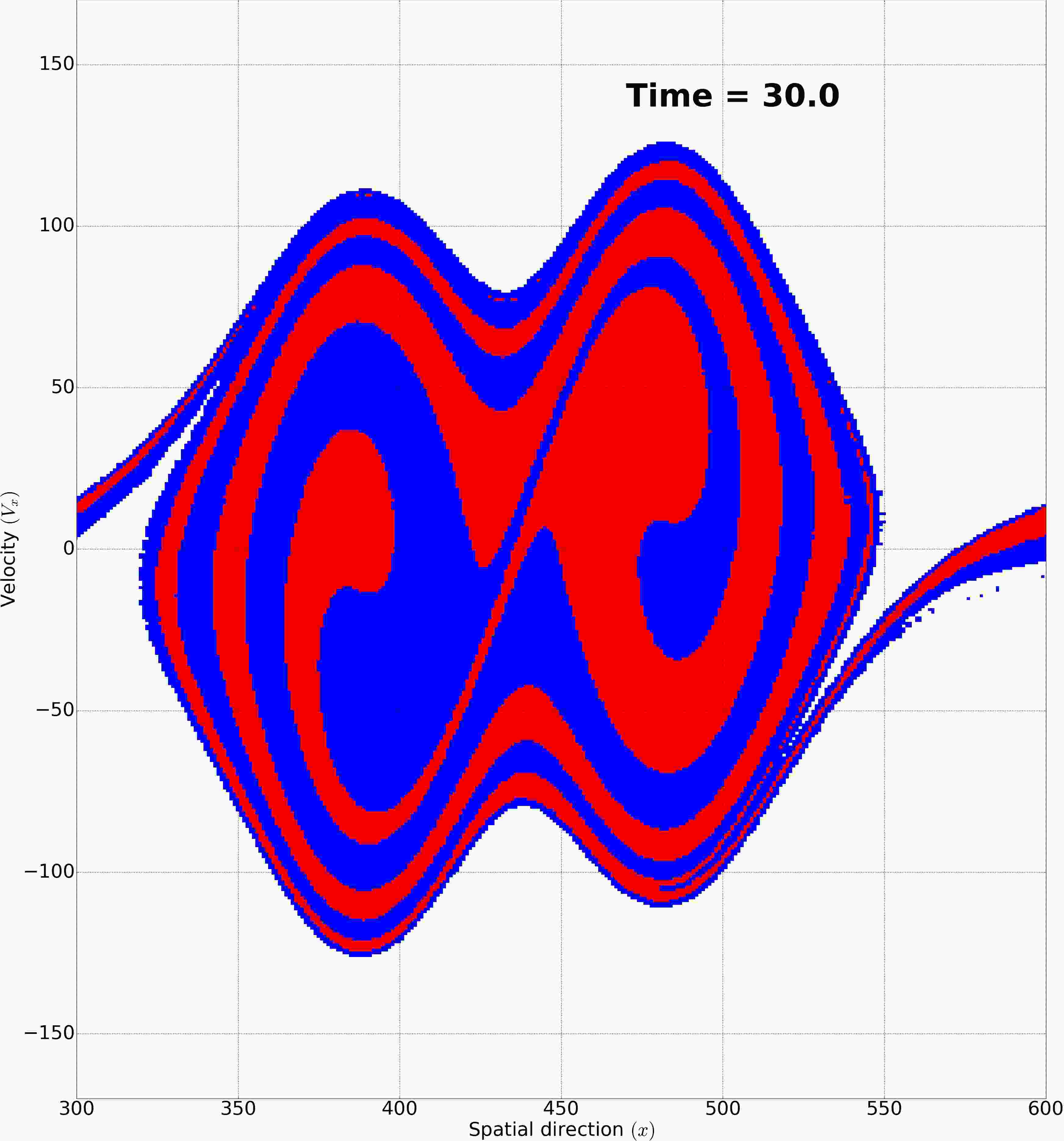}
			\includegraphics[width=0.25\textwidth]{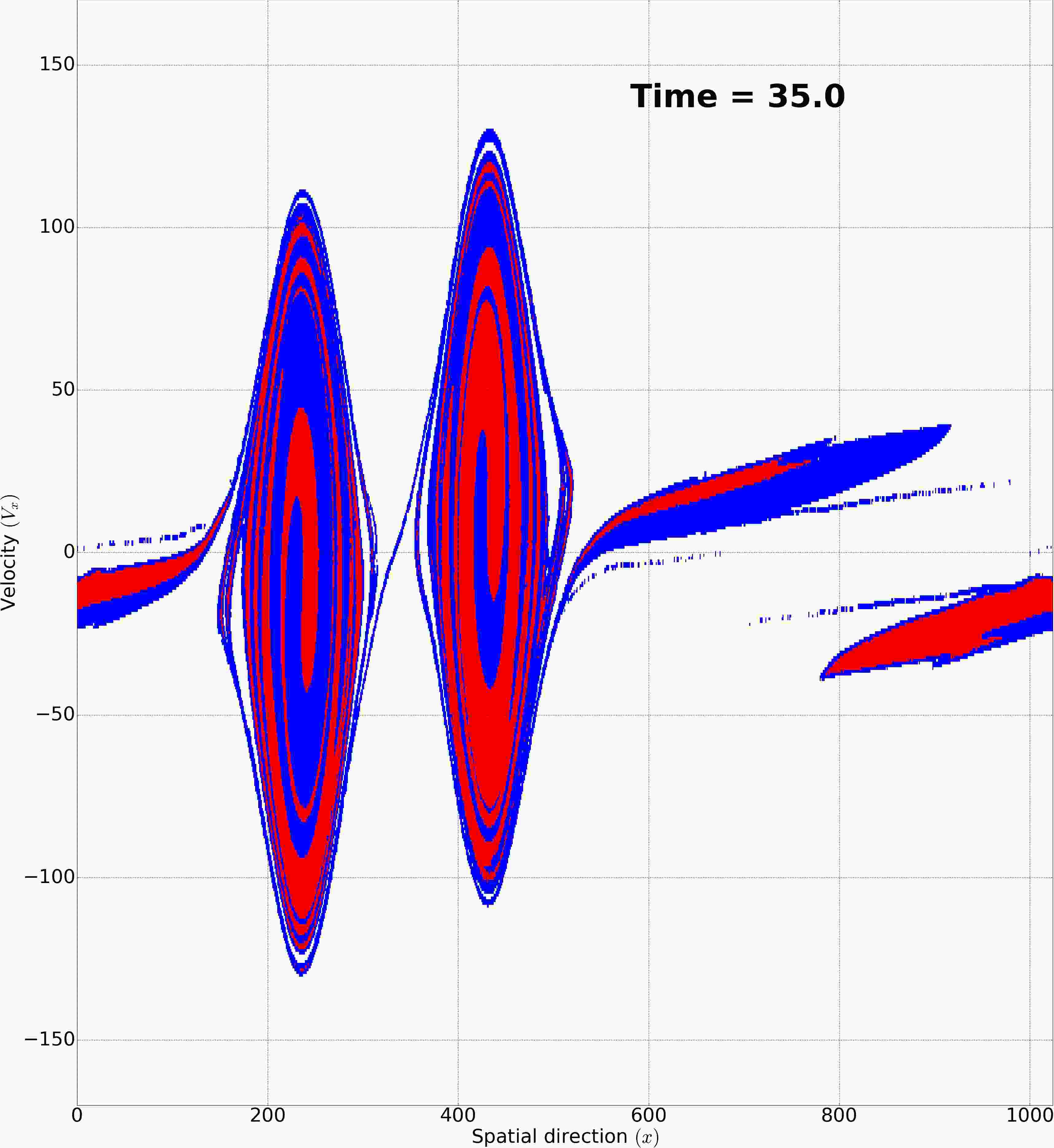}
			}\\
			\subfloat[$M=1.5, \beta = -2.5$]{
			\includegraphics[width=0.25\textwidth]{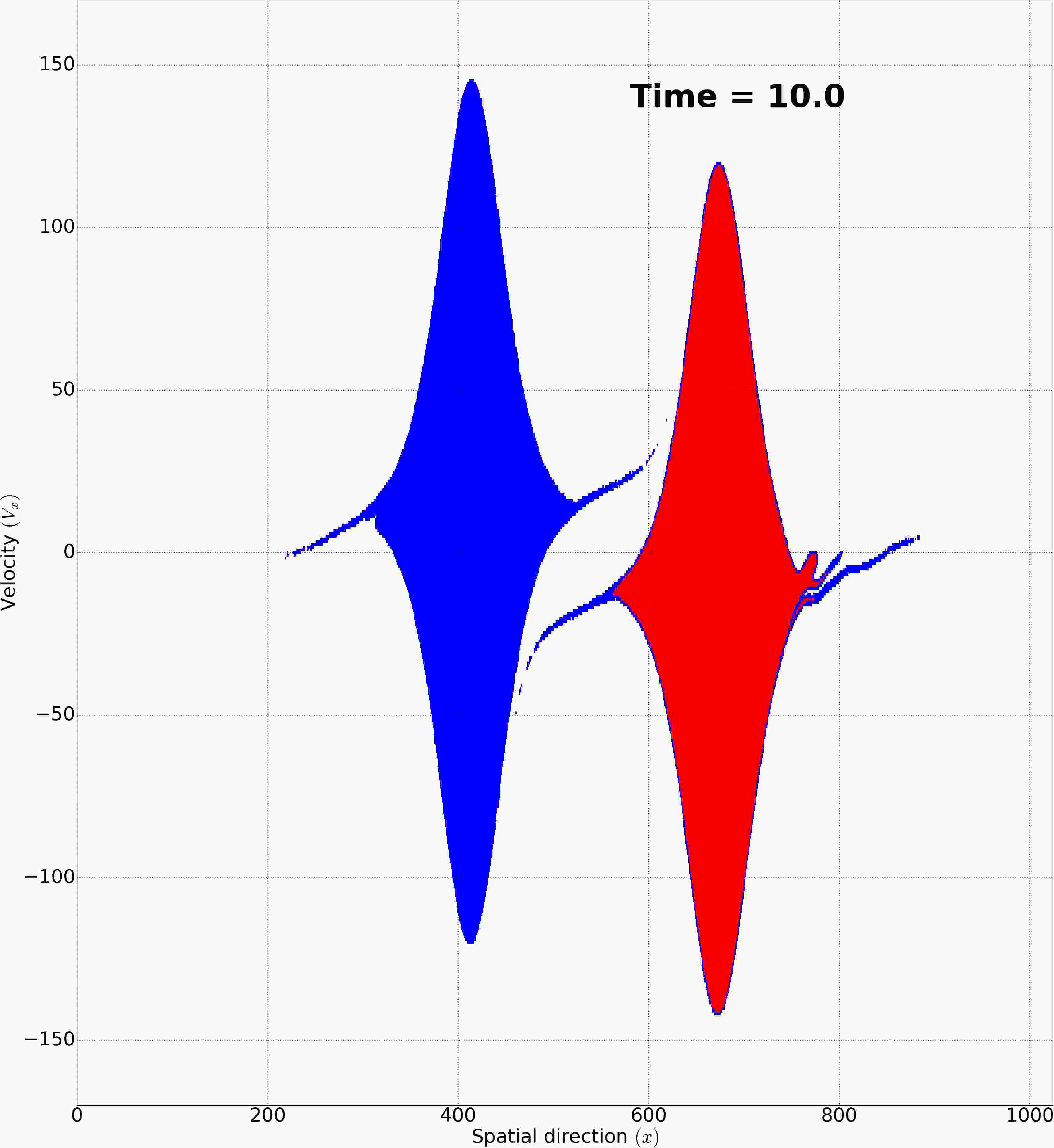}
			\includegraphics[width=0.25\textwidth]{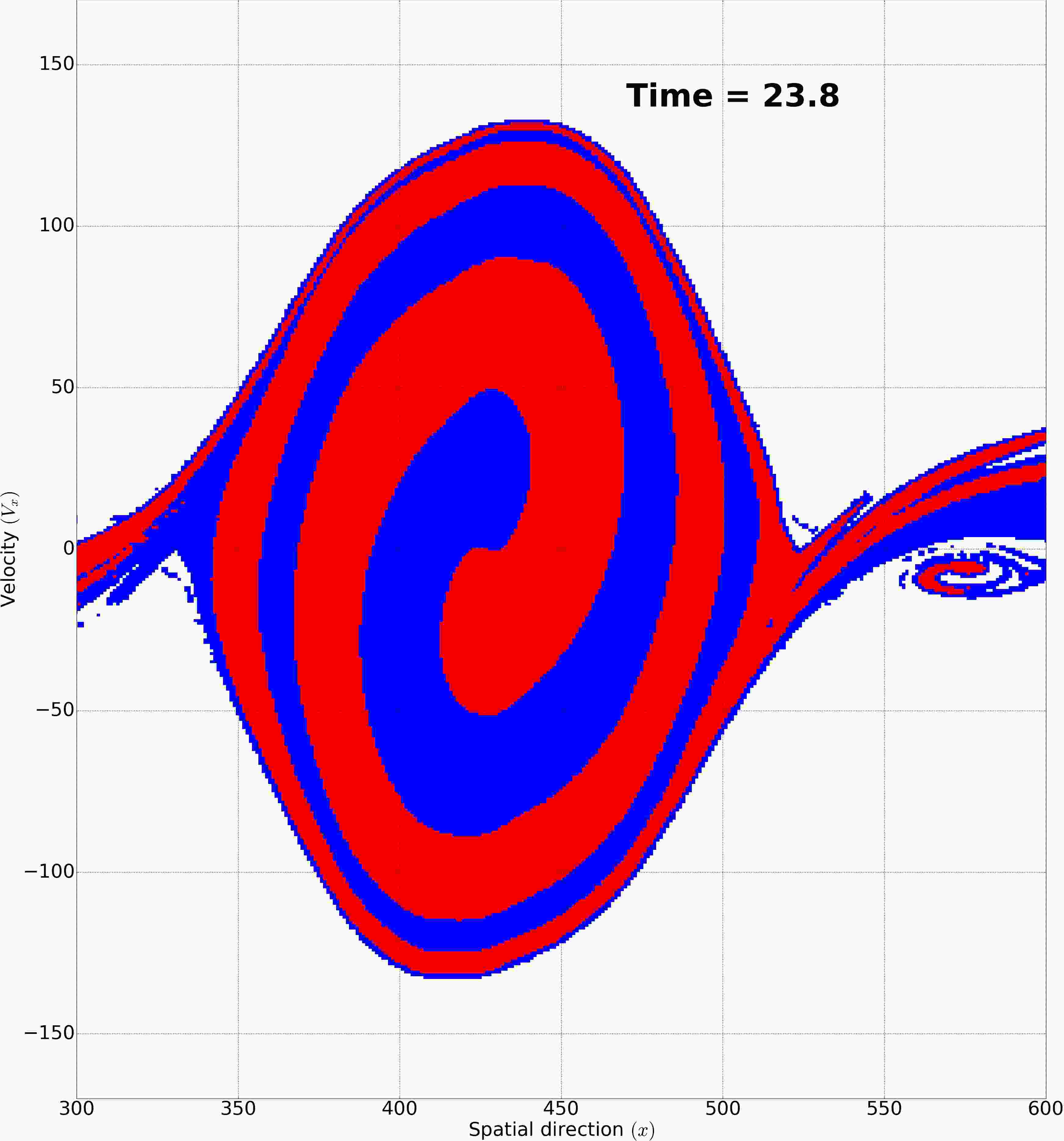}
			\includegraphics[width=0.25\textwidth]{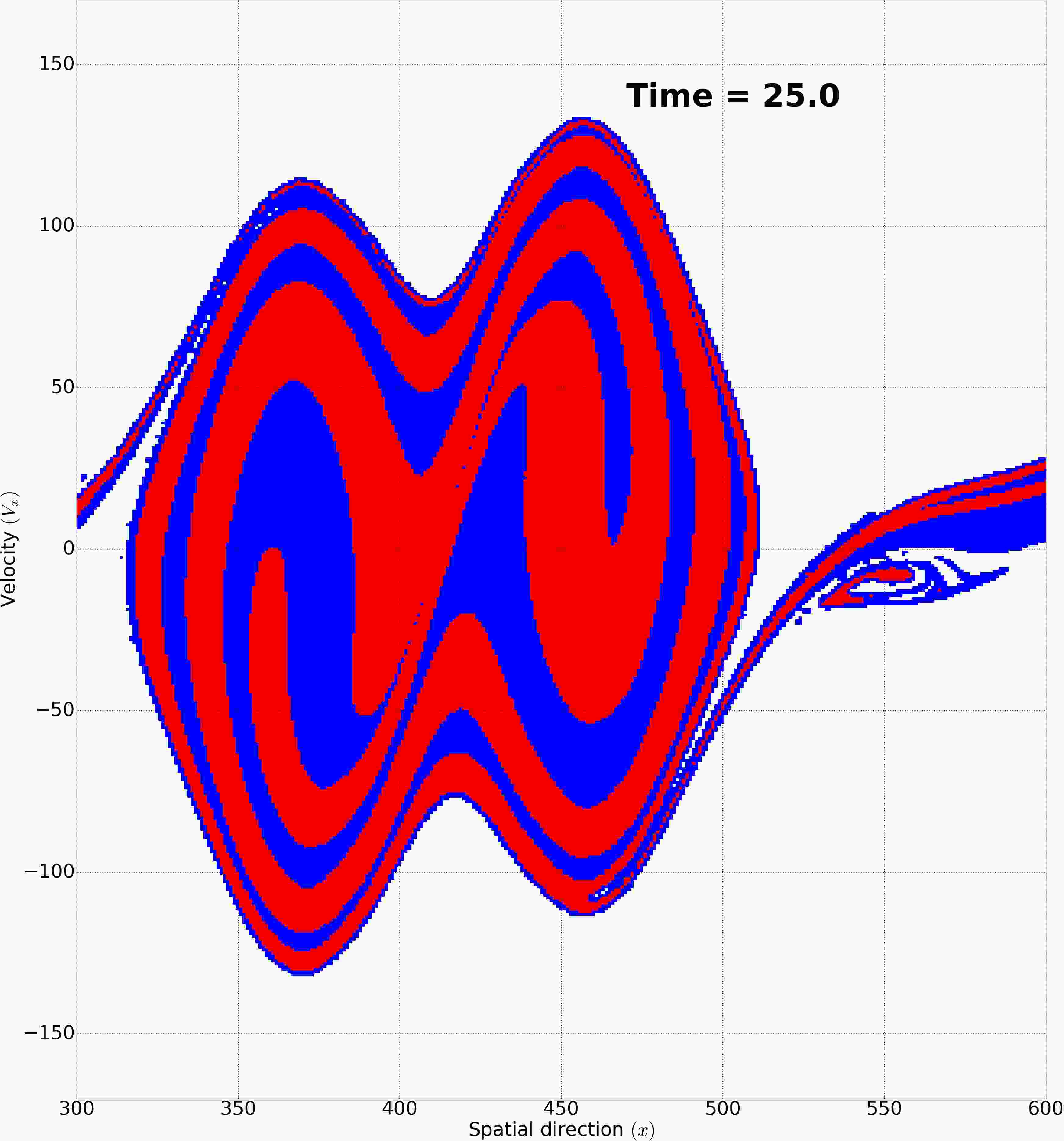}
			\includegraphics[width=0.25\textwidth]{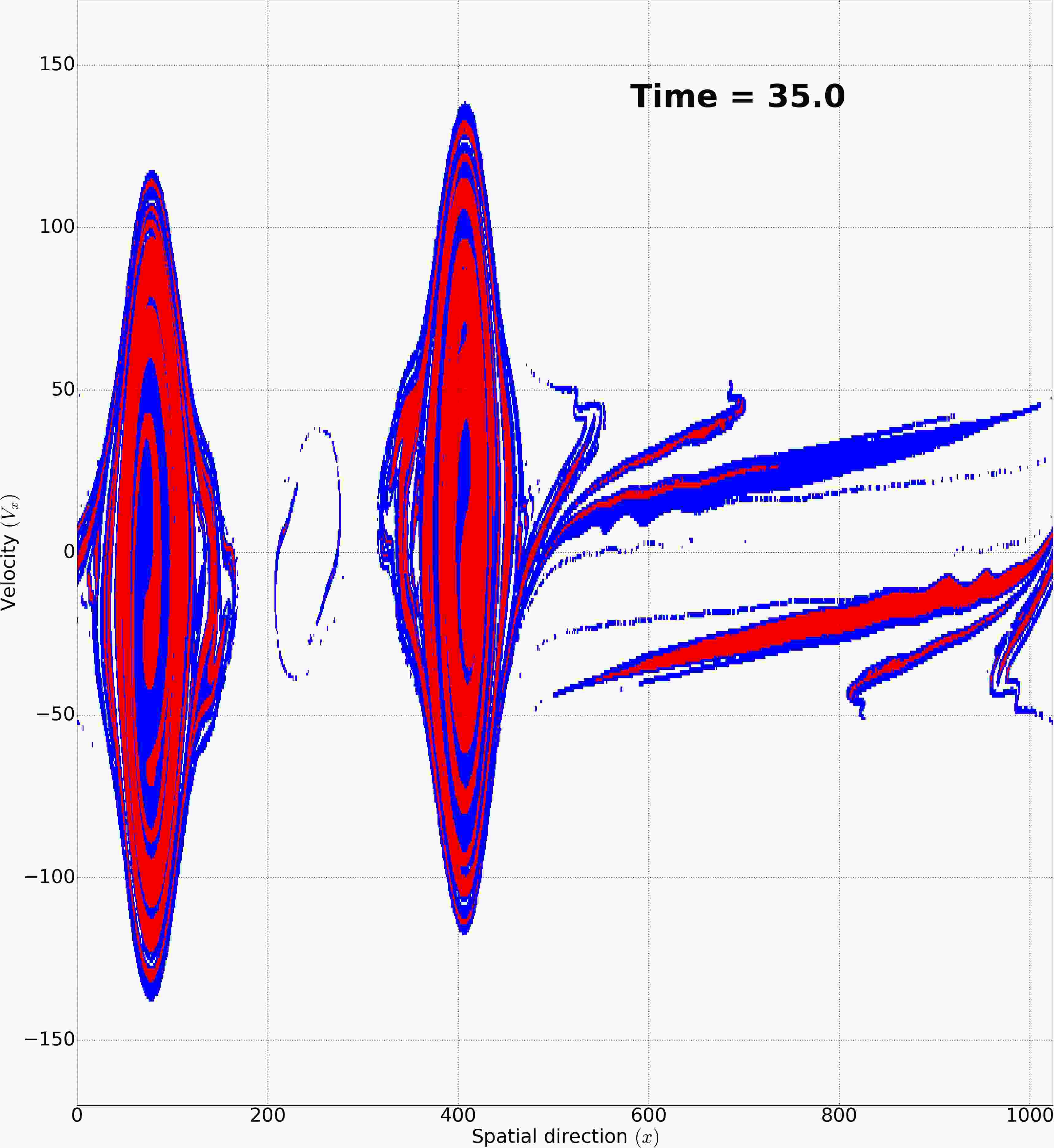}
			} \\
			\subfloat[$M=1.75, \beta = -2.75$]{
			\includegraphics[width=0.25\textwidth]{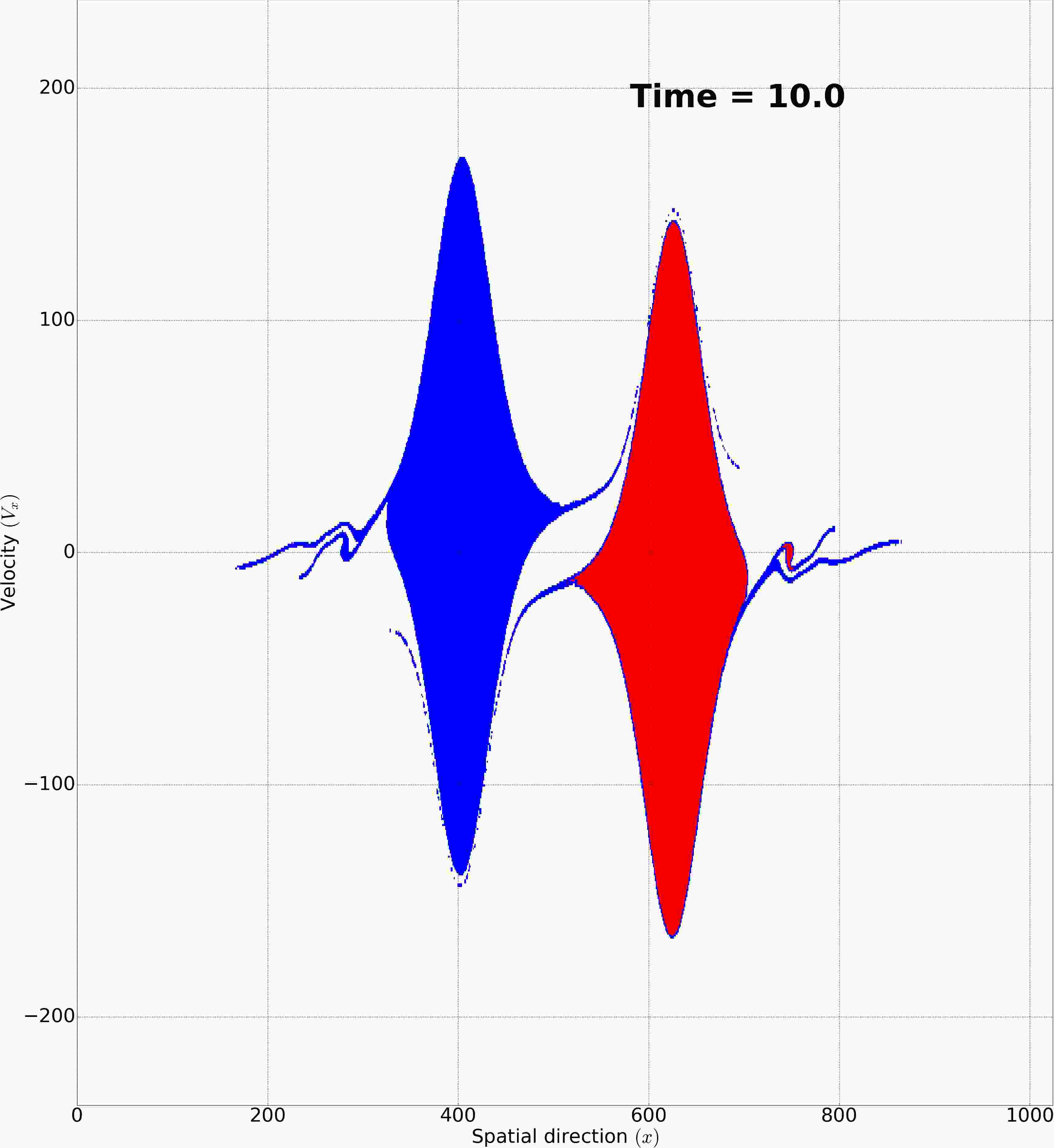}
			\includegraphics[width=0.25\textwidth]{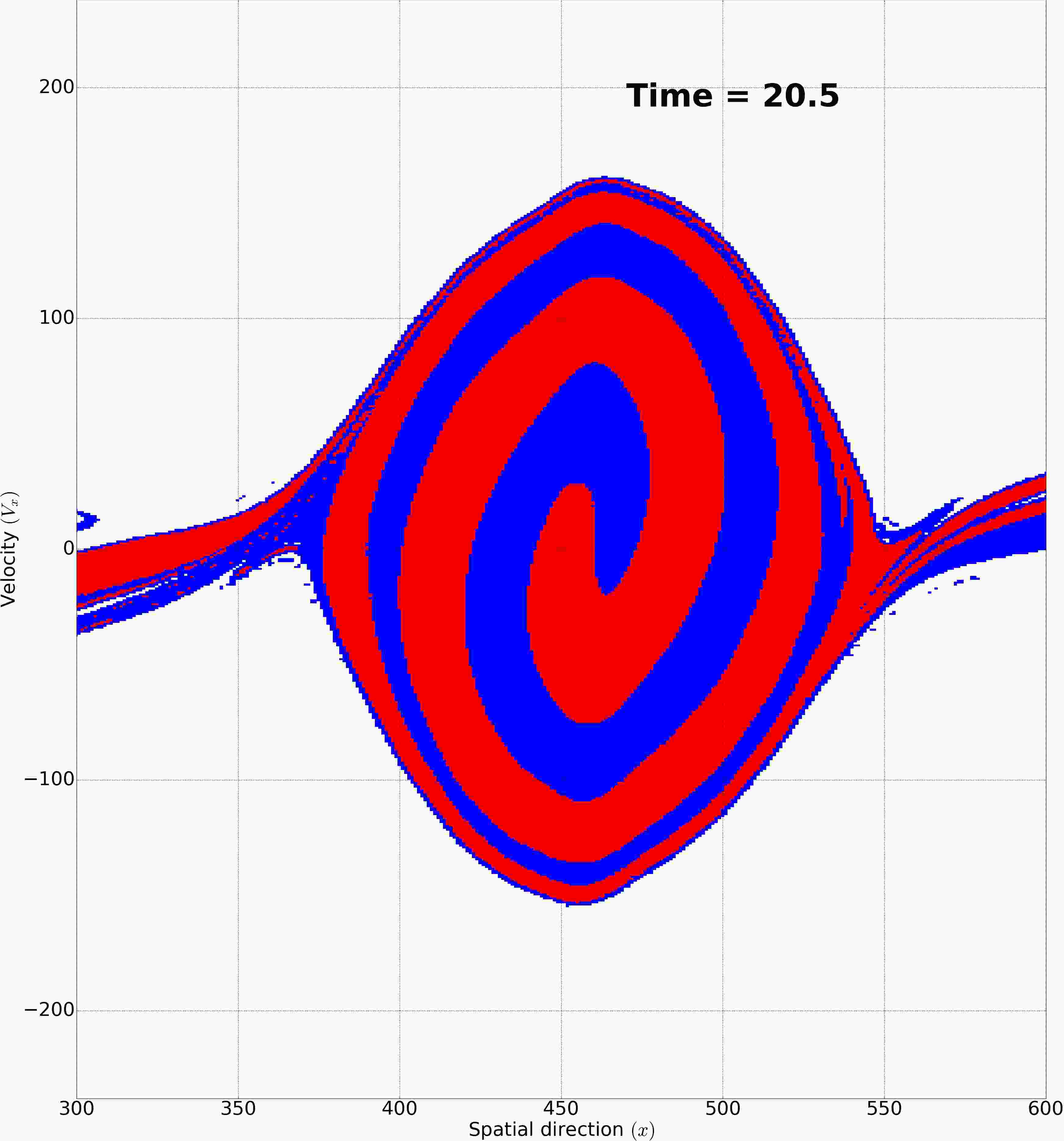}
			\includegraphics[width=0.25\textwidth]{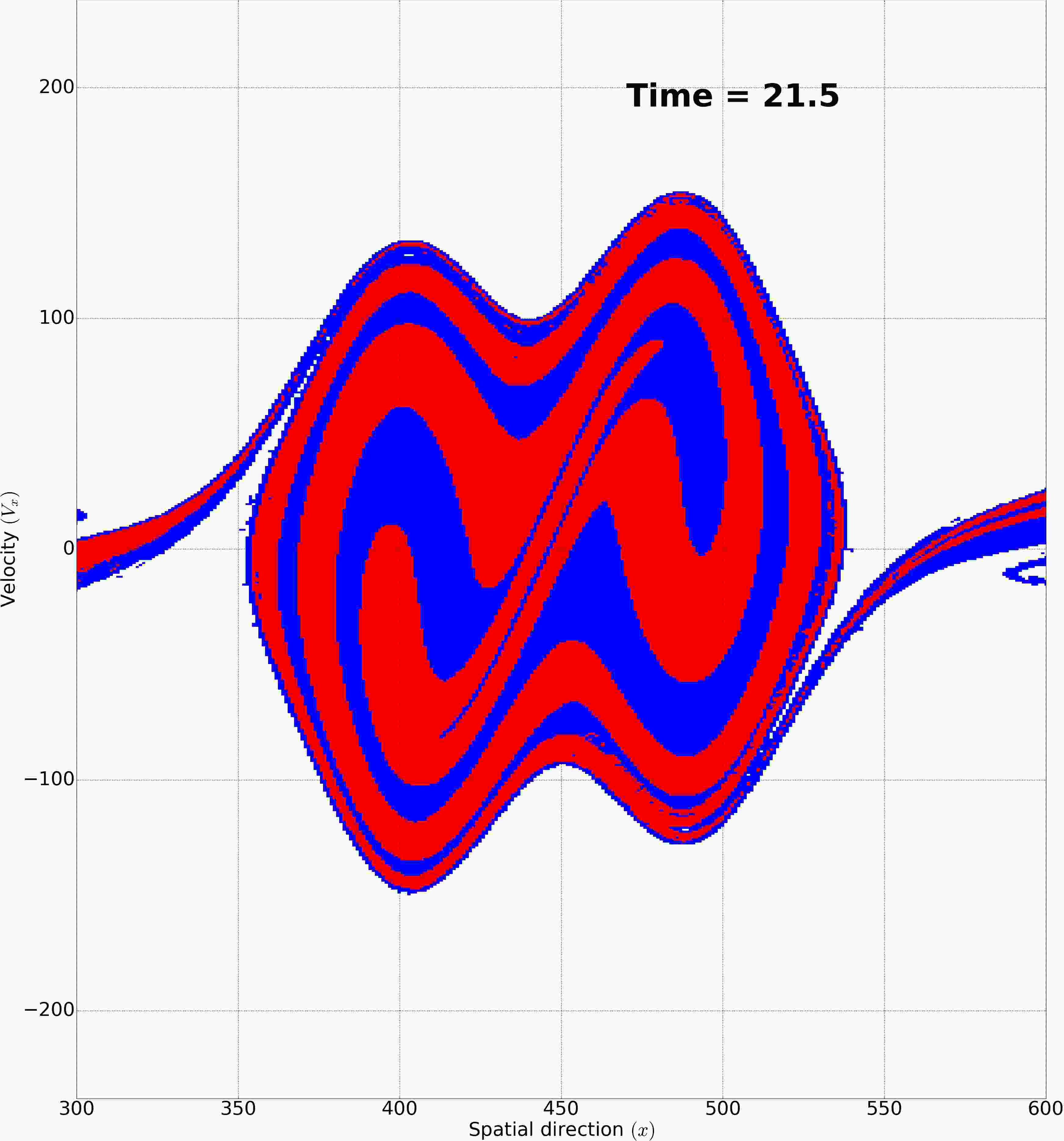}
			\includegraphics[width=0.25\textwidth]{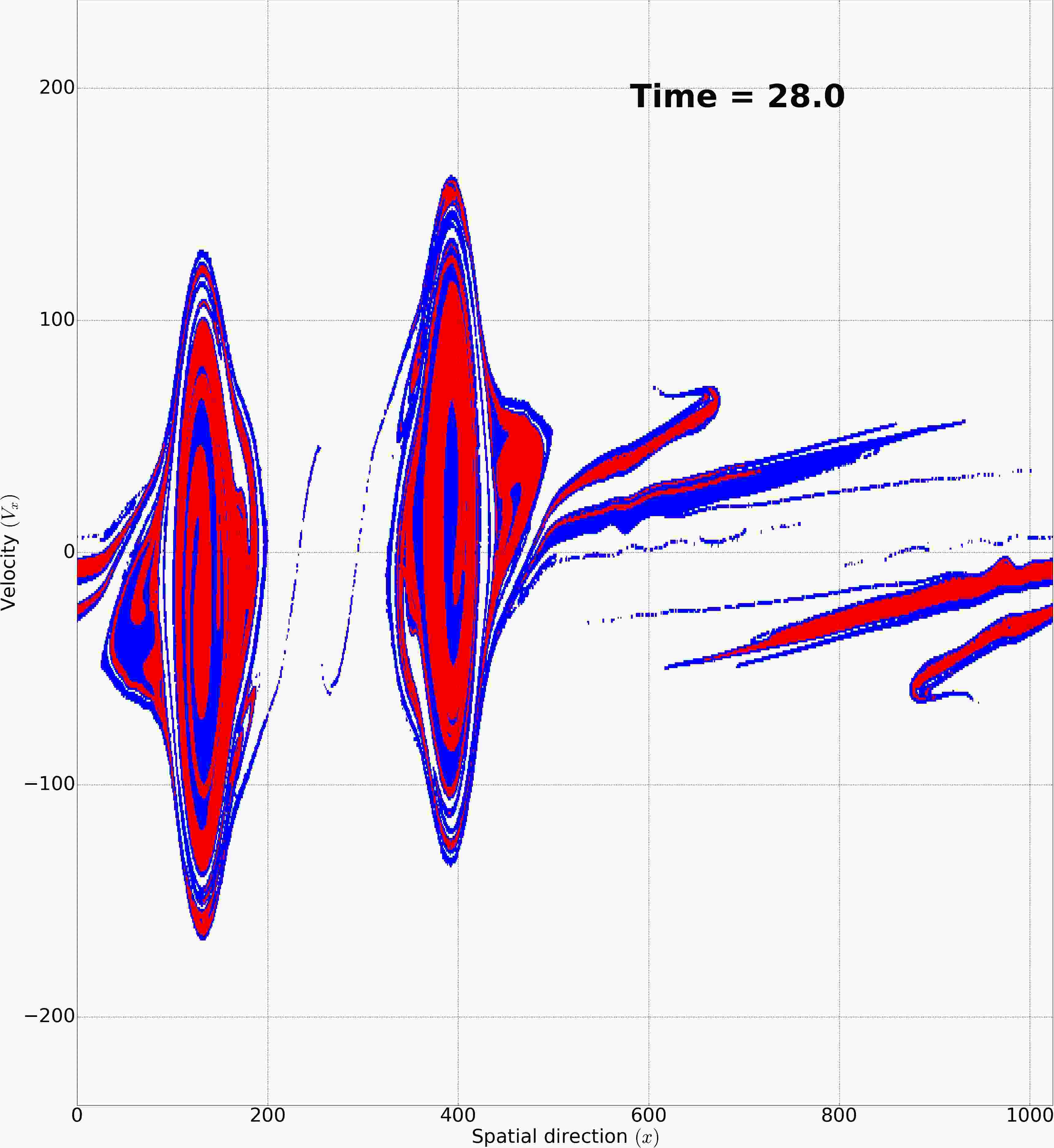}

			} 
			\caption{Details of the electron phase space are presented before the collision (first column),
			during the merging phase (second column), 
			during the splitting phase (third column) and after the splitting (fourth column) of the first collision.
			The same four cases as presented in Fig. 
			\ref{Fig_collision_Mach_effect} are shown.
			The number of rotations changes by increasing the relative speed of the colliding electron holes.
			This can be easily observed by comparing the figures in second column with each other. }
			\label{Fig_hole_dynamics}
		\end{figure*}
		
		Fig.~\ref{Fig_hole_dynamics} shows the temporal evolution of the electron holes during collisions for different cases of
		relative speed. 
		During the merging phase of a collision, two electron holes rotate around the collective center of mass. 
		Our results show that the number of rotations depends on the relative speed of the electron holes.
		The expectation would be that as the relative speed increases, the time span of collision decreases and hence 
		a fewer number of rotations can take place.
		However, it can be observed that as the relative speed increases, the number of rotations grow.
		In other words, electron holes spin faster when their relative speed is higher.
		
		As the Mach number increases, the instability grows and hence the electron hole after collisions show a larger 
		deviation from the stable shape. 
		In cases of $(M=1.5,\beta=-2.5)$ and $(M=1.75,\beta=-2.75)$, it can be seen that the electron holes 
		lose a rather large chunk of their trapped population after the collision. 
		The instability originates from the bipolar structure in electron flux, that pushes the 
		trapped population in opposite direction on the two ends of the electron hole.
		Hence, the electron hole loses parts of the trapped population on both ends of its domain.
		At the front of the electron hole (right/left side if the electron hole propagate to right/left), 
		the trapped population is pushed upwards and leave the electron hole as a stretched arms in front of it.
		While on the rear side, the trapped population leaks out in the form of smaller electron holes leaving the 
		trapped area. 
		
\begin{figure*}
			\subfloat[First collision]{\includegraphics[width=0.35\textwidth]{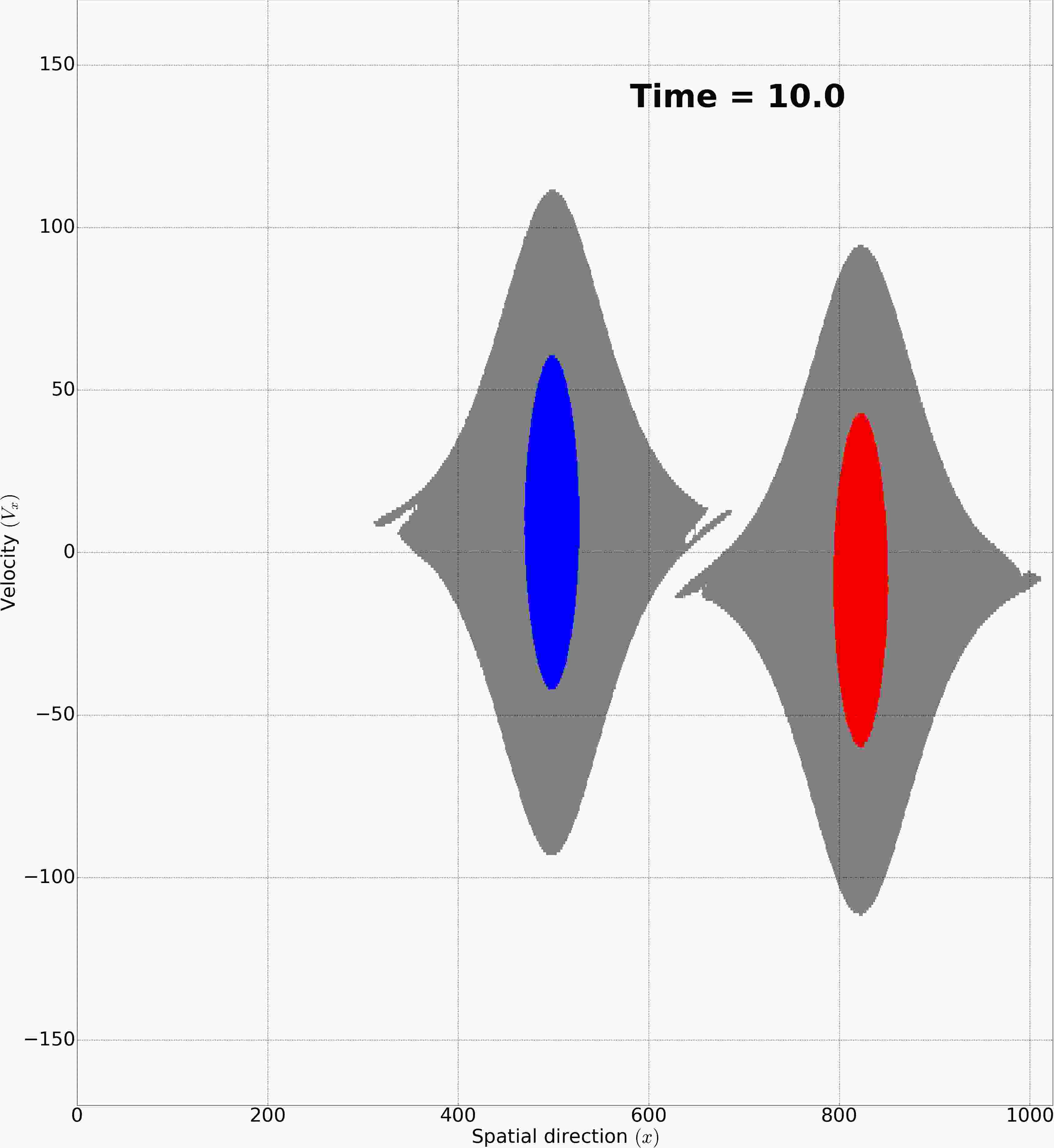}\hspace{1.5cm}
						   \includegraphics[width=0.35\textwidth]{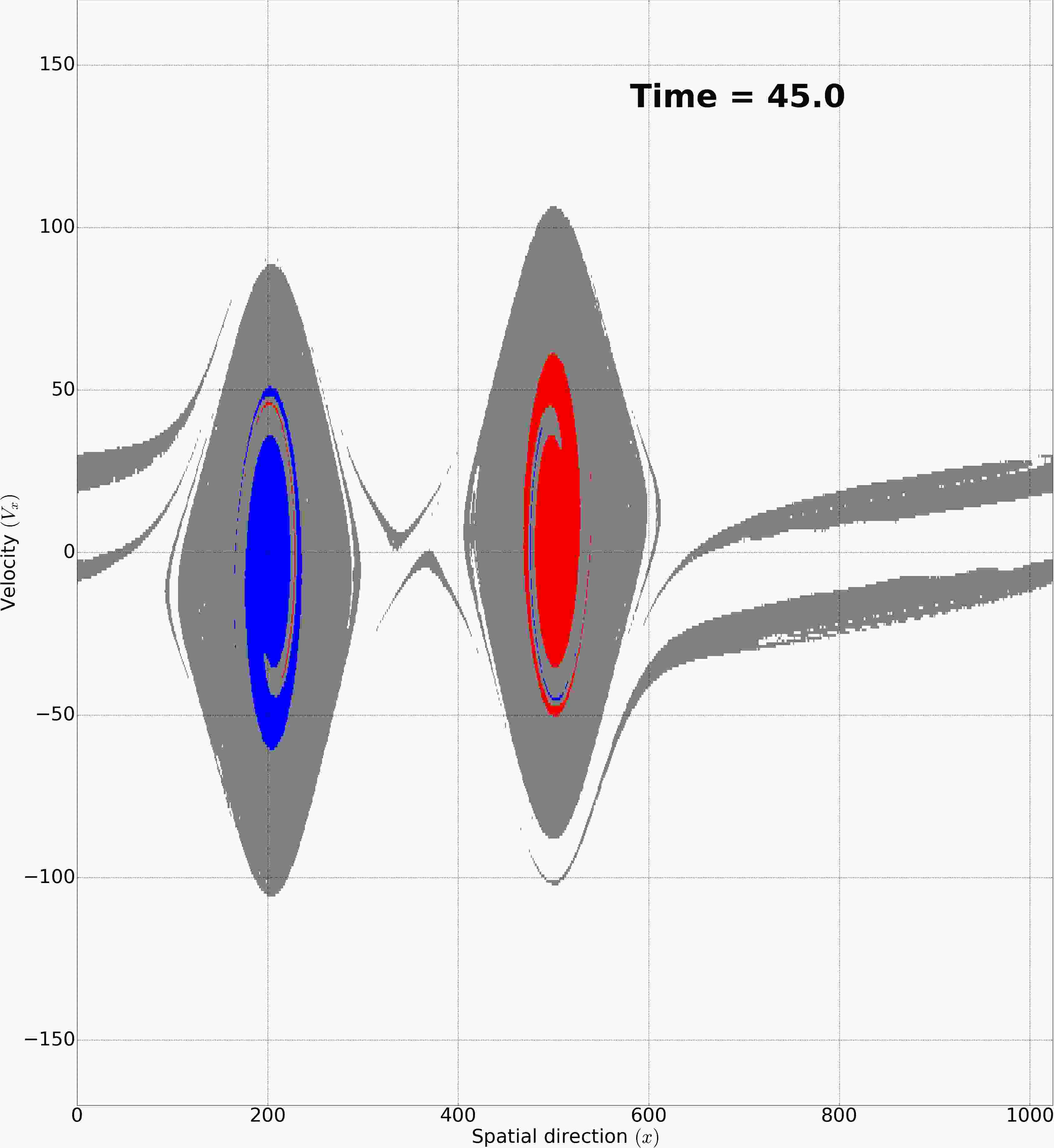}} \\
			\subfloat[Second collision]{\includegraphics[width=0.35\textwidth]{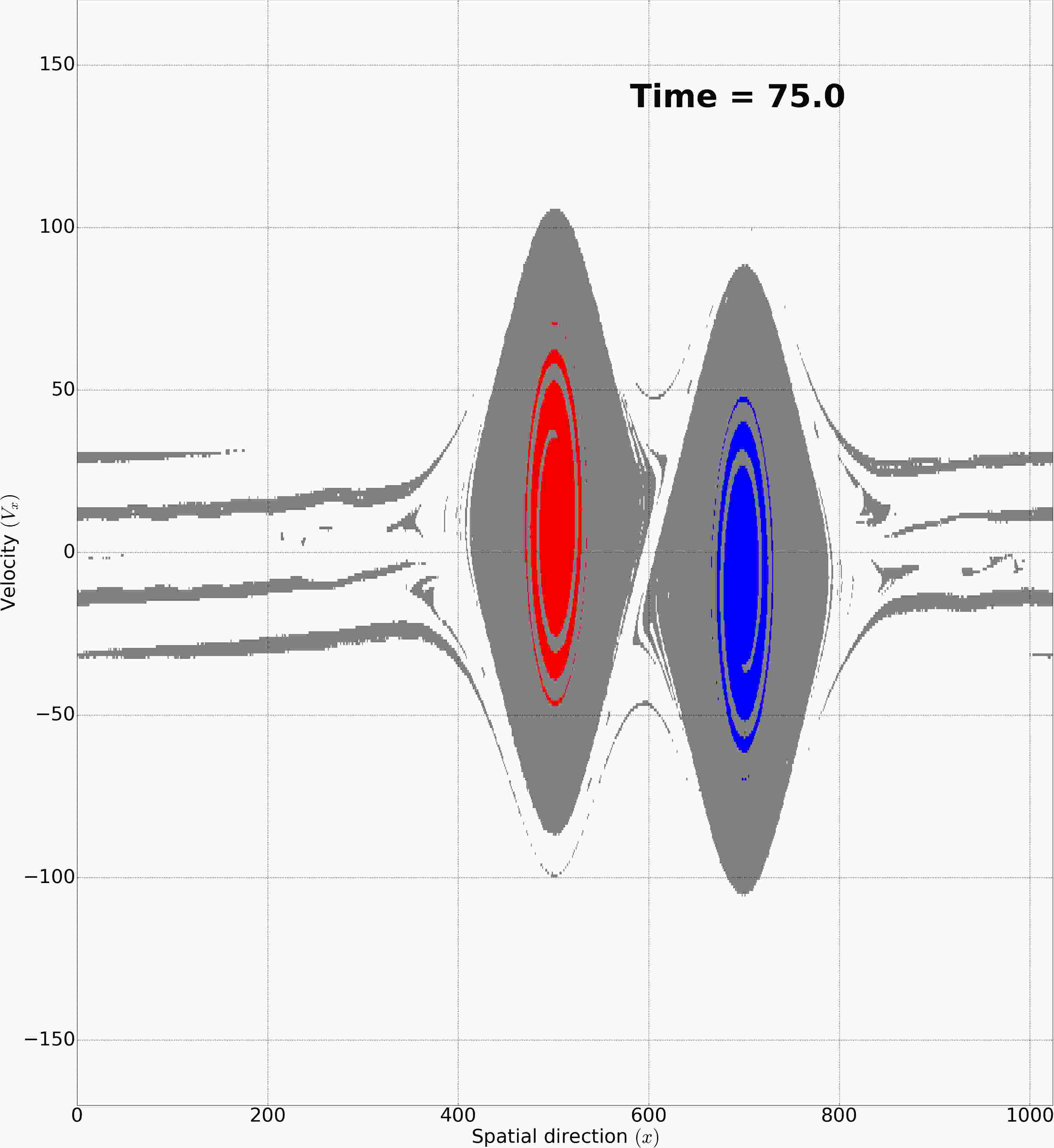}\hspace{1.5cm}
						    \includegraphics[width=0.35\textwidth]{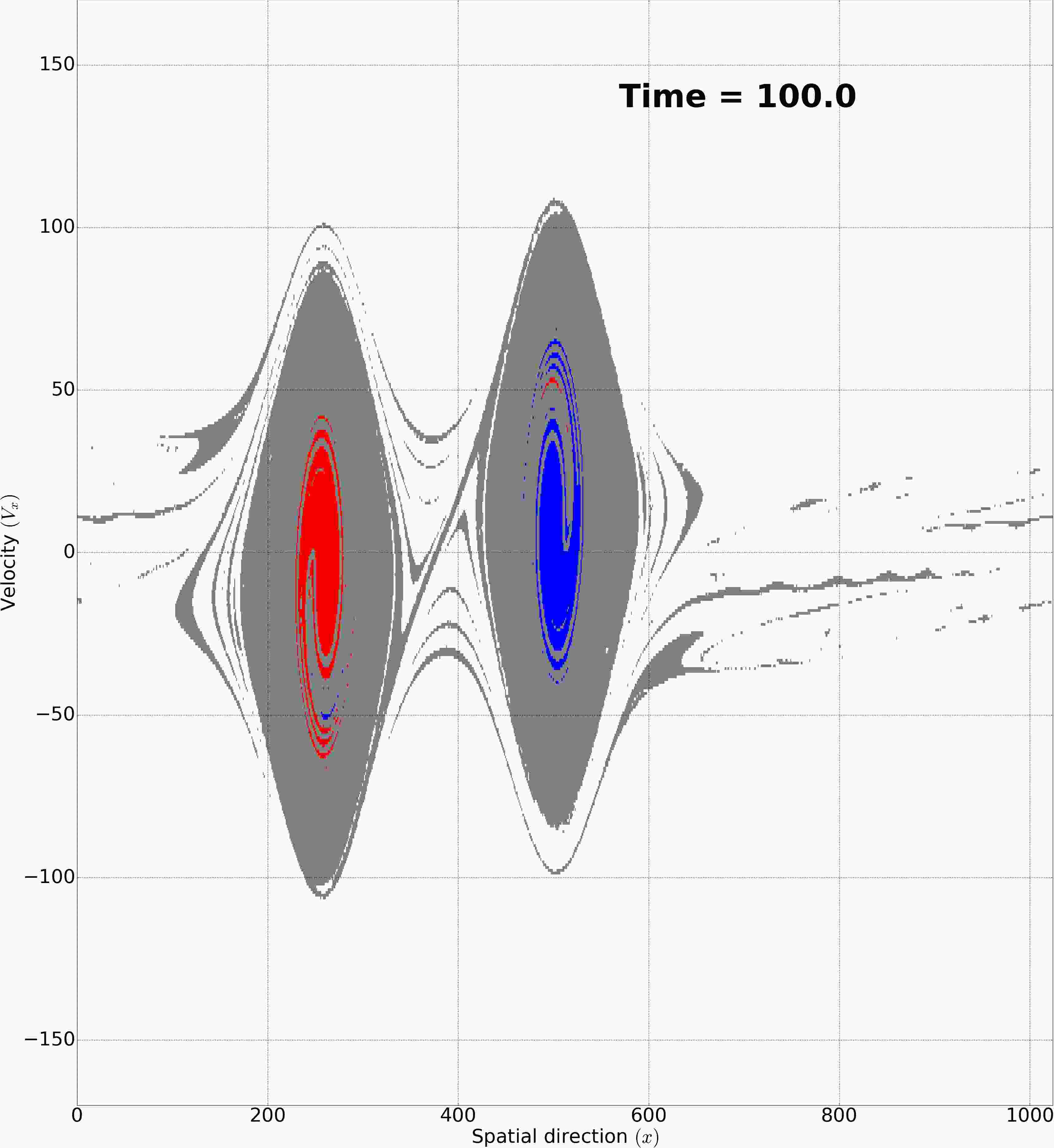}}
			\caption{A symmetrical exchange of the cores during each collision is shown for the case of $M=1.1, \beta=0$.
			Four snapshots of the phase space of electrons is presented including a)first collision  before ($\tau=10$) and after ($\tau=45$),
			b) second collision, before ($\tau=75$) and after ($\tau=100$). 
			The cores are marked with red and blue while the rest of electron holes are shown with gray color.
			Note that the frame of plots is fixed on the right-propagating hole. }
			\label{Fig_exchage_core}
		\end{figure*}

\begin{figure*}
			\begin{minipage}[c]{0.65\textwidth}
				\includegraphics[width=\textwidth]{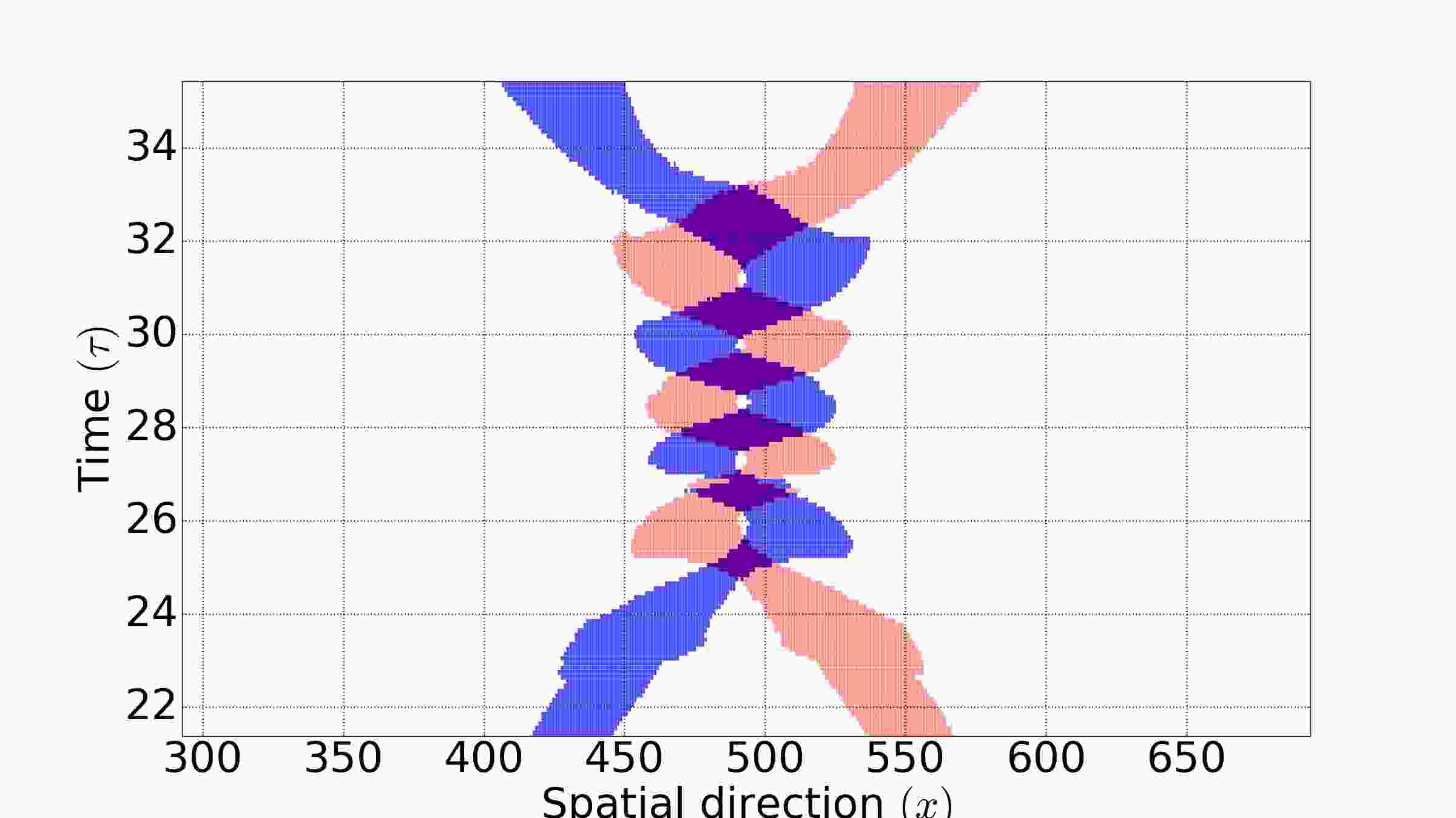}
			\end{minipage}\hfill
			\begin{minipage}[c]{0.35\textwidth}
				\caption{The trace of the cores of two nonlinear solutions in phase space are shown for 
				the case of  $M=1.1, \beta=0$, during the first collision ($20<\tau<40$).
				The core of the left-propagating hole (shown by red color) 
				rotates around the center of collective mass a few times during the collision.
				After collision, its place is exchanged with the right-propagating hole and it moves to the right. 
				The core of the right-propagating hole (shown by blue color) experience the exact opposite.
				A phase shift occurs which can interpreted as the time lag originating from the rotations. } \label{Fig_tracing_of_core}
			\end{minipage}
		\end{figure*}

		The number of rotations during the merging phase plays a crucial role in the internal structures of the 
		electron holes after the collision as well as during   
		the collision process itself. 
		Fig.~\ref{Fig_exchage_core} shows that in case of a non-integer number of rotations, 
		the two electron holes exchange
		the cores with each other.
		Fig.~\ref{Fig_tracing_of_core} presents the trace of cores for the case of $M=1.1, \beta=0$.
		During collision, each core makes $2.5$ rotations and then they change their cores.
		Phase shift during collision of solitons has been predicted from a few theoretical studies
		of two-soliton solutions in KdV regime. 
		Phase shift refers to the discontinuity in the time-line of soliton's propagation before and after collision. 
		This discontinuity can be observed in Fig.~\ref{Fig_tracing_of_core}, as the time line shifts up during collision.
		In other words, on a kinetic level, the phase shift happens due to the rotations during the collision and the time lag
		that it produces in the time-line of propagations. 
		
\section{Conclusions} \label{Sec_Conclusions}
		In this work, nonlinear solutions for the Vlasov-Poisson set of equations 
		were studied using a fully kinetic simulation approach. The solutions has been provided 
		by modeling the trapped population based on the Schamel approach. 
		Simulations are carried out in a plasma consisting of electrons and ions with 
		Maxwellian distribution functions. 
		The main point of inquiry focused on the collisional stability of these solutions and its
		dependency on the propagation velocity (Mach number) as a major parameter.
		The Mach number plays a crucial role in the nonlinear dispersion relation (NDR) of the nonlinear solutions.
		
		The kinetic energy of the plasma constituents are presented for different cases of stability. 
		The exchange of energy between electron and ions provides a reliable measure of the stability. 
		It is found that the solutions can be divided into three categories.
		(I) Stable solutions which are basically solitons of the Vlasov-Poisson equations.
		These are the solutions which can survive multiple mutual head-on collisions 
		(at least nine successive collisions are considered in our simulations).
		(II) Semi-stable solutions which can hold their shapes for a few collisions. 
		(III) Unstable solutions which get destroyed during the first collision. 
		It is demonstrated here that the solutions smoothly transform from stable to semi-stable and then to unstable forms by
		increasing the propagation velocity, i.e. increasing the Mach number from 1. 		

		Details of the exchange of energy between electrons and ions present 
		two different natures of dynamics. 
		Electron dynamics follows a linear superposition of kinetic energy during the collisions.
		However, ions do not follow a linear superposition principle. 
		The combined dynamics result in an exchange pattern of energy between electron and ions. 
		This exchange pattern turns out to be a reliable criterion to determine the stability of the solutions. 
		
		Our investigation shows that there exist a violation of the net-neutrality for the nonlinear solutions studied here.
		Violation of the net-neutrality results in a break-up of the energy balance of the nonlinear solutions. 
		This violation appears as a bipolar structure in the flux of electrons. 
		By increasing the Mach number of the solutions, the strength of this bipolar structure
		grows in comparison to the flux of ions.
		Therefore, it causes a stronger break-up of the energy balance sooner in the nonlinear solutions.
		Hence, the stability of the solutions decreases. 
		
		Furthermore, we have presented details of the electron phase space dynamics during collisions.
		These details show that the process of collision is built up of two major steps, namely merging and splitting. 
		During the merging process electron holes rotate around their collective center of mass. 
		The number of rotations depends on their relative speed. 
		As the relative speed increases, the time span of this phase of collision decreases. 
		However, the number of rotations increases simultaneously.
		Hence, different internal dynamics are possible.
		As a special case, it is shown that the collision of electron holes can be approximately 
		considered as an exchange of their cores with each other. 
		Furthermore, simulation results demonstrate a phase shift in the hole's propagation time-line before and after collision. 
		The phase shift can be interpreted as a time lag during collision in which the rotations of holes around
		their collective center of mass takes place. 

		Unstable solutions can be considered as a warning sign of over-interpreting the meaning of Sagdeev solutions. 
		Our simulation show that despite that the solutions originate from the Sagdeev approach, 
		not all of them are solitons or solitary waves, i.e. stable against collisions. 
		Hence, the stability of the Sagdeev solution has to be considered as a non-granted problem in itself
		when the Sagdeev pseudo-potential is employed. 
		There are many studies in multi-species plasmas which basically are attempts
		to find Sagdeev solutions to the set of fluid or kinetic equations of the system. 
		It seems that this approach has taken for granted that the Sagdeev solutions
		can be considered as stable solutions. 
		
		Our simulation results shows that balance of energy and momentum should be considered as 
		an important factor in the stability studies. 
		This study also hints that in constructing BGK solutions
		one should pay attention to the flux profile. 
		In order to construct BGK modes which are stable against mutual collisions, 
		initial conditions should be adjusted in order to avoid the formation of bipolar structures
		in the electron flux, or at least to decrease its amplitude. 

\acknowledgments
	G. Brodin and S. M. Hosseini Jenab would like to acknowledgment financial support 
	by the Swedish Research Council, grant number 2016-03806. 
	This work is supported by the Knut and Alice Wallenberg Foundation through the PLIONA project. 


\begin{thebibliography}{42}%
\makeatletter
\providecommand \@ifxundefined [1]{%
 \@ifx{#1\undefined}
}%
\providecommand \@ifnum [1]{%
 \ifnum #1\expandafter \@firstoftwo
 \else \expandafter \@secondoftwo
 \fi
}%
\providecommand \@ifx [1]{%
 \ifx #1\expandafter \@firstoftwo
 \else \expandafter \@secondoftwo
 \fi
}%
\providecommand \natexlab [1]{#1}%
\providecommand \enquote  [1]{``#1''}%
\providecommand \bibnamefont  [1]{#1}%
\providecommand \bibfnamefont [1]{#1}%
\providecommand \citenamefont [1]{#1}%
\providecommand \href@noop [0]{\@secondoftwo}%
\providecommand \href [0]{\begingroup \@sanitize@url \@href}%
\providecommand \@href[1]{\@@startlink{#1}\@@href}%
\providecommand \@@href[1]{\endgroup#1\@@endlink}%
\providecommand \@sanitize@url [0]{\catcode `\\12\catcode `\$12\catcode
  `\&12\catcode `\#12\catcode `\^12\catcode `\_12\catcode `\%12\relax}%
\providecommand \@@startlink[1]{}%
\providecommand \@@endlink[0]{}%
\providecommand \url  [0]{\begingroup\@sanitize@url \@url }%
\providecommand \@url [1]{\endgroup\@href {#1}{\urlprefix }}%
\providecommand \urlprefix  [0]{URL }%
\providecommand \Eprint [0]{\href }%
\providecommand \doibase [0]{http://dx.doi.org/}%
\providecommand \selectlanguage [0]{\@gobble}%
\providecommand \bibinfo  [0]{\@secondoftwo}%
\providecommand \bibfield  [0]{\@secondoftwo}%
\providecommand \translation [1]{[#1]}%
\providecommand \BibitemOpen [0]{}%
\providecommand \bibitemStop [0]{}%
\providecommand \bibitemNoStop [0]{.\EOS\space}%
\providecommand \EOS [0]{\spacefactor3000\relax}%
\providecommand \BibitemShut  [1]{\csname bibitem#1\endcsname}%
\let\auto@bib@innerbib\@empty
\bibitem [{\citenamefont {Abbasi}, \citenamefont {Jenab},\ and\ \citenamefont
  {Pajouh}(2011)}]{jenab2011preventing}%
  \BibitemOpen
  \bibfield  {author} {\bibinfo {author} {\bibnamefont {Abbasi}, \bibfnamefont
  {H.}}, \bibinfo {author} {\bibnamefont {Jenab}, \bibfnamefont {M.}}, \ and\
  \bibinfo {author} {\bibnamefont {Pajouh}, \bibfnamefont {H.~H.}},\
  }\href@noop {} {\bibfield  {journal} {\bibinfo  {journal} {Physical Review
  E}\ }\textbf {\bibinfo {volume} {84}},\ \bibinfo {pages} {036702} (\bibinfo
  {year} {2011})}\BibitemShut {NoStop}%
\bibitem [{\citenamefont {Aravindakshan}, \citenamefont {Kakad},\ and\
  \citenamefont {Kakad}(2018{\natexlab{a}})}]{aravindakshan2018bernstein}%
  \BibitemOpen
  \bibfield  {author} {\bibinfo {author} {\bibnamefont {Aravindakshan},
  \bibfnamefont {H.}}, \bibinfo {author} {\bibnamefont {Kakad}, \bibfnamefont
  {A.}}, \ and\ \bibinfo {author} {\bibnamefont {Kakad}, \bibfnamefont {B.}},\
  }\href@noop {} {\bibfield  {journal} {\bibinfo  {journal} {Physics of
  Plasmas}\ }\textbf {\bibinfo {volume} {25}},\ \bibinfo {pages} {052901}
  (\bibinfo {year} {2018}{\natexlab{a}})}\BibitemShut {NoStop}%
\bibitem [{\citenamefont {Aravindakshan}, \citenamefont {Kakad},\ and\
  \citenamefont {Kakad}(2018{\natexlab{b}})}]{aravindakshan2018effects}%
  \BibitemOpen
  \bibfield  {author} {\bibinfo {author} {\bibnamefont {Aravindakshan},
  \bibfnamefont {H.}}, \bibinfo {author} {\bibnamefont {Kakad}, \bibfnamefont
  {A.}}, \ and\ \bibinfo {author} {\bibnamefont {Kakad}, \bibfnamefont {B.}},\
  }\href@noop {} {\bibfield  {journal} {\bibinfo  {journal} {Physics of
  Plasmas}\ }\textbf {\bibinfo {volume} {25}},\ \bibinfo {pages} {122901}
  (\bibinfo {year} {2018}{\natexlab{b}})}\BibitemShut {NoStop}%
\bibitem [{\citenamefont {Bale}\ \emph {et~al.}(1998)\citenamefont {Bale},
  \citenamefont {Kellogg}, \citenamefont {Larsen}, \citenamefont {Lin},
  \citenamefont {Goetz},\ and\ \citenamefont {Lepping}}]{bale1998bipolar}%
  \BibitemOpen
  \bibfield  {author} {\bibinfo {author} {\bibnamefont {Bale}, \bibfnamefont
  {S.}}, \bibinfo {author} {\bibnamefont {Kellogg}, \bibfnamefont {P.}},
  \bibinfo {author} {\bibnamefont {Larsen}, \bibfnamefont {D.}}, \bibinfo
  {author} {\bibnamefont {Lin}, \bibfnamefont {R.}}, \bibinfo {author}
  {\bibnamefont {Goetz}, \bibfnamefont {K.}}, \ and\ \bibinfo {author}
  {\bibnamefont {Lepping}, \bibfnamefont {R.}},\ }\href@noop {} {\bibfield
  {journal} {\bibinfo  {journal} {Geophysical research letters}\ }\textbf
  {\bibinfo {volume} {25}},\ \bibinfo {pages} {2929} (\bibinfo {year}
  {1998})}\BibitemShut {NoStop}%
\bibitem [{\citenamefont {Bernstein}, \citenamefont {Greene},\ and\
  \citenamefont {Kruskal}(1957)}]{bernstein1957exact}%
  \BibitemOpen
  \bibfield  {author} {\bibinfo {author} {\bibnamefont {Bernstein},
  \bibfnamefont {I.~B.}}, \bibinfo {author} {\bibnamefont {Greene},
  \bibfnamefont {J.~M.}}, \ and\ \bibinfo {author} {\bibnamefont {Kruskal},
  \bibfnamefont {M.~D.}},\ }\href@noop {} {\bibfield  {journal} {\bibinfo
  {journal} {Physical Review}\ }\textbf {\bibinfo {volume} {108}},\ \bibinfo
  {pages} {546} (\bibinfo {year} {1957})}\BibitemShut {NoStop}%
\bibitem [{\citenamefont {Bohm}\ and\ \citenamefont
  {Gross}(1949)}]{bohm1949theory}%
  \BibitemOpen
  \bibfield  {author} {\bibinfo {author} {\bibnamefont {Bohm}, \bibfnamefont
  {D.}}\ and\ \bibinfo {author} {\bibnamefont {Gross}, \bibfnamefont {E.~P.}},\
  }\href@noop {} {\bibfield  {journal} {\bibinfo  {journal} {Physical Review}\
  }\textbf {\bibinfo {volume} {75}},\ \bibinfo {pages} {1851} (\bibinfo {year}
  {1949})}\BibitemShut {NoStop}%
\bibitem [{\citenamefont {Davis}, \citenamefont {Lust},\ and\ \citenamefont
  {Schl{\"u}ter}(1958)}]{davis1958structure}%
  \BibitemOpen
  \bibfield  {author} {\bibinfo {author} {\bibnamefont {Davis}, \bibfnamefont
  {L.}}, \bibinfo {author} {\bibnamefont {Lust}, \bibfnamefont {R.}}, \ and\
  \bibinfo {author} {\bibnamefont {Schl{\"u}ter}, \bibfnamefont {A.}},\
  }\href@noop {} {\bibfield  {journal} {\bibinfo  {journal} {Zeitschrift
  f{\"u}r Naturforschung A}\ }\textbf {\bibinfo {volume} {13}},\ \bibinfo
  {pages} {916} (\bibinfo {year} {1958})}\BibitemShut {NoStop}%
\bibitem [{\citenamefont {Deng}\ \emph {et~al.}(2006)\citenamefont {Deng},
  \citenamefont {Tang}, \citenamefont {Matsumoto}, \citenamefont {Pickett},
  \citenamefont {Fazakerley}, \citenamefont {Kojima}, \citenamefont
  {Baumjohann}, \citenamefont {Coates}, \citenamefont {Nakamura}, \citenamefont
  {Gurnett} \emph {et~al.}}]{deng2006observations}%
  \BibitemOpen
  \bibfield  {author} {\bibinfo {author} {\bibnamefont {Deng}, \bibfnamefont
  {X.}}, \bibinfo {author} {\bibnamefont {Tang}, \bibfnamefont {R.}}, \bibinfo
  {author} {\bibnamefont {Matsumoto}, \bibfnamefont {H.}}, \bibinfo {author}
  {\bibnamefont {Pickett}, \bibfnamefont {J.}}, \bibinfo {author} {\bibnamefont
  {Fazakerley}, \bibfnamefont {A.}}, \bibinfo {author} {\bibnamefont {Kojima},
  \bibfnamefont {H.}}, \bibinfo {author} {\bibnamefont {Baumjohann},
  \bibfnamefont {W.}}, \bibinfo {author} {\bibnamefont {Coates}, \bibfnamefont
  {A.}}, \bibinfo {author} {\bibnamefont {Nakamura}, \bibfnamefont {R.}},
  \bibinfo {author} {\bibnamefont {Gurnett}, \bibfnamefont {D.}},  \emph
  {et~al.},\ }\href@noop {} {\bibfield  {journal} {\bibinfo  {journal}
  {Advances in Space Research}\ }\textbf {\bibinfo {volume} {37}},\ \bibinfo
  {pages} {1373} (\bibinfo {year} {2006})}\BibitemShut {NoStop}%
\bibitem [{\citenamefont {Eliasson}\ and\ \citenamefont
  {Shukla}(2006)}]{eliasson2006formation}%
  \BibitemOpen
  \bibfield  {author} {\bibinfo {author} {\bibnamefont {Eliasson},
  \bibfnamefont {B.}}\ and\ \bibinfo {author} {\bibnamefont {Shukla},
  \bibfnamefont {P.~K.}},\ }\href@noop {} {\bibfield  {journal} {\bibinfo
  {journal} {Physics reports}\ }\textbf {\bibinfo {volume} {422}},\ \bibinfo
  {pages} {225} (\bibinfo {year} {2006})}\BibitemShut {NoStop}%
\bibitem [{\citenamefont {Elskens}, \citenamefont {Escande},\ and\
  \citenamefont {Doveil}(2014)}]{elskens2014vlasov}%
  \BibitemOpen
  \bibfield  {author} {\bibinfo {author} {\bibnamefont {Elskens}, \bibfnamefont
  {Y.}}, \bibinfo {author} {\bibnamefont {Escande}, \bibfnamefont {D.~F.}}, \
  and\ \bibinfo {author} {\bibnamefont {Doveil}, \bibfnamefont {F.}},\
  }\href@noop {} {\bibfield  {journal} {\bibinfo  {journal} {arXiv preprint
  arXiv:1403.0056}\ } (\bibinfo {year} {2014})}\BibitemShut {NoStop}%
\bibitem [{\citenamefont {Ergun}\ \emph {et~al.}(1998)\citenamefont {Ergun},
  \citenamefont {Carlson}, \citenamefont {McFadden}, \citenamefont {Mozer},
  \citenamefont {Delory}, \citenamefont {Peria}, \citenamefont {Chaston},
  \citenamefont {Temerin}, \citenamefont {Elphic}, \citenamefont {Strangeway}
  \emph {et~al.}}]{ergun1998fast}%
  \BibitemOpen
  \bibfield  {author} {\bibinfo {author} {\bibnamefont {Ergun}, \bibfnamefont
  {R.}}, \bibinfo {author} {\bibnamefont {Carlson}, \bibfnamefont {C.}},
  \bibinfo {author} {\bibnamefont {McFadden}, \bibfnamefont {J.}}, \bibinfo
  {author} {\bibnamefont {Mozer}, \bibfnamefont {F.}}, \bibinfo {author}
  {\bibnamefont {Delory}, \bibfnamefont {G.}}, \bibinfo {author} {\bibnamefont
  {Peria}, \bibfnamefont {W.}}, \bibinfo {author} {\bibnamefont {Chaston},
  \bibfnamefont {C.}}, \bibinfo {author} {\bibnamefont {Temerin}, \bibfnamefont
  {M.}}, \bibinfo {author} {\bibnamefont {Elphic}, \bibfnamefont {R.}},
  \bibinfo {author} {\bibnamefont {Strangeway}, \bibfnamefont {R.}},  \emph
  {et~al.},\ }\href@noop {} {\bibfield  {journal} {\bibinfo  {journal}
  {Geophysical research letters}\ }\textbf {\bibinfo {volume} {25}},\ \bibinfo
  {pages} {2025} (\bibinfo {year} {1998})}\BibitemShut {NoStop}%
\bibitem [{\citenamefont {Fijalkow}\ and\ \citenamefont
  {Nocera}(2003)}]{fijalkow2003phase}%
  \BibitemOpen
  \bibfield  {author} {\bibinfo {author} {\bibnamefont {Fijalkow},
  \bibfnamefont {E.}}\ and\ \bibinfo {author} {\bibnamefont {Nocera},
  \bibfnamefont {L.}},\ }\href@noop {} {\bibfield  {journal} {\bibinfo
  {journal} {Journal of plasma physics}\ }\textbf {\bibinfo {volume} {69}},\
  \bibinfo {pages} {93} (\bibinfo {year} {2003})}\BibitemShut {NoStop}%
\bibitem [{\citenamefont {Franz}, \citenamefont {Kintner},\ and\ \citenamefont
  {Pickett}(1998)}]{franz1998polar}%
  \BibitemOpen
  \bibfield  {author} {\bibinfo {author} {\bibnamefont {Franz}, \bibfnamefont
  {J.~R.}}, \bibinfo {author} {\bibnamefont {Kintner}, \bibfnamefont {P.~M.}},
  \ and\ \bibinfo {author} {\bibnamefont {Pickett}, \bibfnamefont {J.~S.}},\
  }\href@noop {} {\bibfield  {journal} {\bibinfo  {journal} {Geophysical
  research letters}\ }\textbf {\bibinfo {volume} {25}},\ \bibinfo {pages}
  {1277} (\bibinfo {year} {1998})}\BibitemShut {NoStop}%
\bibitem [{\citenamefont {Ghizzo}\ \emph {et~al.}(1988)\citenamefont {Ghizzo},
  \citenamefont {Izrar}, \citenamefont {Bertrand}, \citenamefont {Fijalkow},
  \citenamefont {Feix},\ and\ \citenamefont {Shoucri}}]{ghizzo1988stability}%
  \BibitemOpen
  \bibfield  {author} {\bibinfo {author} {\bibnamefont {Ghizzo}, \bibfnamefont
  {A.}}, \bibinfo {author} {\bibnamefont {Izrar}, \bibfnamefont {B.}}, \bibinfo
  {author} {\bibnamefont {Bertrand}, \bibfnamefont {P.}}, \bibinfo {author}
  {\bibnamefont {Fijalkow}, \bibfnamefont {E.}}, \bibinfo {author}
  {\bibnamefont {Feix}, \bibfnamefont {M.}}, \ and\ \bibinfo {author}
  {\bibnamefont {Shoucri}, \bibfnamefont {M.}},\ }\href@noop {} {\bibfield
  {journal} {\bibinfo  {journal} {Physics of Fluids (1958-1988)}\ }\textbf
  {\bibinfo {volume} {31}},\ \bibinfo {pages} {72} (\bibinfo {year}
  {1988})}\BibitemShut {NoStop}%
\bibitem [{\citenamefont {Hosseini~Jenab}\ and\ \citenamefont
  {Spanier}(2016)}]{jenab2016IASWs}%
  \BibitemOpen
  \bibfield  {author} {\bibinfo {author} {\bibnamefont {Hosseini~Jenab},
  \bibfnamefont {S.~M.}}\ and\ \bibinfo {author} {\bibnamefont {Spanier},
  \bibfnamefont {F.}},\ }\href@noop {} {\bibfield  {journal} {\bibinfo
  {journal} {Physics of Plasmas}\ }\textbf {\bibinfo {volume} {23}},\ \bibinfo
  {pages} {102306} (\bibinfo {year} {2016})}\BibitemShut {NoStop}%
\bibitem [{\citenamefont {Hosseini~Jenab}\ and\ \citenamefont
  {Spanier}(2017)}]{jenab2017fully}%
  \BibitemOpen
  \bibfield  {author} {\bibinfo {author} {\bibnamefont {Hosseini~Jenab},
  \bibfnamefont {S.~M.}}\ and\ \bibinfo {author} {\bibnamefont {Spanier},
  \bibfnamefont {F.}},\ }\href@noop {} {\bibfield  {journal} {\bibinfo
  {journal} {Physical Review E}\ }\textbf {\bibinfo {volume} {95}},\ \bibinfo
  {pages} {053201} (\bibinfo {year} {2017})}\BibitemShut {NoStop}%
\bibitem [{\citenamefont {Hutchinson}(2017)}]{hutchinson2017electron}%
  \BibitemOpen
  \bibfield  {author} {\bibinfo {author} {\bibnamefont {Hutchinson},
  \bibfnamefont {I.}},\ }\href@noop {} {\bibfield  {journal} {\bibinfo
  {journal} {Physics of Plasmas}\ }\textbf {\bibinfo {volume} {24}},\ \bibinfo
  {pages} {055601} (\bibinfo {year} {2017})}\BibitemShut {NoStop}%
\bibitem [{\citenamefont {Infeld}\ and\ \citenamefont
  {Rowlands}(2000)}]{infeld2000nonlinear}%
  \BibitemOpen
  \bibfield  {author} {\bibinfo {author} {\bibnamefont {Infeld}, \bibfnamefont
  {E.}}\ and\ \bibinfo {author} {\bibnamefont {Rowlands}, \bibfnamefont {G.}},\
  }\href@noop {} {\emph {\bibinfo {title} {Nonlinear waves, solitons and
  chaos}}}\ (\bibinfo  {publisher} {Cambridge university press},\ \bibinfo
  {year} {2000})\BibitemShut {NoStop}%
\bibitem [{\citenamefont {Kakad}, \citenamefont {Kakad},\ and\ \citenamefont
  {Omura}(2017)}]{kakad2017formation}%
  \BibitemOpen
  \bibfield  {author} {\bibinfo {author} {\bibnamefont {Kakad}, \bibfnamefont
  {A.}}, \bibinfo {author} {\bibnamefont {Kakad}, \bibfnamefont {B.}}, \ and\
  \bibinfo {author} {\bibnamefont {Omura}, \bibfnamefont {Y.}},\ }\href@noop {}
  {\bibfield  {journal} {\bibinfo  {journal} {Physics of Plasmas}\ }\textbf
  {\bibinfo {volume} {24}},\ \bibinfo {pages} {060704} (\bibinfo {year}
  {2017})}\BibitemShut {NoStop}%
\bibitem [{\citenamefont {Kazeminezhad}, \citenamefont {Kuhn},\ and\
  \citenamefont {Tavakoli}(2003)}]{kazeminezhad2003vlasov}%
  \BibitemOpen
  \bibfield  {author} {\bibinfo {author} {\bibnamefont {Kazeminezhad},
  \bibfnamefont {F.}}, \bibinfo {author} {\bibnamefont {Kuhn}, \bibfnamefont
  {S.}}, \ and\ \bibinfo {author} {\bibnamefont {Tavakoli}, \bibfnamefont
  {A.}},\ }\href@noop {} {\bibfield  {journal} {\bibinfo  {journal} {Physical
  Review E}\ }\textbf {\bibinfo {volume} {67}},\ \bibinfo {pages} {026704}
  (\bibinfo {year} {2003})}\BibitemShut {NoStop}%
\bibitem [{\citenamefont {Kilian}, \citenamefont {Schreiner},\ and\
  \citenamefont {Spanier}(2018)}]{kilian2018afterlive}%
  \BibitemOpen
  \bibfield  {author} {\bibinfo {author} {\bibnamefont {Kilian}, \bibfnamefont
  {P.}}, \bibinfo {author} {\bibnamefont {Schreiner}, \bibfnamefont {C.}}, \
  and\ \bibinfo {author} {\bibnamefont {Spanier}, \bibfnamefont {F.}},\
  }\href@noop {} {\bibfield  {journal} {\bibinfo  {journal} {Computer Physics
  Communications}\ } (\bibinfo {year} {2018})}\BibitemShut {NoStop}%
\bibitem [{\citenamefont {Kojima}\ \emph {et~al.}(1997)\citenamefont {Kojima},
  \citenamefont {Matsumoto}, \citenamefont {Chikuba}, \citenamefont {Horiyama},
  \citenamefont {Ashour-Abdalla},\ and\ \citenamefont
  {Anderson}}]{kojima1997geotail}%
  \BibitemOpen
  \bibfield  {author} {\bibinfo {author} {\bibnamefont {Kojima}, \bibfnamefont
  {H.}}, \bibinfo {author} {\bibnamefont {Matsumoto}, \bibfnamefont {H.}},
  \bibinfo {author} {\bibnamefont {Chikuba}, \bibfnamefont {S.}}, \bibinfo
  {author} {\bibnamefont {Horiyama}, \bibfnamefont {S.}}, \bibinfo {author}
  {\bibnamefont {Ashour-Abdalla}, \bibfnamefont {M.}}, \ and\ \bibinfo {author}
  {\bibnamefont {Anderson}, \bibfnamefont {R.}},\ }\href@noop {} {\bibfield
  {journal} {\bibinfo  {journal} {Journal of Geophysical Research: Space
  Physics}\ }\textbf {\bibinfo {volume} {102}},\ \bibinfo {pages} {14439}
  (\bibinfo {year} {1997})}\BibitemShut {NoStop}%
\bibitem [{\citenamefont {Lotekar}, \citenamefont {Kakad},\ and\ \citenamefont
  {Kakad}(2017)}]{lotekar2017generation}%
  \BibitemOpen
  \bibfield  {author} {\bibinfo {author} {\bibnamefont {Lotekar}, \bibfnamefont
  {A.}}, \bibinfo {author} {\bibnamefont {Kakad}, \bibfnamefont {A.}}, \ and\
  \bibinfo {author} {\bibnamefont {Kakad}, \bibfnamefont {B.}},\ }\href@noop {}
  {\bibfield  {journal} {\bibinfo  {journal} {Physics of Plasmas}\ }\textbf
  {\bibinfo {volume} {24}},\ \bibinfo {pages} {102127} (\bibinfo {year}
  {2017})}\BibitemShut {NoStop}%
\bibitem [{\citenamefont {Lynov}\ \emph {et~al.}(1980)\citenamefont {Lynov},
  \citenamefont {Michelsen}, \citenamefont {P{\'e}cseli},\ and\ \citenamefont
  {Rasmussen}}]{lynov1980interaction}%
  \BibitemOpen
  \bibfield  {author} {\bibinfo {author} {\bibnamefont {Lynov}, \bibfnamefont
  {J.-P.}}, \bibinfo {author} {\bibnamefont {Michelsen}, \bibfnamefont {P.}},
  \bibinfo {author} {\bibnamefont {P{\'e}cseli}, \bibfnamefont {H.}}, \ and\
  \bibinfo {author} {\bibnamefont {Rasmussen}, \bibfnamefont {J.~J.}},\
  }\href@noop {} {\bibfield  {journal} {\bibinfo  {journal} {Physics Letters
  A}\ }\textbf {\bibinfo {volume} {80}},\ \bibinfo {pages} {23} (\bibinfo
  {year} {1980})}\BibitemShut {NoStop}%
\bibitem [{\citenamefont {Lynov}\ \emph {et~al.}(1985)\citenamefont {Lynov},
  \citenamefont {Michelsen}, \citenamefont {P{\'e}cseli}, \citenamefont
  {Rasmussen},\ and\ \citenamefont {S{\o}rensen}}]{lynov1985phase}%
  \BibitemOpen
  \bibfield  {author} {\bibinfo {author} {\bibnamefont {Lynov}, \bibfnamefont
  {J.-P.}}, \bibinfo {author} {\bibnamefont {Michelsen}, \bibfnamefont {P.}},
  \bibinfo {author} {\bibnamefont {P{\'e}cseli}, \bibfnamefont {H.}}, \bibinfo
  {author} {\bibnamefont {Rasmussen}, \bibfnamefont {J.~J.}}, \ and\ \bibinfo
  {author} {\bibnamefont {S{\o}rensen}, \bibfnamefont {S.}},\ }\href@noop {}
  {\bibfield  {journal} {\bibinfo  {journal} {Physica Scripta}\ }\textbf
  {\bibinfo {volume} {31}},\ \bibinfo {pages} {596} (\bibinfo {year}
  {1985})}\BibitemShut {NoStop}%
\bibitem [{\citenamefont {Matsumoto}\ \emph {et~al.}(1994)\citenamefont
  {Matsumoto}, \citenamefont {Kojima}, \citenamefont {Miyatake}, \citenamefont
  {Omura}, \citenamefont {Okada}, \citenamefont {Nagano},\ and\ \citenamefont
  {Tsutsui}}]{matsumoto1994electrostatic}%
  \BibitemOpen
  \bibfield  {author} {\bibinfo {author} {\bibnamefont {Matsumoto},
  \bibfnamefont {H.}}, \bibinfo {author} {\bibnamefont {Kojima}, \bibfnamefont
  {H.}}, \bibinfo {author} {\bibnamefont {Miyatake}, \bibfnamefont {T.}},
  \bibinfo {author} {\bibnamefont {Omura}, \bibfnamefont {Y.}}, \bibinfo
  {author} {\bibnamefont {Okada}, \bibfnamefont {M.}}, \bibinfo {author}
  {\bibnamefont {Nagano}, \bibfnamefont {I.}}, \ and\ \bibinfo {author}
  {\bibnamefont {Tsutsui}, \bibfnamefont {M.}},\ }\href@noop {} {\bibfield
  {journal} {\bibinfo  {journal} {Geophysical Research Letters}\ }\textbf
  {\bibinfo {volume} {21}},\ \bibinfo {pages} {2915} (\bibinfo {year}
  {1994})}\BibitemShut {NoStop}%
\bibitem [{\citenamefont {Montgomery}\ and\ \citenamefont
  {Joyce}(1969)}]{montgomery1969shock}%
  \BibitemOpen
  \bibfield  {author} {\bibinfo {author} {\bibnamefont {Montgomery},
  \bibfnamefont {D.}}\ and\ \bibinfo {author} {\bibnamefont {Joyce},
  \bibfnamefont {G.}},\ }\href@noop {} {\bibfield  {journal} {\bibinfo
  {journal} {Journal of Plasma Physics}\ }\textbf {\bibinfo {volume} {3}},\
  \bibinfo {pages} {1} (\bibinfo {year} {1969})}\BibitemShut {NoStop}%
\bibitem [{\citenamefont {Muschietti}\ \emph {et~al.}(1999)\citenamefont
  {Muschietti}, \citenamefont {Roth}, \citenamefont {Ergun},\ and\
  \citenamefont {Carlson}}]{muschietti1999analysis}%
  \BibitemOpen
  \bibfield  {author} {\bibinfo {author} {\bibnamefont {Muschietti},
  \bibfnamefont {L.}}, \bibinfo {author} {\bibnamefont {Roth}, \bibfnamefont
  {I.}}, \bibinfo {author} {\bibnamefont {Ergun}, \bibfnamefont {R.}}, \ and\
  \bibinfo {author} {\bibnamefont {Carlson}, \bibfnamefont {C.}},\ }\href@noop
  {} {\bibfield  {journal} {\bibinfo  {journal} {Nonlinear Processes in
  Geophysics}\ }\textbf {\bibinfo {volume} {6}},\ \bibinfo {pages} {211}
  (\bibinfo {year} {1999})}\BibitemShut {NoStop}%
\bibitem [{\citenamefont {Nunn}(1993)}]{nunn1993novel}%
  \BibitemOpen
  \bibfield  {author} {\bibinfo {author} {\bibnamefont {Nunn}, \bibfnamefont
  {D.}},\ }\href@noop {} {\bibfield  {journal} {\bibinfo  {journal} {Journal of
  Computational Physics}\ }\textbf {\bibinfo {volume} {108}},\ \bibinfo {pages}
  {180} (\bibinfo {year} {1993})}\BibitemShut {NoStop}%
\bibitem [{\citenamefont {Rasmussen}(1982)}]{rasmussen1982effects}%
  \BibitemOpen
  \bibfield  {author} {\bibinfo {author} {\bibnamefont {Rasmussen},
  \bibfnamefont {J.~J.}},\ }\href@noop {} {\bibfield  {journal} {\bibinfo
  {journal} {Physica Scripta}\ }\textbf {\bibinfo {volume} {1982}},\ \bibinfo
  {pages} {29} (\bibinfo {year} {1982})}\BibitemShut {NoStop}%
\bibitem [{\citenamefont {Saeki}\ and\ \citenamefont
  {Genma}(1998)}]{saeki1998electron}%
  \BibitemOpen
  \bibfield  {author} {\bibinfo {author} {\bibnamefont {Saeki}, \bibfnamefont
  {K.}}\ and\ \bibinfo {author} {\bibnamefont {Genma}, \bibfnamefont {H.}},\
  }\href@noop {} {\bibfield  {journal} {\bibinfo  {journal} {Physical review
  letters}\ }\textbf {\bibinfo {volume} {80}},\ \bibinfo {pages} {1224}
  (\bibinfo {year} {1998})}\BibitemShut {NoStop}%
\bibitem [{\citenamefont {Sagdeev}(1966)}]{sagdeev1966cooperative}%
  \BibitemOpen
  \bibfield  {author} {\bibinfo {author} {\bibnamefont {Sagdeev}, \bibfnamefont
  {R.}},\ }\href@noop {} {\bibfield  {journal} {\bibinfo  {journal} {Rev plasma
  phys}\ }\textbf {\bibinfo {volume} {4}},\ \bibinfo {pages} {23} (\bibinfo
  {year} {1966})}\BibitemShut {NoStop}%
\bibitem [{\citenamefont {Schamel}(1971)}]{schamel1971stationary}%
  \BibitemOpen
  \bibfield  {author} {\bibinfo {author} {\bibnamefont {Schamel}, \bibfnamefont
  {H.}},\ }\href@noop {} {\bibfield  {journal} {\bibinfo  {journal} {Plasma
  Physics}\ }\textbf {\bibinfo {volume} {13}},\ \bibinfo {pages} {491}
  (\bibinfo {year} {1971})}\BibitemShut {NoStop}%
\bibitem [{\citenamefont {Schamel}(1972{\natexlab{a}})}]{schamel_2}%
  \BibitemOpen
  \bibfield  {author} {\bibinfo {author} {\bibnamefont {Schamel}, \bibfnamefont
  {H.}},\ }\href@noop {} {\bibfield  {journal} {\bibinfo  {journal} {Journal of
  Plasma Physics}\ }\textbf {\bibinfo {volume} {7}},\ \bibinfo {pages} {1}
  (\bibinfo {year} {1972}{\natexlab{a}})}\BibitemShut {NoStop}%
\bibitem [{\citenamefont {Schamel}(1972{\natexlab{b}})}]{schamel_3}%
  \BibitemOpen
  \bibfield  {author} {\bibinfo {author} {\bibnamefont {Schamel}, \bibfnamefont
  {H.}},\ }\href@noop {} {\bibfield  {journal} {\bibinfo  {journal} {Plasma
  Physics}\ }\textbf {\bibinfo {volume} {14}},\ \bibinfo {pages} {905}
  (\bibinfo {year} {1972}{\natexlab{b}})}\BibitemShut {NoStop}%
\bibitem [{\citenamefont {Schamel}(2000)}]{schamel2000hole}%
  \BibitemOpen
  \bibfield  {author} {\bibinfo {author} {\bibnamefont {Schamel}, \bibfnamefont
  {H.}},\ }\href@noop {} {\bibfield  {journal} {\bibinfo  {journal} {Physics of
  Plasmas}\ }\textbf {\bibinfo {volume} {7}},\ \bibinfo {pages} {4831}
  (\bibinfo {year} {2000})}\BibitemShut {NoStop}%
\bibitem [{\citenamefont {Schamel}(2012)}]{schamel2012cnoidal}%
  \BibitemOpen
  \bibfield  {author} {\bibinfo {author} {\bibnamefont {Schamel}, \bibfnamefont
  {H.}},\ }\href@noop {} {\bibfield  {journal} {\bibinfo  {journal} {Physics of
  Plasmas}\ }\textbf {\bibinfo {volume} {19}},\ \bibinfo {pages} {020501}
  (\bibinfo {year} {2012})}\BibitemShut {NoStop}%
\bibitem [{\citenamefont {Shoucri}(2017)}]{shoucri2017formation}%
  \BibitemOpen
  \bibfield  {author} {\bibinfo {author} {\bibnamefont {Shoucri}, \bibfnamefont
  {M.}},\ }\href@noop {} {\bibfield  {journal} {\bibinfo  {journal} {Laser and
  Particle Beams}\ }\textbf {\bibinfo {volume} {35}},\ \bibinfo {pages} {706}
  (\bibinfo {year} {2017})}\BibitemShut {NoStop}%
\bibitem [{\citenamefont {Smith}(1970{\natexlab{a}})}]{smith1970exact}%
  \BibitemOpen
  \bibfield  {author} {\bibinfo {author} {\bibnamefont {Smith}, \bibfnamefont
  {A.}},\ }\href@noop {} {\bibfield  {journal} {\bibinfo  {journal} {Journal of
  Plasma Physics}\ }\textbf {\bibinfo {volume} {4}},\ \bibinfo {pages} {549}
  (\bibinfo {year} {1970}{\natexlab{a}})}\BibitemShut {NoStop}%
\bibitem [{\citenamefont {Smith}(1970{\natexlab{b}})}]{smith1970steady}%
  \BibitemOpen
  \bibfield  {author} {\bibinfo {author} {\bibnamefont {Smith}, \bibfnamefont
  {A.}},\ }\href@noop {} {\bibfield  {journal} {\bibinfo  {journal} {Journal of
  Plasma Physics}\ }\textbf {\bibinfo {volume} {4}},\ \bibinfo {pages} {511}
  (\bibinfo {year} {1970}{\natexlab{b}})}\BibitemShut {NoStop}%
\bibitem [{\citenamefont {Turikov}(1984)}]{turikov1984electron}%
  \BibitemOpen
  \bibfield  {author} {\bibinfo {author} {\bibnamefont {Turikov}, \bibfnamefont
  {V.}},\ }\href@noop {} {\bibfield  {journal} {\bibinfo  {journal} {Physica
  Scripta}\ }\textbf {\bibinfo {volume} {30}},\ \bibinfo {pages} {73} (\bibinfo
  {year} {1984})}\BibitemShut {NoStop}%
\bibitem [{\citenamefont {Zhou}\ and\ \citenamefont
  {Hutchinson}(2018)}]{zhou2018dynamics}%
  \BibitemOpen
  \bibfield  {author} {\bibinfo {author} {\bibnamefont {Zhou}, \bibfnamefont
  {C.}}\ and\ \bibinfo {author} {\bibnamefont {Hutchinson}, \bibfnamefont
  {I.~H.}},\ }\href@noop {} {\bibfield  {journal} {\bibinfo  {journal} {Physics
  of Plasmas}\ }\textbf {\bibinfo {volume} {25}},\ \bibinfo {pages} {082303}
  (\bibinfo {year} {2018})}\BibitemShut {NoStop}%
\end{thebibliography}
%


\end{document}